\documentclass[aps,prd,twocolumn,superscriptaddress,nofootinbib]{revtex4-1}

\usepackage{amsmath}
\usepackage{graphicx}
\usepackage{hyperref}
\usepackage{siunitx}
\usepackage{kbordermatrix}
\usepackage{relsize}
\usepackage{natbib}
\usepackage{multirow}
\usepackage{float}
\usepackage{colortbl}
\restylefloat{figure}

\newcolumntype{A}{ >{$} r <{$} @{} >{${}} l <{$} }

\begin{document}

\title{Endowing $\mathbf{\Lambda}$ with a dynamic nature: constraints in a spatially curved Universe}

\author{Christine R. Farrugia}
\email[]{christine.r.farrugia@um.edu.mt}
\affiliation{Department of Mathematics, Faculty of Science, University of Malta, Msida MSD 2080, Malta}
\author{Joseph Sultana}
\email[]{joseph.sultana@um.edu.mt}
\affiliation{Department of Mathematics, Faculty of Science, University of Malta, Msida MSD 2080, Malta}
\author{Jurgen Mifsud}
\email[]{jurgenmifsud@kasi.re.kr}
\affiliation{Consortium for Fundamental Physics, School of Mathematics and Statistics, University of Sheffield, Hounsfield Road, Sheffield S3 7RH, UK}
\affiliation{Korea Astronomy and Space Science Institute, 776 Daedeokdae--ro, Yuseong--gu, Daejeon 34055, Republic of Korea}

\date{\today}

\begin{abstract}
In this study, we consider three dark energy models in which $\Lambda$ is not constant, but has a dynamic nature that depends on the Hubble parameter $H$ and/or its time derivative $\dot{H}$. We analyze the generalized running vacuum model, for which $\Lambda(H)=A+BH^2+C\dot{H}$, along with the two models obtained by setting $B$ or $C$ equal to zero. A null value for $C$ yields the classical running vacuum model (RVM), while $B=0$ corresponds to what we term the generalized running vacuum sub-case, or GRVS. Our main aim is to investigate whether these models can accommodate non-zero spatial curvature. To this end, we carry out a Markov Chain Monte Carlo analysis using data for the observables associated with Type-Ia supernovae, cosmic chronometers, the cosmic microwave background and baryon acoustic oscillations, as well as two values for the Hubble constant. Then we include data relating to the growth of large-scale structure (LSS) and repeat the procedure. Our results indicate that taking LSS observations into account helps to tighten constraints and determine a definite sign for the model parameters. In the case of the RVM and GRVS, the addition of growth data results in dynamical vacuum energy being preferred to a cosmological constant at a little over $1\sigma$. This happens in both the flat and non-flat scenarios -- there are only a few exceptions -- but comes at the cost of an extra parameter which can degrade the performance of the models (as assessed by model selection criteria). Of special relevance is the fact that the inclusion of LSS data appears to increase compatibility with a flat geometry. It also brings the constraints on the Hubble constant closer to the range of values established by \emph{Planck}.
\end{abstract}

\maketitle
\section{Introduction}
\label{sec:intro}
Twenty years after the scientific community collectively acknowledged the existence of dark energy, its nature is still as elusive as ever. This despite the plethora of models that have been put forward \cite{Miao2011, Yoo2012} in an attempt to explain why the Universe seems to be expanding at an accelerated rate \cite{Riess1998, Perlmutter1999, Schmidt1998}, the phenomenon that first brought dark energy -- whose negative pressure is supposed to be responsible for the said acceleration -- to the forefront of cosmological research. 

The initial tentative explanation of dark energy took the form of a cosmological constant $\Lambda$ included in the field equations that underlie General Relativity. This is not to say that the concept of a cosmological constant emerged two decades ago. Indeed, $\Lambda$ had been introduced into General Relativity by Einstein himself to ensure a quasi-static distribution of matter \cite{Einstein1917}. Once the Universe was discovered to be expanding at an increasing rate, however, $\Lambda$ seemed to provide the means by which cosmic acceleration could be accounted for \cite{Riess1998, Tegmark1998}. The resulting cosmology is known as $\Lambda$CDM ($\Lambda + $Cold Dark Matter).  In it, the role of dark energy is played by the energy of the vacuum, whose density is hypothesized to remain constant as the Universe expands. Consequently, it begins to dominate the energy budget of the cosmos when the densities of matter and radiation have been sufficiently diluted. 

$\Lambda$CDM is arguably still the most popular among the many dark energy models that have been proposed. Most of these can be classified as either modified matter or modified gravity models \cite{Yoo2012}; the former explain the accelerated expansion of the Universe by introducing a new matter component with negative pressure, such as a scalar field. On the other hand, modified gravity models are based on the view that dark energy is a relic of the inaccuracies in the $\Lambda$CDM description of the space-time geometry. As yet, however, the available evidence is not sufficient for $\Lambda$CDM to be discarded in favor of one of the alternatives \cite{Ade2016, Joudaki2017}. And for good reason: theoretically, its framework is appealingly simple, and when it comes to observations, $\Lambda$CDM has not only turned out to be compatible with local gravity constraints \cite{Einstein1915, Shapiro1990}, but it also successfully predicted the baryon acoustic oscillations (BAOs) imprint on galaxy clustering \cite{Eisenstein2005} and the existence of gravitational waves \cite{Abbott2016}. Additionally, it can properly describe the cosmology at the redshifts probed by cosmic microwave background (CMB) data \cite{Planck2018, Planck2015, Bennett2013}. This list is by no means complete, but it serves to illustrate why $\Lambda$CDM is considered the standard model of cosmology. On the other hand, it has a number of shortcomings that cannot be overlooked, prominent among which are the cosmic coincidence and smallness problems \cite{Weinberg2001}. Another case in point is the tension between the local value of the Hubble constant \cite{Riess2018} and the result obtained by the \emph{Planck} collaboration \cite{Planck2018} in the context of a $\Lambda$CDM cosmology. There is also the challenge posed by the `small-scale crisis' (see Ref.~\cite{Nakama2017} and works cited therein), which refers to the discrepancies between sub-galactic-scale observations and the predictions resulting from $N$-body simulations of structure formation in the standard model. 

A sound alternative model of dark energy, therefore, is expected to emulate the successes of $\Lambda$CDM while bridging the existing gaps between theory and observation (or some of them, at least). Consequently, such models should mimic $\Lambda$CDM at the high redshifts where it is well-tested by CMB data, and give a comparable expansion history at low redshifts, albeit without invoking a true cosmological constant \cite{Hu2007}. Furthermore, on Solar-System scales their behavior must be in accordance with experimentally-supported General Relativistic predictions \cite{Will2006}. One way of achieving this is by means of screening mechanisms, which depend on the density contrast between the local environment and the cosmic fluid to suppress small-scale deviations from the standard model (see, for instance, Ref.~\cite{Khoury2013} and works cited therein). 

In view of all this, and keeping in mind that the successes of $\Lambda$CDM have not been eclipsed, the simplest -- and perhaps most natural -- extension of the standard model is a scenario characterized by a mildly-evolving cosmological `constant'. We therefore consider three dynamical-$\Lambda$ models: the Running Vacuum Model (RVM), in which $\Lambda$ varies with the Hubble parameter $H$ according to the relation $\Lambda(H)=A+B H^2$ ($A$ and $B$ being constants), a generalization of the RVM (GRVM) with $\Lambda(H)=A+B H^2+ C(\text{d}H/\text{d}t)$, where $t$ is cosmic time and $C$ another constant, and a second sub-case of the GRVM: $\Lambda(H)=A+C(\text{d}H/\text{d}t)$. We shall refer to the last as the `generalized running vacuum sub-case', or GRVS. The GRVM and RVM were introduced in Refs.~\cite{Basilakos2012} and \cite{Shapiro2003}, respectively, and have been analyzed in works such as Refs.~\cite{Basilakos2012, Valent2015, Peracaula2018, Basilakos2009, Grande} and \cite{Geng2017}, while the GRVS was investigated in Ref.~\cite{Karimkhani2015} as a model with a variable dark energy equation-of-state parameter. 

These models are especially appealing due to the fact that they are motivated by Quantum Field Theory (QFT) considerations \cite{Shapiro2000, Shapiro2002, Shapiro2004, Sola2008, MorenoPulido}. Additionally, the RVM can properly account for cosmic dynamics at both the linear perturbation and background levels \cite{Geng2017} -- and in certain cases has been shown to outperform $\Lambda$CDM \cite{JSola2017, PeracaulaJ, Sola2017, Peracaula2018}. Likewise, the GRVM is compatible with observations \cite{Basilakos2012, Valent2015, Karimkhani2015}, and it, too, has been reported to receive greater support from cosmological data than $\Lambda$CDM \cite{Sola2017}. There is also the fact that both the RVM and GRVM appear to provide a better fit to structure formation data \cite{JSola2017, Sola2017, Peracaula2018}. 

To our knowledge, however, the GRVM, RVM and GRVS [with $\Lambda(H)$ taking the exact forms specified above] have not been analyzed in the context of a spatially curved space-time -- although extended versions have, as discussed in subsection \ref{subsec:GRVM}. Indeed, a great number of works in the literature are based on the premise of spatial flatness. We find this practice rather concerning, because although it is true that observational data appears to favor a flat geometry, the evidence comes mainly from studies which assume a flat $\Lambda$CDM cosmology \cite{Bennett2013, Planck2018, Sanchez, Moresco}. Our primary aim, therefore, will be to investigate whether the GRVM, RVM and GRVS can accommodate spatial curvature while remaining compatible with the data available. To this end, we will briefly introduce dynamical-$\Lambda$ models in section \ref{sec:dynamical}, with special emphasis on the ones we shall be considering. The relevant likelihoods and statistics are reviewed in section \ref{sec:observational}, while results are presented and discussed in section \ref{sec:results} and the study is concluded in section \ref{sec:conclusion}. We use units in which $c=8\pi G/3=1$.

\section{Dynamical-$\mathbf{\Lambda}$ models}
\label{sec:dynamical}
The literature contains many examples of models in which dynamical dark energy takes the form of a varying $\Lambda$. In most cases, $\Lambda$ is allowed to have a large value at early times, and this then decays to the much smaller one observed at present. Therefore, such models go some way in addressing the smallness problem  \cite{Berman1991, Beesham1993, Perico2013}, which refers to the fact that in $\Lambda$CDM, the observed value of $\rho_\Lambda$ (where $\rho_\Lambda$ is the vacuum energy density) happens to be around a factor of $10^{120}$ smaller than the theoretical estimate \cite{Weinberg2001}.

Endowing $\Lambda$ with a dynamic nature may be achieved in two ways. One can either model $\Lambda$ as an explicit function of time, or else take an implicit approach and express it in terms of appropriate cosmic parameters. In the former case, the most popular choice is undoubtedly the inverse power relation given by\footnote{The parameters $n$, $m$, $A$ and $B$ shall henceforth represent constants.} $\Lambda(t) \propto t^{-n}$. The inverse power-law model features in works such as Refs.~\cite{Bertolami1986, Kalligas1992, Lopez1996, Arbab1997, Fujii1990} -- the list is by no means exhaustive -- and has additionally been investigated in differing scenarios, including a Bianchi Type-I cosmology with variable gravitational coupling \cite{Beesham1994} and the Brans-Dicke (BD) theory \cite{Berman1990, Endo1977}. Albeit less popular, exponential decay has also been proposed \cite{Beesham1993, Spindel1994}.

In the category of implicit time dependence, one finds studies in which $\Lambda$ is a function of the scale factor $a$, with expressions such as $\Lambda(a)=Aa^n+Ba^m$ \cite{Dussattar1996, John1997, Carvalho1992, Waga1993} and $\Lambda(a)=A+Ba^{-n}, A\neq 0$ \cite{Sistero1991}. Models having $\Lambda(a) \propto a^{-n}$ are very popular -- the reader is referred to Refs.~\cite{Overduin1998, Arbab1994, AbdelRahman1992, Mendez1996, Calvao1992} -- and under certain conditions may be seen as equivalent to standard cosmology with matter, radiation and an additional component: an exotic fluid characterized by an equation of state parameter $w=n/3-1$ \cite{Silveira1997}. The case $n=2$ is of particular interest. It may not only have its foundations in quantum cosmology \cite{Chen1990}, but has also been shown -- in the framework of a closed geometry -- to result from the assumption that the energy density of the Universe is equal to its critical value at all times, not just at present \cite{Ozer1986}. This assumption would ensure that the current epoch is not special in any way and effectively solve the cosmic coincidence problem. 

Another notable study is Ref.~\cite{Freese1987}. Here, the authors present a model in which the vacuum couples with radiation (during the radiation-dominated epoch), and has an associated energy density that scales as $a^{-4(1-x)}$, where $x$ depends on the balance between the energy densities of radiation and dark energy. Meanwhile, the innovative approach detailed in Ref.~\cite{Wang2005} is based on the ansatz that the energy density of cold dark matter (CDM) varies as $a^{-3+y}$, rather than the customary $a^{-3}$. The small positive constant $y$ results from the interaction with dark energy and quantifies the decrease in the rate at which CDM gets diluted. It is interesting to note that provided\footnote{$\Omega_\text{m}^0$ denotes the current value of the matter density parameter.} $\Omega_\text{m}^0\geq0.2$ and $n\geq1.6$, spatially flat cosmologies having $\Lambda(a) \propto a^{-n}$ show consistency with lensing data \cite{Silveira1997}. 

A third popular class of expressions for $\Lambda$ is based on the Hubble parameter $H$ and functions thereof. Prominent among these is again the power law: $\Lambda(H) \propto H^n$ \cite{Waga1993, Overduin1998, Lima1994, Wetterich1995, Ray2007, Ray2011}. Other interesting possibilities include combinations of $H$ or $H^n$ with either $a^m$ \cite{Carvalho1992, Arbab1994} or $\text{d}H/\text{d}t$, or even the total energy density \cite{Maia1994}. A case in point is the entropic acceleration model \cite{Easson2011}. This model is characterized by an entropic force which acts at the apparent horizon of the Universe,\footnote{The apparent horizon is determined by the quantity $\left(H^2+k/a^2\right)^{-1/2}$, where $k$ is the spatial curvature parameter and $a$ the scale factor, both in normalized form \cite{Easson2011}. In the absence of spatial curvature, the apparent and Hubble horizons are equivalent.} and which behaves essentially like a dark energy component whose density varies as $A(\text{d}H/\text{d}t)+BH^2$ ($B\neq 0$) in flat space. According to the authors of Ref.~\cite{Basilakos2014}, however, the entropic model is problematic in that the sign of its deceleration parameter never changes. Additionally, the possibility that it describes the late-time behavior of a more complete model is ruled out by its failure to reconcile recent cosmic growth data with an accelerated expansion \cite{Basilakos2014}. An alternative entropic model in which $\Lambda(H)=AH+BH^2$ also has this shortcoming, while putting $\Lambda(H)\propto H$ results in a scenario that is disqualified by CMB data \cite{Basilakos2014}. It has in fact been proposed that when $\Lambda(H)$ is a simple function of terms from the set $\{H,\text{d}H/\text{d}t, H^2\}$, the addition of a constant to the said function is crucial to get a valid model \cite{Basilakos2014}. 

\subsection{The Generalized Running Vacuum Model}
\label{subsec:GRVM}

The inspiration for the GRVM comes from the interpretation of $\Lambda$ as a running parameter in the curved space-time version of QFT. The associated energy density, $\rho_\Lambda$, is thus expected to evolve according to a renormalization group equation of the form \cite{Sola2011}
\begin{equation}
\frac{\text{d}\rho_\Lambda}{\text{d}\ln{\beta}}=\frac{1}{(4\pi)^2}\sum_{n=1}^\infty S_n\beta^{2n}~,
\label{RGeq}
\end{equation}
where the dynamical variable $\beta$ represents some characteristic infrared-cutoff scale. In a cosmological context, the role of $\beta$ may be played by the Hubble parameter $H$, since the latter is of the order of the energy scale associated with the Friedmann-Lema\^{i}tre-Robertson-Walker (FLRW) cosmology \cite{Valent2015}. We furthermore note that the coefficients $S_n$ result from loop contributions of fields having different masses and spins \cite{Sola2011}. Meanwhile, the absence of odd powers of $\beta$ reflects the general covariance of the effective action \cite{Valent2015, Shapiro}. 

Given that $\beta\sim H$, the small present-day value of $H$  $(\sim\SI{e-27}{m^{-1}})$ implies that terms in Eq.~($\ref{RGeq}$) with $n\geq 2$ would be suppressed in the current epoch. An expression for $\rho_{\Lambda}$ $\left(\text{henceforth}~\rho_{\Lambda(H)}\right)$ may hence be obtained by integrating the remaining term on the right-hand side. One gets the relation $(4\pi)^2\rho_{\Lambda(H)}= S_0+S_1\beta^2/2$, with $S_0$ denoting the constant of integration. Consequently, if $\beta^2$ is identified with a linear combination of\footnote{These two quantities represent independent degrees of freedom \cite{Basilakos2012}.} $H^2$ and $\text{d}H/\text{d}t$, the expression for $\rho_{\Lambda(H)}$ becomes $(4\pi)^2\rho_{\Lambda(H)} = S_0+\tilde{S}_2H^2+\tilde{S}_3(\text{d}H/\text{d}t)$, where $\tilde{S}_2$ and $\tilde{S}_3$ are constants \cite{Valent2015}.

In conclusion, we shall be investigating a model in which the cosmological constant is replaced with
\begin{equation}
\Lambda(H)=A+BH^2+C\dot{H}~.
\label{Lambda}
\end{equation}
The leading constant and the coefficients of $H^2$ and $\dot{H}$ have been written as $A$, $B$ and $C$ for the sake of simplicity, $B$ and $C$ being dimensionless and $A$ having units of $\text{length}^{-2}$. An overdot denotes differentiation with respect to cosmic time $t$. The model specified by Eq.~($\ref{Lambda}$) is none other than the GRVM, introduced in section \ref{sec:intro} -- the RVM and GRVS follow as special cases by setting $C=0$ and $B=0$, respectively. One notes that $\Lambda(H)$ is not an \emph{explicit} function of time (the dependence on $t$ is established implicitly, via $H$) and it is in fact this property that the name `running vacuum model' is meant to reflect \cite{Valent2015}.

The dynamic nature of $\Lambda(H)$ means that the Bianchi identity may be satisfied in one of two ways \cite{Sola2017, Shapiro2005, Sola2015}. Let us see how such a possibility comes about. If we restore the constant $8\pi G/3$ and do not incorporate $\Lambda(H)$ into the energy-momentum tensor $T_{\mu\nu}$, Einstein's field equations read 
\begin{equation}
G_{\mu\nu}=8\pi G T_{\mu\nu}-\Lambda(H)g_{\mu\nu}~,
\end{equation}
so that the twice-contracted Bianchi identity, $\nabla^{\mu}G_{\mu\nu}=0$ \cite{Carroll2004} ($G_{\mu\nu}$ being the Einstein tensor), implies that 
\begin{equation}
\nabla^\mu(8\pi G T_{\mu\nu})=g_{\mu\nu}\nabla^\mu\Lambda(H)~,
\end{equation}
where we have made use of the fact that $\nabla_\mu$ is constructed from a metric-compatible connection (i.e. $\nabla^\rho g_{\mu\nu}=0$ at all points) \cite{Carroll2004}. In the presence of a time-varying $G$, $T_{\mu\nu}$ may be conserved separately, in which case the above equation becomes $8\pi T_{\mu\nu}\nabla^\mu G(t)=g_{\mu\nu}\nabla^\mu\Lambda(H)$. If $G$ is constant, however, the Bianchi identity requires that $8\pi G \nabla^\mu T_{\mu\nu}=g_{\mu\nu}\nabla^\mu \Lambda(H)$. In other words, we have the choice of either a `running' gravitational coupling $G(t)$ or of energy transfer between the vacuum and any other component/s of the cosmic fluid (combinations of the two options are also possible). We shall take the constant-$G$ approach. Assuming that the densities of baryonic matter and radiation evolve as in the standard model, it may be deduced that dark energy interacts with cold dark matter (whose energy density is denoted by $\rho_{\text{cdm}}$) according to the equation
\begin{equation}
\dot{\rho}_{\text{cdm}}^{~}+3H \rho_\text{cdm}^{~}=-\dot{\rho}_{\Lambda(H)}^{~} ~.
\label{conservation1}
\end{equation}
In the system of units with $c=8\pi G/3=1$, $\rho_{\Lambda(H)}=\Lambda(H)/3$, and hence Eq.~($\ref{Lambda}$) translates into
\begin{equation}
\rho_{\Lambda(H)}^{~}=\frac{1}{3}\left(A+BH^2+C\dot{H}\right)~,
\label{rhoLambda}
\end{equation}
which, when inserted into Eq.~($\ref{conservation1}$), yields the relation
\begin{equation}
\dot{\rho}_{\text{cdm}}=-\frac{1}{3}\left[H\left(9\rho_\text{cdm}+2B\dot{H}\right)+C \ddot{H}\right]~.
\label{conservation2}
\end{equation}
To obtain an expression for $\dot{H}$, we use Eq.~($\ref{Lambda}$) with the second Friedmann equation. The latter reads as follows:
\begin{equation}
\dot{H}+H^2=-\frac{1}{2}\left(\rho+3p\right)+\frac{\Lambda(H)}{3}~.
\label{2ndFried}
\end{equation}
Here, $\rho$ denotes the sum of the energy densities of cold dark matter ($\rho_\text{cdm}$), baryons ($\rho_\text{b}$) and radiation ($\rho_\text{r}$), while $p$ represents the total of the corresponding pressures. Dark energy is modeled with an equation of state parameter $w_{\Lambda(H)}$ fixed at $-1$, as in $\Lambda$CDM. If $w_{\Lambda(H)}$ is instead allowed to vary, it would be possible for dark energy to be conserved independently of the other cosmic components. Such a scenario has been investigated in Ref.~\cite{Karimkhani2015}. 

As stated previously, it is assumed that neither radiation (i.e.~photons and massless neutrinos) nor baryons interact with dark energy. Consequently, cosmic expansion dilutes the respective energy densities in accordance with the familiar $\Lambda$CDM relations: 
\begin{equation}
\rho_\text{b}^{\phantom{0}}=\rho_\text{b}^0 a^{-3}~, \qquad \rho_\text{r}^{\phantom{0}}=\rho_\text{r}^0 a^{-4}~,
\label{rhoBrhoR}
\end{equation}
$a$ being the normalized scale factor. A 0-superscript indicates present-day quantities.

Let us now return to Eq.~(\ref{2ndFried}). It provides us with an expression for $\dot{H}$, and we proceed by differentiating it with respect to $t$ to find $\ddot{H}$. Then we substitute for the first and second time derivatives of $H$ in Eq.~($\ref{conservation2}$), getting that
\begin{widetext}
\begin{align}
\bigg\{&\frac{3}{2}(C-2)(C-3){\rho}'^{}_\text{cdm}(a)\,a+3\left[9-B+C(C-5)\right]\rho^{\phantom{0}}_\text{cdm}-\frac{3}{2}[2B+(C-5)C]\rho_\text{b}^0 a^{-3}-2[3B+C(2C-9)]\rho_\text{r}^0 a^{-4}\notag\\[0.3em]&+2(B-C)\left[A+(B-3)H^2\right]\bigg\}\frac{1}{(C-3)^2}=0~,
\label{conservation3}
\end{align}
where a prime denotes differentiation with respect to the argument and we have replaced $\text{d}/\text{d}t$ with $a H (\text{d}/\text{d}a)$. The next step is to solve the differential equation ($\ref{conservation3}$), but before attempting to do so, the Hubble parameter must be expressed in terms of $a$. To this end, we make use of the Friedmann equation:
\begin{equation}
H^2=\rho_\text{cdm}^{~}+\rho_\text{b}^{~}+\rho_\text{r}^{~}+\rho_{\Lambda(H)}^{~}-\kappa a^{-2}~.
\label{1stFried}
\end{equation}
Here, $\kappa$ represents the curvature of the spatial hypersurfaces in an FLRW Universe. It is a scaled version of the normalized curvature parameter $k$, and is defined as the ratio of $k$ to $R_0^2$, $R_0$ being the value that the (non-normalized) scale factor $R$ takes at present. 

The energy densities in Eq.~($\ref{1stFried}$) may be replaced with the corresponding relations given by Eqs.~($\ref{rhoLambda}$) and ($\ref{rhoBrhoR}$). This allows us to determine $H$ as a function of $a$:

\begin{equation}
H=\frac{\left[3(C-2)\rho_\text{cdm}^{\phantom{0}}a^4+3(C-2)\rho_\text{b}^0a+2(2C-3)\rho_\text{r}^0-2(C-3)\kappa a^2-2Aa^4\right]^{1/2}}{a^2\sqrt{2(B-3)}}~.
\label{H}
\end{equation}
~~~~~~~\\
Inserting the above into Eq.~$(\ref{conservation3})$ yields the final version of Eq.~($\ref{conservation1}$):
\begin{equation}
3(C-2)\rho'_\text{cdm}(a)\,a+6(B-3)\rho_\text{cdm}+3(2B-3C)\rho_\text{b}^0a^{-3}+8(B-2C)\rho_\text{r}^0a^{-4}-4(B-C)\kappa a^{-2}=0~,
\end{equation}
which can readily be solved for $\rho_\text{cdm}$. We find that
\begin{align}
&\rho_\text{cdm}=\notag\\[0.3em]&H_0^2\left[a^{\frac{2(3-B)}{C-2}}\Omega_\text{cdm}^0-\left(a^{-3}-a^{\frac{2(3-B)}{C-2}}\right)\Omega_\text{b}^0-\frac{4(B-2C)}{3(B-2C+1)}\left(a^{-4}-a^{\frac{2(3-B)}{C-2}}\right)\Omega_\text{r}^0-\frac{2(B-C)}{3(B-C-1)}\left(a^{-2}-a^{\frac{2(3-B)}{C-2}}\right)\Omega_k^0\right]~.
\label{rhocdm}
\end{align}
\end{widetext}
In the above equation, $H_0$ is the Hubble constant, and we have utilized the fact that each energy density $\rho_i$ has an associated density parameter $\Omega_i$, defined as the ratio of  $\rho_i$ to\footnote{The subscript $i$ represents a generic component, and is replaced by `cdm' for cold dark matter, `b' for baryons, `r' for radiation, `$\Lambda(H)$' for dark energy and `$k$' for spatial curvature. $\Omega_k$ is equivalent to $-\kappa/(Ha)^2$. Present-day quantities are denoted by a 0-sub/superscript.} the critical energy density $\rho_\text{c}$. In the unit system we have adopted, $\rho_\text{c}=H^2$, and hence $\Omega_i=\rho_i/H^2$. 

Now we require an expression for $\rho_{\Lambda(H)}$. Equipped with Eq.~($\ref{rhocdm}$), we first eliminate $\rho_\text{cdm}$ from Eq.~($\ref{H}$). Next, Eq.~($\ref{rhocdm}$) is used in conjunction with Eq.~($\ref{rhoBrhoR}$), the updated version of Eq.~$(\ref{H})$, and Eq.~($\ref{Lambda}$); they are inserted into Eq.~($\ref{2ndFried}$) to find an expression for $\dot{H}$. In all cases, we write the energy densities in terms of the current values of the density parameters. Finally, we substitute for $H$ [Eq.~(\ref{H})] and $\dot{H}$ in Eq.~$(\ref{rhoLambda})$, which consequently takes the form
\begin{widetext}
\begin{align}
&\rho_{\Lambda(H)}^{~}=\left\{\frac{(2B-3C)}{2(B-3)}\left(1-a^{\frac{2(3-B)}{C-2}}\right)\left(\Omega_\text{cdm}^0+\Omega_\text{b}^0\right)+\frac{B-2C}{3(B-3)(B-2C+1)}\bigg[(B-3)a^{-4}+2(3C-2B)a^{\frac{2(3-B)}{C-2}}\right.\notag\\[0.3em]&\left.+3(B-2C+1)\bigg]\Omega_\text{r}^0-\frac{(B-C)}{3(B-3)(B-C-1)}\bigg[(B-3)a^{-2}+(2B-3C)a^{\frac{2(3-B)}{C-2}}-3(B-C-1)\bigg]\Omega_k^0+\Omega_{\Lambda(H)}^0\right\}H_0^2~.
\label{rhoLambdaFinal}
\end{align}
\end{widetext}
The requirement that $\rho_{\Lambda(H)}$ is currently equal to $H_0^2\Omega_{\Lambda(H)}^0$ has been used to fix the value of $A$ at $H_0^2\left(3\Omega_{\Lambda(H)}^0-B\right)$.

A few comments about the role of spatial curvature are in order before we proceed. In Ref.~\cite{Perico2013}, the RVM is represented as the late-time limit of a model that can describe the complete cosmic history. Its generalized version takes spatial curvature into account \cite{Lima2015}, and is based on the following expression for $\Lambda$:
\begin{align}
&\Lambda(H, a)=\notag\\[0.3em]&\Lambda_\infty+3\nu\left(H^2-H_\text{F}^2+\frac{\kappa}{a^2}\right)+3\tau\left(\frac{H}{H_\text{I}}\right)^n\left(H^2+\frac{\kappa}{a^2}\right)~,
\label{longLambda}
\end{align}
where the integer $n$ satisfies $n\geq 1$ \cite{Perico2013} and $\Lambda_\infty$ is the limit of $\Lambda(H, a)$ as $a\rightarrow\infty$. $H_\text{I}$ and $H_\text{F}$ stand for the Hubble parameter in two different epochs. The former characterizes inflation, while the latter denotes the `final' value of $H$ (or the limit of $H$ as $a\rightarrow\infty$) \cite{Lima2015}. Lastly, $\nu$ and $\tau$ correspond to dimensionless constants \cite{Lima2015}. The quantity $3\nu$ is the counterpart of the model parameter $B$ we have introduced in Eq.~($\ref{Lambda}$). 

The reason why we limit ourselves to the RVM, instead of analyzing the extended version just described, is twofold. Firstly, $H$ is expected to be already much smaller than $H_\text{I}$ at the start of the adiabatic radiation phase \cite{Lima2015}. Since we are not concerned with inflation, but rather with the late-time behavior of dark energy models, the term in $(H/H_\text{I})^n$ may therefore be dropped. Secondly, the explicit inclusion of $\kappa$ in Eq.~($\ref{longLambda}$) is motivated by phenomenological considerations \cite{Lima2015}. In light of this, we think it would be interesting to study how the RVM -- in its original simple form -- behaves if $\Omega_k^0$ is allowed to vary.

\section{Observational data and corresponding likelihoods}
\label{sec:observational}
If a model is to be considered a candidate in the dark energy contest, one must first of all determine whether it is compatible with observational data. To this end, we employ Bayesian statistics, and perform a Markov Chain Monte Carlo (MCMC) analysis using the Cosmic Linear Anisotropy Solving System (\textsc{CLASS}) v.2.6.3 \cite{Blas2011} in conjunction with \textsc{Monte Python} v.2.2.2 \cite{Audren2013}. The plots presented in this work have been constructed using the MCMC analysis package \textsc{GetDist} v.0.2.8 \cite{Lewis2015}.

In this section, we briefly introduce the likelihoods with which we constrain model parameters.

\subsection{The JLA likelihood for SNeIa}
Type-Ia supernovae (SNeIa) make it possible to probe the expansion history of the Universe by looking at how the luminosity distance to an object varies with redshift $z$. Whenever this relation departs from a pure Hubble law \cite{Schmidt1998}, the difference (to lowest order in $z$) depends on just the deceleration parameter, and can thus yield important information about the rate of cosmic expansion. SNeIa are ideal in this regard because they act as standard candles -- in the sense that their homogeneity as a group means their intrinsic luminosity (or absolute magnitude) can be calibrated \cite{Katz}, and hence astronomers may readily find how distant they are by measuring their observed luminosity (called the apparent magnitude) \cite{Carroll}. 

The Joint Light-Curve Analysis (JLA) data set is based on a sample of 740 SNeIa \cite{Betoule2014}. The observable relevant to us is the distance modulus $\mu_\text{obs}$, whose theoretical counterpart is given by:
\begin{equation}
\mu_\text{th}=5\log_{10}\left(\frac{d_\text{L}}{\text{Mpc}}\right)+25~,
\end{equation}
where the luminosity distance $d_\text{L}$ should be quoted in Mpc, and is in turn determined from the equation\footnote{In a flat Universe, $d_\text{L}=H_0^{-1}(1+z)~\mathcal{F}\left(\int^z_0 H_0 H(\bar{z})^{-1}\text{d}\bar{z}\right)$.}
\begin{equation}
d_\text{L}=\frac{1+z}{H_0\sqrt{|\Omega_k^0|}}~\mathcal{F}\left(\sqrt{|\Omega_k^0|}\int^z_0\frac{H_0\text{d}\bar{z}}{H(\bar{z})}\right)~.
\label{lumdist}
\end{equation}
The form of the function $\mathcal{F}(x)$ depends on the spatial geometry:
\begin{equation}
\mathcal{F}(x)=
\begin{cases}
x&\text{if}~\kappa=0\,;\\
\sin{(x)}&\text{if}~\kappa>0\,;\\
\sinh{(x)}&\text{if}~\kappa<0\,.
\end{cases}
\end{equation}
We are now in a position to construct the associated $\chi^2$. This may be expressed as
\begin{equation}
\chi^2_\text{JLA}=\Delta \mu^\text{T}C_\text{JLA}^{-1}\Delta\mu~,
\label{chi2JLA}
\end{equation}
where $\Delta\mu$ is a vector whose $i^{\text{th}}$ entry is the difference between the observed and theoretical distance moduli $\left(\mu_\text{obs}^i-\mu_\text{th}^i\right)$ of the $i^{\text{th}}$ supernova \cite{Zou2018}. $\Delta \mu^\text{T}$ represents its transpose.

The inverted covariance matrix for the observational values of $\mu$ is denoted in Eq.~($\ref{chi2JLA}$) by $C_\text{JLA}^{-1}$. Details about its construction are provided in Ref.~\cite{Betoule2014}.

\subsection{The cosmic chronometer (clocks) likelihood}
{
\renewcommand{\arraystretch}{1.2}
\begin{table}[h]
\caption{\label{Tcc} Cosmic chronometer data. Each value of $H(z)$ in the third column is measured at an effective redshift $z$ given in the second column, and has a corresponding error $\sigma$ (fourth column).}
\vspace{1.5mm}
\begin{tabular}{@{\hskip 0.2cm}c@{\hskip 0.5cm} S[table-format=2.5]@{\hskip 0.6cm} S[table-format=3.2]@{\hskip 0.5cm} S[table-format=3.2]}
\hline
\hline
Ref.  & {$z$} & {$~H(z)$} & {$\sigma$}\\
~ & ~ & \multicolumn{2}{l}{$\left(\si{km.s^{-1}.Mpc^{-1}}\right)$}\\
\hline
\cite{Zhang2014} & 0.0700 & 69.0 & 19.6\\
\cite{Zhang2014} & 0.1200 & 68.6 & 26.2\\
\cite{Simon2005} & 0.1700 & 83.0 & 8.0\\
\cite{Moresco2012} & 0.1791 & 75.0 &  4.0\\
\cite{Moresco2012} & 0.1993 & 75.0 & 5.0\\
\cite{Zhang2014} & 0.2000 & 72.9 & 29.6\\
\cite{Simon2005} & 0.2700 & 77.0 & 14.0\\
\cite{Zhang2014} & 0.2800 & 88.8 & 36.6\\ 
\cite{Moresco2012} & 0.3519 & 83.0 & 14.0\\
\cite{Moresco2016} & 0.3802 & 83.0 & 13.6\\
\cite{Simon2005} & 0.4000 & 95.0 & 17.0\\
\cite{Moresco2016} & 0.4004 & 77.0 & 10.2\\
\cite{Moresco2016} & 0.4247 & 87.1 & 11.2\\
\cite{Moresco2016} & 0.4497 & 92.8 & 12.9\\
\cite{Ratsimbazafy2017} & 0.4700 & 89.0 & 49.6\\
\cite{Moresco2016} & 0.4783 & 80.9 & 9.0\\
\cite{Stern2010} & 0.4800 & 97.0 & 62.0\\
\cite{Moresco2012} & 0.5929 & 104.0 & 13.0\\
\cite{Moresco2012} & 0.6797 & 92.0 & 8.0\\
\cite{Moresco2012} & 0.7812 & 105.0 & 12.0\\
\cite{Moresco2012} & 0.8754 & 125.0 & 17.0\\
\cite{Stern2010} & 0.8800 & 90.0 & 40.0\\
\cite{Simon2005} & 0.9000 & 117.0 & 23.0\\
\cite{Moresco2012} & 1.0370 & 154.0 & 20.0\\
\cite{Simon2005} & 1.3000 & 168.0 & 17.0\\
\cite{Moresco2015} & 1.3630 & 160.0 & 33.6\\
\cite{Simon2005} & 1.4300 & 177.0 & 18.0\\
\cite{Simon2005} & 1.5300 & 140.0 & 14.0\\
\cite{Simon2005} & 1.7500 & 202.0 & 40.0\\
\cite{Moresco2015} & 1.9650 & 186.5 & 50.4\\
\hline
\hline
\end{tabular}
\end{table}
}
The Hubble parameter is defined in terms of the scale factor as the ratio $\dot{a}/a$, and the relation $a=1/(1+z)$ allows us to express it as a function of the redshift $z$: 
\begin{equation}
H(z)=-\frac{1}{1+z}\frac{\text{d}z}{\text{d}t}~.
\end{equation}

The differential age (or cosmic chronometer/clocks) method entails measuring $\text{d}z/\text{d}t$ to directly arrive at $H(z)$. This approach, first put forward in Ref.~\cite{Jimenez2002}, effectively involves determining the age difference between two cosmic `chronometers' \cite{Jimenez2002} located in a given redshift interval. The best chronometers are massive early-type galaxies which acquired more than $90$ percent of their stellar mass very rapidly at high redshifts, and have been evolving passively since then, without major episodes of star formation \cite{Moresco2012} that would otherwise dominate the emission spectrum \cite{Jimenez2002}. The age of such a galaxy can consequently be inferred from the differential dating of its stellar population \cite{Moresco2012}.

Table \ref{Tcc} lists the cosmic chronometer data employed in this work.\footnote{In the case of the Ratsimbazafy et al.~data point \cite{Ratsimbazafy2017}, $\sigma$ was calculated by summing the systematic and statistical errors in quadrature.} Where possible, we chose results that were obtained using the Bruzual and Charlot 2003 (BC03) stellar population synthesis (SPS) model \cite{Bruzual2003}. It should be pointed out, however, that the values of $H(z)$ are expected to be largely unaffected by the choice of SPS \cite{Moresco2012, Stern2010}.

The $\chi^2$ for the cosmic chronometer likelihood reads:
\begin{equation}
\chi^2_{H(z)}=\mathlarger{\mathlarger{\sum}}_i\left(\frac{H^\text{obs}_i-H^\text{th}\left(z_i^{~}\right)}{\sigma_{H(z),i}^{~}}\right)^2~.
\label{chi2Hz}
\end{equation}
Here, each $H^\text{obs}_ i$ is the observed value from Table \ref{Tcc} corresponding to $z=z_i$, $\sigma_{i}$ represents the associated error, and $H^\text{th}(z_i)$ stands for the theoretical prediction at the same redshift.

\subsection{The CMB likelihood}
Anisotropies present in the temperature and polarization power spectra of the CMB can yield a wealth of information when used as cosmological probes. We shall work with two main distance priors: the shift parameter $\mathcal{R}$ and the acoustic scale $l_\text{A}$. These are related to the amplitude and distribution of the temperature anisotropy peaks. The shift parameter $\mathcal{R}$ characterizes the temperature power spectrum in the line-of-sight direction and is defined as follows \cite{Komatsu2009}:
\begin{equation}
\mathcal{R}(z_*^{~})=\sqrt{\Omega_\text{m}^0}H_0^{~}(1+z_*^{~})d_\text{A}^{~}(z_*^{~})~,
\end{equation}
where $z_*$ denotes the redshift of the photon decoupling epoch.  The angular diameter distance $d_\text{A}$ may be expressed via the distance-duality relation as $d_\text{L}/(1+z)^2$, $d_\text{L}$ being the luminosity distance from Eq.~($\ref{lumdist}$). 

The acoustic scale $l_\text{A}$, on the other hand, relates to attributes of the CMB temperature power spectrum in the transverse direction \cite{Huang2015}. It, too, depends on $d_\text{A}$  \cite{Komatsu2009}:
\begin{equation}
l_\text{A}^{~}(z_*^{~})=(1+z_*^{~})\frac{\pi d_\text{A}^{~}(z_*^{~})}{r_\text{s}^{~}(z_*^{~})}~.
\end{equation}
Here, $r_\text{s}(z_*)$ is the comoving sound horizon evaluated at $z_*$. In our case, it shall be determined numerically by \textsc{CLASS}, although it is worth noting that in general, the function $r_\text{s}(z)$ takes the form
\begin{equation}
r_\text{s}(z)=\int_z^\infty \frac{c_\text{s}(\bar{z})}{H(\bar{z})}\text{d}\bar{z}~,
\label{rs}
\end{equation}
where $c_\text{s}(z)$ is the sound speed in the photon-baryon fluid and equates to $1/\sqrt{3[1+\eta(z)]}$. The function $\eta(z)$ is given by $0.75\rho_\text{b}/\rho_\gamma$ in the standard scenario \cite{Huang2015, Komatsu2009, Aubourg2015} ($\rho_\gamma$ stands for the energy density of photons), but should be modified when considering cosmological models in which  $\rho_\text{b}$ and $\rho_\gamma$ scale differently with $z$. More details may be found in Ref.~\cite{Valent2015}. 
{
\renewcommand{\arraystretch}{1.3}
\begin{table}
\caption{\label{TCMB} Mean values and corresponding errors for the CMB distance priors \cite{Huang2015}.}
\vspace{2mm}
\begin{tabular}{ c | c } 
\hline
\hline
$\mathcal{R}$ & $1.7448\pm0.0054$\\
$l_\text{A}$ & $301.460 \pm 0.094$\\
$\Omega_\text{b}^0 h^2$ & $0.02240 \pm 0.00017$\\
$n_s$ & $0.9680 \pm 0.0051$\\
\hline
\hline
\end{tabular}
\end{table}
}

It is interesting to note that while differences in $\mathcal{R}$ affect the amplitude of the acoustic peaks, changes in $l_\text{A}$ are instead reflected in the distribution of peaks and troughs \cite{Huang2015}.

The data we use to constrain our model parameters is taken from Ref.~\cite{Huang2015} and shown in Table \ref{TCMB}. It was obtained in the context of a flat $\Lambda$CDM cosmology with $A_\text{L}$ as a free parameter ($A_\text{L}$ being the amplitude of the lensing power spectrum). The fact that a particular cosmological model had to be assumed is, however, only a minor disadvantage, since $\mathcal{R}(z_*)$ and $l_\text{A}(z_*)$ are effective observables, while the quantities $\Omega_\text{b}^0 h^2$ and $n_s$ -- which serve as a third and fourth distance prior\footnote{The dimensionless constant $h$ is equivalent to $H_0/(\SI{100}{km.s^{-1}.Mpc^{-1}})$, and $n_s$ represents the index of the primordial scalar power spectrum.} -- do not appear to be affected significantly by the choice of cosmology \cite{Huang2015, Mukherjee}.

The $\chi^2$ associated with this likelihood reads:
\begin{equation}
\chi^2_\text{CMB}=\Delta x^\text{T}C_\text{CMB}^{-1}\Delta x~.
\label{chi2CMB}
\end{equation}
In the above, $\Delta x$ is the vector $\{\mathcal{R}_\text{obs}(z_*)-\mathcal{R}_\text{th}(z_*), l_\text{A}^\text{obs}(z_*)-l_\text{A}^\text{th}(z_*), (\Omega_\text{b}^0 h^2)_\text{obs}- (\Omega_\text{b}^0 h^2)_\text{th}, n_s^\text{obs}-n_s^\text{th}\}$. We use the notation `obs' to indicate the observed values listed in Table \ref{TCMB}, while `th' denotes theoretical quantities. The covariance matrix $C_\text{CMB}$ may be obtained in normalized form from Ref.~\cite{Huang2015}. We reproduce it below for ease of reference:
\begin{equation}
\begin{kbordermatrix}
{~&\mathcal{R}&l_\text{A}^{\phantom{~}}&\Omega_\text{b}^0h^2&n_s\\
\mathcal{R}&\phantom{-}1.00&\phantom{-}0.53&-0.73&-0.80~\\
l_\text{A}^{\phantom{~}}&\phantom{-}0.53&\phantom{-}1.00&-0.42&-0.43~\\
\Omega_\text{b}^0h^2&-0.73&-0.42&\phantom{-}1.00&\phantom{-}0.59~\\
n_s&-0.80&-0.43&\phantom{-}0.59&\phantom{-}1.00~}
\end{kbordermatrix}~.
\label{normcov}
\end{equation}
It should be noted that $n_s$ is only included as a distance prior [in Table \ref{TCMB} and Eq.~(\ref{normcov})] when measurements related to the growth of large-scale structure are added to the data set.

\subsection{The BAO likelihood}
The physics of BAOs is centered around the imprint left by pre-recombination acoustic waves on clusters of matter \cite{Alam2017}. Simply put, galaxies clustered with a preferred separation equal to  $r_\text{s}(z_\text{d})$, the sound horizon at the drag epoch [$r_\text{s}(z)$ is given by Eq.~($\ref{rs}$), and $z_\text{d}$ denotes the redshift of the drag epoch]. A prominent signature of BAOs is the presence of a localized peak in the galaxy correlation function. Another characteristic feature takes the form of a damped series of oscillations in the CMB power spectrum (see Ref.~\cite{Alam2017} and works cited therein), and so $r_\text{s}(z_\text{d})$ may first be inferred from CMB data, then combined with measurements of angular and redshift separations between clusters. As a result, it becomes possible to calculate the Hubble parameter at the redshift of the said clusters and the angular diameter distance to them \cite{Bonvin2014}. It is common practice, however, to use a distance measure that depends on both $H$ and $d_\text{A}$ -- and this is where the volume distance (or dilation scale) $d_\text{v}$ comes in. It is defined as follows \cite{Eisenstein2005}: 
\begin{equation}
d_\text{v}^{\phantom{~}}=\left(D_\text{A}^2\frac{z}{H}\right)^{1/3}~.
\end{equation}
In the above, $D_\text{A}$ stands for the comoving angular diameter distance and is equivalent to $(1+z)d_\text{A}$.

{
\renewcommand{\arraystretch}{1.3}
\begin{table}
\caption{\label{TBAO1} Uncorrelated BAO data measured at different effective redshifts, $z_\text{eff}$. Column 4 gives the error in each quantity.}
\vspace{1.5mm}
\begin{tabular}{c S[table-format=2.4] S[table-format=2.4] S[table-format=2.4] c} 
\hline
\hline
Ref. & {$z_\text{eff}$} & {Quantity} & {$\sigma$} & Type \\ 
\hline
\cite{Beutler2011} & 0.106 & 0.323 & 0.014 & 1 \\ 
\cite{Ross2015} & 0.150 & 4.490 & 0.170 & 2\\
\cite{Ata2018} & 1.520 & 26.005 & 0.995 & 2\\
\cite{Bautista2017} & 2.330 & 1.031 & 0.026 & 3\\
\hline
\multicolumn{5}{l}{1: $r_s(z_\text{d})/d_\text{v}$\,;~~~2: $d_\text{v}/r_s(z_\text{d})$\,;}\\
\multicolumn{5}{l}{3: $\alpha_\parallel^{0.7}\alpha_\perp^{0.3}$\,;} \\
\multicolumn{5}{l}{$r_\text{s, fid}(z_\text{d})=\SI{147.78}{Mpc}$ \cite{Alam2017}.}\\
\hline
\hline
\end{tabular}
\end{table}
}
{
\renewcommand{\arraystretch}{1.3}
\begin{table}
\caption{\label{TBAO2} BAO data. In the case of the first six data points, the associated errors -- displayed in column 4 -- were derived from the corresponding covariance matrix. The value of $\sigma$ for the last two entries was estimated by constructing the covariance matrix for the quantities numbered 4 and 5.}
\vspace{2mm}
\begin{tabular}{c S[table-format=2.4] S[table-format=5.4] S[table-format=3.4] c} 
\hline
\hline
Ref. & {$z_\text{eff}$} & {Quantity} & {$\sigma$} &Type \\ 
\hline
\cite{Alam2017} & 0.380 & 1512.390 & 24.994 & 4\\
\cite{Alam2017} & 0.380 & 81.209 & 2.368 & 5\\
\cite{Alam2017} & 0.510 & 1975.220 & 30.096 & 4\\
\cite{Alam2017} & 0.510 & 90.903 & 2.329 & 5\\
\cite{Alam2017} & 0.610 & 2306.680 & 37.083 & 4\\
\cite{Alam2017} & 0.610 & 98.965 & 2.502 & 5\\
\cite{Bourboux2017} & 2.400 & 5277.480 & 246.091 & 4\\
\cite{Bourboux2017} & 2.400 & 225.067 & 8.750 & 5\\
\hline
\multicolumn{5}{l}{4: $D_\text{A}\times r_\text{s, fid}(z_\text{d})/r_\text{s}(z_\text{d})(\si{Mpc})$\,;} \\
\multicolumn{5}{l}{5: $H\times r_\text{s}(z_\text{d})/r_\text{s, fid}(z_\text{d})$(\si{km.s^{-1}.Mpc^{-1}})\,;} \\
\multicolumn{5}{l}{$r_\text{s, fid}(z_\text{d})=\SI{147.78}{Mpc}$ \cite{Alam2017}.}\\
\hline
\hline
\end{tabular}
\end{table}
}

The data used in our analysis is summarized in Tables \ref{TBAO1} and \ref{TBAO2}. We introduce $r_\text{s, fid}(z_\text{d})$ to represent the sound horizon as evaluated at the drag epoch in the fiducial cosmology (quantities pertaining to this cosmology shall henceforth be indicated by a sub/superscript `fid'). As for the dimensionless parameters $\alpha_\perp$ and $\alpha_\parallel$, these describe how the BAO peak is displaced with respect to its position in the fiducial model, and correspond to shifts perpendicular and parallel to the line of sight, respectively \cite{Alam2017}:
\begin{equation}
\alpha_\perp=\frac{D_\text{A}^{\phantom{~}}r_\text{s, fid}^{\phantom{~}}\left(z_\text{d}^{\phantom{~}}\right)}{D_\text{A}^\text{fid}\,r_\text{s}^{\phantom{.}}\left(z_\text{d}^{\phantom{~}}\right)}~,\qquad\alpha_\parallel=\frac{H^\text{fid}\,r_\text{s,fid}^{\phantom{~}}\left(z_\text{d}^{\phantom{~}}\right)}{H\,r_\text{s}^{\phantom{.}}\left(z_\text{d}^{\phantom{~}}\right)}~.
\end{equation}
The choice of a fiducial cosmology is necessary to convert redshifts into comoving distances. The problem is that this may inadvertently distort the data. In Ref.~\cite{Alam2017}, therefore, constraints on distances are scaled by the ratio $r_\text{s, fid}(z_\text{d})/r_\text{s}(z_\text{d})$, the aim being to make a conversion of length scales and thus erase any bias potentially resulting from the fiducial model \cite{Alam2017}. We have used the same fiducial value of $r_\text{s}(z_\text{d})$ to scale any data points obtained from other studies. Accordingly, the values listed under `Quantity' in Table \ref{TBAO1} and in the last two rows of Table \ref{TBAO2} are scaled versions of the original.

The $\chi^2$ for the BAO likelihood may be expressed in the usual way:
\begin{equation}
\chi^2_\text{BAO}=\Delta x^\text{T}C_\text{BAO}^{-1}\Delta x~.
\label{chi2BAO}
\end{equation}
Here, the vector $\Delta x$ gives the difference between the observed quantities from Tables \ref{TBAO1} and \ref{TBAO2} (in that order) and their theoretical counterparts, while $C_\text{BAO}^{-1}$ is the inverse covariance matrix and takes the form indicated below:
\begin{equation}
C^{-1}_\text{BAO} =
\begin{pmatrix}
\sigma_1^{-2}&0&0&0&0&0\\
0&\sigma_2^{-2}&0&0&0&0\\
0&0&\sigma_3^{-2}&0&0&0\\
0&0&0&\sigma_4^{-2}&0&0\\
0&0&0&0&C^{-1}_\text{A}&0\\
0&0&0&0&0&C^{-1}_\text{B}
\end{pmatrix}~,
\label{invcovBAO}
\end{equation}
$\sigma_1$ to $\sigma_4$ being the standard deviations listed in column 4 of Table \ref{TBAO1}. As for the submatrices, $C^{-1}_\text{A}$ is the inverse covariance matrix for the Alam et al.~observations (the first six data points in Table \ref{TBAO2}), and $C^{-1}_\text{B}$ corresponds to the quantities reported by des Bourboux et al.~(the last two entries in Table \ref{TBAO2}). These matrices may be constructed from data available in Refs.~\cite{Alam2017} and \cite{Bourboux2017}, respectively.

\subsection{The LSS likelihood}
%The gradient field of the gravitational potential results in galaxies having peculiar velocities superimposed on the Hubble flow. This velocity field serves to trace perturbations in the density of the matter distribution, since these are what makes the gravitational potential fluctuate in the first place.  

The redshift of a galaxy depends on its velocity relative to us, and is hence affected by any peculiar velocity the galaxy might have. If only the Hubble recession is taken into account when converting redshifts into distances, therefore, the recovered over-density field is characterized by redshift space distortions (RSDs) \cite{Percival} along the line of sight. The anisotropies that RSDs introduce into the galaxy power spectrum encode information about the growth of large-scale structure (LSS) \cite{Kwan}. 

In this work we shall be using LSS data in the form of $f\sigma_8$ measurements. The growth rate $f$ and the quantity $\sigma_8$ are defined as follows \cite{Sagredo}:
\begin{equation}
f = \dfrac{\text{d}(\ln{\delta_\text{m}})}{\text{d}\ln{a}}~,\quad \sigma_8 = \sigma_{8,0}\,\frac{\delta_\text{m}(a)}{\delta_\text{m}(1)}~,
\label{f_sigma8}
\end{equation}
where $\delta_\text{m}$ denotes the matter density contrast function, $\delta_\text{m}(1)=\delta_\text{m}(a=1)$ and $\sigma_{8,0}^2$ is the variance of the density field in spheres of radius $R_8=8h^{-1}$ Mpc. It is important to note that $\sigma_{8,0}$ is calculated by linearly evolving the initial power spectrum to the present time, so the square of its value is not necessarily equal to the variance of the current distribution \cite{Houjun}. 

Let us consider $\delta_\text{m}$ and $\sigma_8$ one by one. In both cases, our derivations are closely based on the work presented in Refs.~\cite{Peracaula2018} and \cite{Valent2018}.

\subsubsection{The matter density constrast function ($\delta_\text{m}$)}
\label{subsubsec: delta_m}
%Matter density perturbations are thought to have seeded the growth of large-scale structure. 
The matter density constrast function, $\delta_\text{m}$, is the ratio of the perturbation density to its background analogue \cite{Valent2018}:
\begin{equation}
\delta_\text{m}=\frac{\delta\rho_\text{m}}{\rho_\text{m}}=\frac{\delta \rho_\text{cdm}+\delta\rho_\text{b}}{\rho_\text{cdm}+\rho_\text{b}}=\frac{\rho_\text{cdm}\delta_\text{cdm}+\rho_\text{b}\delta_\text{b}}{\rho_\text{cdm}+\rho_\text{b}}~,
\label{delta_m_v_m}
\end{equation}
and evolves according to the equation:\footnote{The $\delta_\text{m}$ we use here is actually the coefficient of a generic mode in momentum space. More specifically, the perturbation in physical space $[\delta_\text{m}(\tau,\vec{x})]$ may be expressed as an integral over momentum modes, every one of which has a characteristic comoving wave vector $\vec{k}_\dagger$: $\delta_\text{m}(\tau,\vec{x})=\int \delta_\text{m}^{k_\dagger}(\tau)Q(\vec{k}_\dagger,\vec{x})\,\text{d}^3k_\dagger$ \cite{Amendola2010}. Here, $\vec{x}$ is the 3-vector $(x,y,z)$, $k_\dagger=|\vec{k}_\dagger|$ and $Q(\vec{k}_\dagger,\vec{x})$ takes the form $\text{e}^{i\vec{k}_\dagger\boldsymbol{\cdot}\vec{x}}$ if $\Omega_k^0=0$ (the generalization to the non-flat case may be found in Ref.~\cite{Abbott1986}). The linearity of our perturbation equations implies that the individual modes decouple and are each a solution to the said equations. We may thus write $\delta_\text{m}$ in terms of a generic mode: $\delta_\text{m}(\tau,\vec{x})=\delta_\text{m}^{k_\dagger}(\tau)Q(\vec{k}_\dagger,\vec{x})$. However, $Q(\vec{k}_\dagger,\vec{x})$ factors out of the equations, and we are left with the coefficient, $\delta_\text{m}^{k_\dagger}(\tau)$, which we subsequently rename $\delta_\text{m}$.}
\begin{equation}
\delta_\text{m}''(a)+\frac{F_1(a)}{a}\delta_\text{m}'(a)+\frac{F_2(a)}{a^2}\delta_\text{m}(a)=0~.
\label{delta_a}
\end{equation}
We use $\delta\rho_\text{cdm}$ and $\delta\rho_\text{b}$ to represent perturbations in the cold dark matter and baryon components, respectively. The functions $F_1$ and $F_2$ are given by
\begin{align}
F_1(a)&=2+\frac{a\mathcal{H}'(a)+\psi}{\mathcal{H}}~,\notag\\[0.3em]
F_2(a)&=\frac{a\psi'(a)+\psi}{\mathcal{H}} - \frac{3\rho_\text{m}}{2\mathcal{H}^2}a^2~,
\label{F_and_G}
\end{align}
where a prime again denotes differentiation with respect to the argument, $\mathcal{H}$ stands for\footnote{The conformal time $\tau$ is related to the cosmic time $t$ via the scale factor: $\text{d}\tau=\text{d}t/a$.} $a'(\tau)/a$ and $\psi$ equates to $-\rho_{\Lambda(H)}'(\tau)/\rho_\text{m}$. Eqs.~(\ref{delta_a}) and (\ref{F_and_G}) are derived in Ref.~\cite{Peracaula2018} and works cited therein. Their relatively simple form is due to the use of the sub-horizon and quasi-static approximations \cite{Chiu}, as well as the fact that dark energy perturbations -- which reflect the dynamic nature of $\Lambda$ in the models being studied -- are negligible in comparison to $\delta\rho_\text{m}$ on sub-horizon scales \cite{Valent2018}. We have found that changes to Eq.~(\ref{delta_a}) resulting from the presence of curvature are inconsequential. The reason is as follows. The dependence on $\kappa$ is introduced via the time-time component of the perturbed Einstein equation:\footnote{The perturbed metric reads \cite{Bertschinger, Mukhanov} $$\text{d}s^2=a^2\left[-(1+2\Phi)\text{d}\tau^2+\gamma_{ij}(1-2\Phi)\text{d}x^i\text{d}x^j\right]~,$$ $\Phi$ being the metric perturbation and $\gamma_{ij}$ the spatial part of the unperturbed metric: $\gamma_{ij}=\delta_{ij}\left[1+\kappa\left(x^2+y^2+z^2\right)/4\right]^{-2}$. $\delta_{ij}$ denotes the Kronecker delta and $x$, $y$ and $z$ are quasi-Cartesian coordinates.}
\begin{equation}
\left(k_\dagger^2-3\kappa\right)\Phi=-\frac{3}{2}a^2\rho_\text{m}\delta_\text{m}~,
\end{equation}
but for values of the comoving wavenumber $k_\dagger$ in the relevant range -- that is, the range that most contributes to the integral in Eq.~(\ref{sigma8}) \cite{Valent2018} -- $\kappa$ must necessarily be much smaller than $k_\dagger^2$. The above equation may therefore be simplified to $k_\dagger^2\Phi=-3a^2\rho_\text{m}\delta_\text{m}/2$.

%We note that this study focuses exclusively on modes that satisfy $0.01\,h \text{Mpc}^{-1}\leq k\leq 0.2\,h \text{Mpc}^{-1}$ \cite{Valent2018}. The motivation is twofold: \emph{(i)} perturbations with $k\lesssim 0.2\,h\text{Mpc}^{-1}$ may safely be considered linear \cite{Dodelson}, and \emph{(ii)} the mode that crosses the horizon at matter-radiation equality has $k\sim 0.015\,h\text{Mpc}^{-1}$, so we may rest assured that modes with $k\gtrsim 0.01\,h\text{Mpc}^{-1}$ are well within the horizon at redshifts $z\leq 100$ (these being the redshifts of interest). 

We are now in a position to derive initial conditions for $\delta_\text{m}$ and $\delta_\text{m}'(a)$. Let us start by finding the approximate form that $\rho_\text{cdm}$ and $\rho_{\Lambda(H)}$ take deep in the matter-dominated epoch. We return to Eqs.~(\ref{rhocdm}) and (\ref{rhoLambdaFinal}) and look for the dominant terms by taking a number of factors into account, such as the order of magnitude of the ratios $\Omega_\text{m}^0/\Omega_\text{r}^0$ and $\Omega_\text{m}^0/\Omega_k^0$, the value of $a$ at which the initial conditions will be applied ($a=0.01$), and the fact that $B$ and $C$ are expected to be much less than unity -- which makes it possible to expand algebraic functions of $B$ and/or $C$ by using the binomial theorem. To first order in $B$ and $C$, the final expressions read:
\begin{align}
\rho_\text{m}&\approx H_0^2\Omega_\text{m}^0 a^{B-3-3C/2}~,\quad \Omega_\text{m}^0=\Omega_\text{b}^0+\Omega_\text{cdm}^0~,\label{rho_m_approx}\\[0.3em] 
\rho_{\Lambda(H)}&\approx H_0^2\left[\Omega_{\Lambda(H)}^0+\frac{1}{6}(2B-3C)\Omega_\text{m}^0 a^{B-3-3C/2}\right]~.\label{rho_L_approx}
\end{align}
These are used in conjunction with Eq.~(\ref{H}) to obtain an approximation for $\mathcal{H}\,(=aH)$:
\begin{equation}
\mathcal{H}\approx H_0\sqrt{\Omega_\text{m}^0}\left[1+\frac{1}{12}(2B-3C)\right]a^{(2B-3C-2)/4}~,
\label{H_approx}
\end{equation}
while combining Eqs.~(\ref{rho_m_approx})--(\ref{H_approx}) allows us to write $\psi$ as:
\begin{align}
\psi&=-\frac{\rho_{\Lambda(H)}'(\tau)}{\rho_\text{m}}=-\frac{\rho_{\Lambda(H)}'(a)\mathcal{H}a}{\rho_\text{m}}~,\notag\\[0.3em]&\approx \frac{1}{2}(2B-3C)H_0\sqrt{\Omega_\text{m}^0}a^{(2B-3C-2)/4}~.
\end{align}
Next, we insert the above results into Eq.~(\ref{F_and_G}) to estimate $F_1$ and $F_2$, and find that:
\begin{equation}
F_1\approx\frac{1}{4}(6+6B-9C)~,~F_2\approx\frac{1}{2}(-3+2B-3C)~.
\end{equation}
Eq.~(\ref{delta_a}) may finally be solved analytically. The expression we obtain for the density contrast function is a sum of two modes:
\begin{equation}
\delta_\text{m}=A_1a^{1-B+3C/2} + A_2 a^{(-6-2B+3C)/4}~,
\end{equation}
a growing mode and a decaying one. The latter is expected to be subdominant at the redshifts of interest ($z\lesssim100$), and so we only retain the former:
\begin{equation}
\delta_\text{m}=A_1a^{1-B+3C/2}~.
\label{delta-initial}
\end{equation}
This is the `initial' condition that we assign to $\delta_\text{m}$ at $a=0.01$. Taking its derivative with respect to $a$ yields the corresponding initial value for $\delta_\text{m}'(a)$. We shall deal with the constant of integration $(A_1)$ below.

\subsubsection{The standard deviation of density perturbations ($\sigma_{8}$)}
\label{subsubsec: sigma_8}
The variance of the perturbation density field in spheres of radius $R_8=8h^{-1}$ Mpc may be calculated as follows \cite{Sagredo, Eisenstein1999}:
\begin{equation}
\sigma_{8,0}^2=\frac{1}{2\pi^2}\int\limits_0^\infty P(k_\dagger) W^2(k_\dagger) k_\dagger^2 \text{d}k_\dagger~. 
\label{sigma8_today}
\end{equation}
Here, $P(k_\dagger)$ is the present-day power spectrum, and the function $W(k_\dagger)$ represents the Fourier transform of a spherical top-hat window function having radius $R_8$:
\begin{equation}
W(k_\dagger) = \frac{3}{k_\dagger^2R_8^2}\left[\frac{\sin{(k_\dagger R_8)}}{k_\dagger R_8}-\cos{(k_\dagger R_8)}\right]~.
\end{equation}
 We construct an expression for $P(k_\dagger)$ as outlined in Refs.~\cite{Peracaula2018} and \cite{Valent2018} and insert the result into Eq.~(\ref{sigma8_today}), getting that
\begin{equation}
\sigma_8^2(a) = \delta_\text{m}^2(a)\int\limits_0^\infty k_\dagger^{2+n_s}\left[\frac{4 A_s\,k_*^{1-n_s}}{25 H_0^4\left(\Omega_\text{m}^0\right)^2}\right]T^2(k_\dagger)W^2(k_\dagger)\,\text{d}k_\dagger~,
\label{sigma8}
\end{equation}
where use has been made of Eq.~(\ref{f_sigma8}). The absence of the normalizing factor $[\delta_\text{m}(1)]$ is due to the fact that we account for it indirectly by putting the integration constant of Eq.~(\ref{delta-initial}) equal to unity.

The quantities $n_s$ and $A_s$ that appear in Eq.~(\ref{sigma8}) are the index and amplitude of the primordial scalar power spectrum, respectively, defined at a pivot scale $k_*$ of $0.05\,\text{Mpc}^{-1}$  \cite{Aghanim2016}. $A_s$ is either fixed by setting $\ln{\left(10^{10}A_s\right)}$ equal to $3.062$ \cite{Planck2015} or treated as a free parameter. In the latter case, we construct a likelihood for $A_s$ by assuming that it is sampled from a Gaussian distribution whose mean and standard deviation are given by $(2.139\pm 0.063) \times 10^{-9}$ \cite{Planck2015}. As for the primordial spectral index $n_s$, this is allowed to vary subject to the CMB constraint of Ref.~\cite{Huang2015} ($n_s=0.9680\pm0.0051$). Finally, the transfer function $T(k_\dagger)$ -- which describes how perturbations evolve as they cross the horizon and as matter begins to dominate \cite{Dodelson} -- is modeled as specified in the work of Eisenstein and Hu \cite{Eisenstein1998}. This requires that we estimate the wave number $(k_\dagger^\text{eq})$ of the mode that crosses the horizon at matter-radiation equality.  

Our approximation for $k_\dagger^\text{eq}$ is arrived at by following the method of Ref.~\cite{Valent2018}. We discard terms whose order in $B$ and $C$ is higher than linear, and additionally use the fact that the scale factor at equality satisfies $a\sim O(-3)$ to remove subdominant terms. The resulting expression reads:
\begin{align}
k_\dagger^\text{eq}\approx\,&\sqrt{2}H_0\frac{\Omega_\text{m}^0}{\sqrt{\Omega_\text{r}^0}}\Bigg[1-\frac{7B}{6}+\frac{19C}{8}+\frac{2\Omega_k^0}{3\Omega_\text{m}^0}(C-B)+\notag\\[0.3em]&\left(B-\frac{3C}{2}\right)\ln{\left(\frac{\Omega_\text{r}^0}{\Omega_\text{m}^0}\right)}\Bigg]~.
\end{align}

{
\renewcommand{\arraystretch}{1.3}
\begin{table}
\caption{\label{TLSS} LSS data from the compilation presented in Ref.~\cite{Sagredo}. Each $f\sigma_8$ measurement is listed together with the corresponding redshift $z$ and error $\sigma$. Column 5 shows the values of $\Omega_\text{m}^0$ for the respective fiducial cosmologies.} 
\vspace{2mm}
\begin{tabular}{c S[table-format=2.3] S[table-format=2.4] S[table-format=2.4] S[table-format=2.3]} 
\hline
\hline
Ref. & {$z$} & {~~$f\sigma_8(z)$} & {~~$\sigma$} & {\,~$\Omega_{\text{m},0}^\text{fid}$}\\ 
\hline
\cite{Davis, Hudson} & 0.02 & 0.3140 & 0.0480 & 0.266\\
\cite{Song} & 0.17 & 0.5100 & 0.0600 & 0.300\\
\cite{Blake} & 0.18 & 0.3600 & 0.0900 & 0.270\\
\cite{Blake} & 0.38 & 0.4400 & 0.0600 & 0.270\\
\cite{Samushia} & 0.25 & 0.3512 & 0.0583 & 0.250\\
\cite{Samushia} & 0.37 & 0.4602 & 0.0378 & 0.250\\
\cite{Blake2012} & 0.44 & 0.4130 & 0.0800 & 0.270\\
\cite{Blake2012} & 0.60 & 0.3900 & 0.0630 & 0.270\\
\cite{Blake2012} & 0.73 & 0.4370 & 0.0720 & 0.270\\
\cite{Pezzotta} & 0.60 & 0.5500 & 0.1200 & 0.300\\
\cite{Pezzotta} & 0.86 & 0.4000 & 0.1100 & 0.300\\
\cite{Okumura} & 1.40 & 0.4820 & 0.1160 & 0.270\\
\hline
\hline
\end{tabular}
\end{table}
}

\subsubsection{Constructing $\chi^2$}
The data we use (Table \ref{TLSS}) is a subset of the updated Gold-2017 compilation of Nesseris et al.~\cite{Sagredo, Nesseris}. Apart from the $f\sigma_8$ values from Ref.~\cite{Blake2012} -- which are correlated with each other -- all data points are independent. Moreover, measurements derived from the same survey data as any of the observables associated with previous likelihoods (especially the BAO likelihood) are excluded. This is done to avoid potential correlations.
%The issue of correlations is a subtle and complex one. One need only consider the overlaps in the regions of sky probed by different surveys to get an idea. Such details are however beyond the scope of the work presented here, especially since the literature is painfully lacking in calculations of correlations between measurements from different publications, at times even if they are based on the same survey data. The values of $f\sigma_8$ used in this study are listed in Table \ref{TLSS}.  
 
At a given redshift $z_i$, the theoretical prediction $f\sigma_8^{\text{th}}$ is computed by combining  $\sigma_8(z_i)$ with the growth rate $f(z_i)$ returned by \textsc{CLASS}. The quantity $\sigma_8(z_i)$ itself is obtained from Eq.~\ref{sigma8}.
%For the reasons given in subsection \ref{subsubsec: delta_m}, the range of $k$ values over which the integration in Eq. \ref{sigma8} is carried out is reduced to $0.01 h \leq k\,(\text{Mpc}^{-1}) \leq 0.2 h$. 

The $\chi^2$ to be minimised is constructed as follows \cite{Lavrentios}:
 \begin{equation}
 \chi^2=V^iC_{ij}^{-1}V^j~.
 \end{equation}
In the above, $C$ is the covariance matrix, assembled as described in Refs.~\cite{Sagredo} and \cite{Nesseris}, and the vector $V$ contains elements of the form\footnote{There is no summation over $i$ in Eq.~(\ref{Vi}); $z_i$ is simply the redshift of the $i^\text{th}$ data point from Table \ref{TLSS}.}\cite{Lavrentios} 
\begin{equation}
V^i=f\sigma_8^{\text{obs}}(z_i) - \frac{f\sigma_8^{\text{th}}(z_i)}{r_\text{AP}^i}~,
\label{Vi}
\end{equation}
$f\sigma_8^{\text{obs}}(z_i)$ being the $i^\text{th}$ data point from Table \ref{TLSS}. The factor $1/r_\text{AP}^i$ is designed to correct for the Alcock-Paczynski (AP) effect. As was also the case with BAO data, each of the studies from which the values in Table \ref{TLSS} are quoted makes use of a different fiducial cosmology to convert redshifts to distances. If the fiducial cosmology is not the same as the true one, the data incorporates an additional anisotropy that is degenerate with RSDs \cite{Lavrentios}. This is the above-mentioned AP effect \cite{Alcock}. One way of correcting for it involves multiplying the observational value of $f\sigma_8(z_i)$ and its associated error by the ratio \cite{Macaulay}
 \begin{equation}
 r_\text{AP}^i = \frac{H(z_i)d_\text{A}(z_i)}{H^\text{fid}(z_i)d_\text{A}^\text{fid}(z_i)}~.
 \end{equation}
 A superscript `fid' indicates quantities calculated in the framework of the respective fiducial cosmologies (flat $\Lambda$CDM).
 
Alternatively, one may simply rescale $f\sigma_8^{\text{th}}(z_i)$ by $1/r_\text{AP}^i$ \cite{Lavrentios}, as we have done in Eq.~(\ref{Vi}). 
 
\section{Results}
\label{sec:results}
\subsection{Preliminaries}
The joint likelihood on which our analysis is based is specified by the function
\begin{align}
&\mathcal{L}_\text{total} \propto\notag\\[0.3em] &\exp{\left[-\frac{1}{2}\left(\chi^2_\text{JLA}+\chi^2_{H(z)}+\chi^2_\text{CMB}+\chi^2_\text{BAO}+\chi^2_\text{LSS}\right)\right]}~,
\end{align}   
where we have used the relation $\mathcal{L}_i\propto \exp{\left(-\chi^2_i/2\right)}$ for each likelihood considered in section \ref{sec:observational}. The full data set -- consisting of the $\text{JLA}+H(z)+\text{CMB}+\text{BAO}+\text{LSS}$ measurements -- shall be referred to as \textsc{All} + LSS, and as the \textsc{All} data set when LSS observations are excluded.

In order to investigate how our results are affected by the value of the Hubble constant, we constrain the parameters of each model six times, first using the \textsc{All} data set, then extending this to $\text{JLA}+H(z)+\text{CMB}+\text{BAO}+H_0^\text{R}$ $\left(\textsc{All}\,+ H_0^\text{R}\right)$ and to $\text{JLA}+H(z)+\text{CMB}+\text{BAO}+H_0^\text{E}$ $\left(\textsc{All}\,+ H_0^\text{E}\right)$, and finally repeating the whole procedure with the \textsc{All}+LSS data set replacing \textsc{All}. $H_0^\text{R}$ is the value of $H_0$ reported by Riess et al.~\cite{Riess2018} and equates to $73.48\pm1.66~\si{km.s^{-1}.Mpc^{-1}}$. As for $H_0^\text{E}$, this stands for the Hubble constant as inferred by Efstathiou \cite{Efstathiou2014} and amounts to $70.6\pm3.3~\si{km.s^{-1}.Mpc^{-1}}$. We do not include the \emph{Planck} result \cite{Planck2018}, opting instead for values of $H_0$ which were derived independently of any cosmological model. 

One of the topics currently at the forefront of cosmological research is the growing discrepancy between -- on the one hand -- the value of $H_0$ determined locally from Cepheid parallax measurements \cite{Riess2018}, and on the other, that obtained in a $\Lambda$CDM framework using measurements of CMB observables. It turns out that $H_0^\text{R}$ is in a $3.5\sigma$ tension with the \emph{Planck} 2018 value $\left(H_0=67.27\pm0.60\,\text{km\,s}^{-1}\text{Mpc}^{-1}\right)$ \cite{Planck2018}. One reason for the said tension could be the presence of systematic errors in the data used by either group. However, despite the investigative studies carried out, no obvious problem has been identified so far (refer to \cite{Planck2018} and works cited therein). The other possibility is that this discrepancy provides compelling evidence for new physics \cite{Mortsell, DiValentino, Poulin}. Interestingly, a BD-like form of RVM \cite{SolaPeracaula} -- in which the dynamical nature of dark energy arises from the properties of the BD field -- has the potential of alleviating the tension \cite{Peracaula, CruzPerez}.

\renewcommand{\arraystretch}{1.5}
\begin{table}
\caption{\label{priors} The flat priors for the baseline parameters.}
\vspace{4mm}
\begin{tabular}{l@{\hskip 0.5cm} c@{\hskip 0.5cm} c} 
\hline
\hline
Parameter & Min & Max\\ \hline 
$H_0~\left(\mathrm{km~s}^{-1}\mathrm{Mpc}^{-1}\right)$ & 50 & 95\\ 
$\Omega_{\mathrm{b}}^0 h^2$ & 0.005 & 0.100\\ 
$\Omega_{\mathrm{cdm}}^0 h^2$ & 0.01 & 0.99\\ 
$\Omega_{k}^0$ & -0.3 & 0.3\\
$B$ & -1.0 & 1.0\\ 
$C$ & -1.0 & 1.0\\ 
$n_s$ & 0.75 & 1.25\\
\hline
\hline
\end{tabular}
\end{table}

In conclusion, the lack of consensus about the value of the Hubble constant makes it imperative to consider different options for $H_0$. This is especially true since, as we later find out from the posterior probability plots, there is significant correlation between $H_0$ and the model parameters $B$ and $C$.

We use the likelihood combinations just described and run MCMC chains to place constraints on the parameters of the GRVM, RVM and GRVS. The general baseline set $\Theta$ is given by\footnote{The four nuisance parameters associated with the JLA likelihood ($\alpha$, $\beta$, $M$ and $\Delta M$) also form part of $\Theta$.} $\{H_0, \Omega_\text{b}^0 h^2, \Omega_\text{cdm}^0 h^2, \Omega_k^0, B,\,C,\,n_s\}$, although $\Omega_k^0$ is set to zero when we assume a flat space-time, and so is $C$ ($B$) when we study the RVM (GRVS). Furthermore, the primordial spectral index is only considered a free parameter if LSS observations are included in the data sets. The reason is that $n_s$ features explicitly in the LSS likelihood but is of minimal importance otherwise. We also differentiate between two scenarios in relation to the amplitude of the primordial power spectrum: the case with a fixed value of $A_s$, and that in which $A_s$ is incorporated into $\Theta$ and allowed to vary freely. However, the two approaches yield very similar results, and consequently we shall not distinguish between them when discussing the effects of introducing growth data. 

The flat priors for the main baseline parameters are listed in Table \ref{priors}. We additionally note that the reionization redshift $z_\text{reio}$ is set to 8.8 \cite{Planck2015}, while all other parameters take the CLASS default values.\footnote{The exception is $A_s$, and only when LSS data is included in the analysis. More details are provided in subsection \ref{subsubsec: sigma_8}.} In particular, this implies that the effective number of relativistic neutrino species ($N_\text{eff}$) is fixed at $3.046$ \cite{Mangano2005} and the current CMB temperature ($T_\text{CMB}$) at \SI{2.7255}{K} \cite{Fixsen2009}.  

\subsection{The GRVM}
\begin{figure}[ht!]
\hspace*{-0.5cm}
\includegraphics[width=0.47\textwidth]{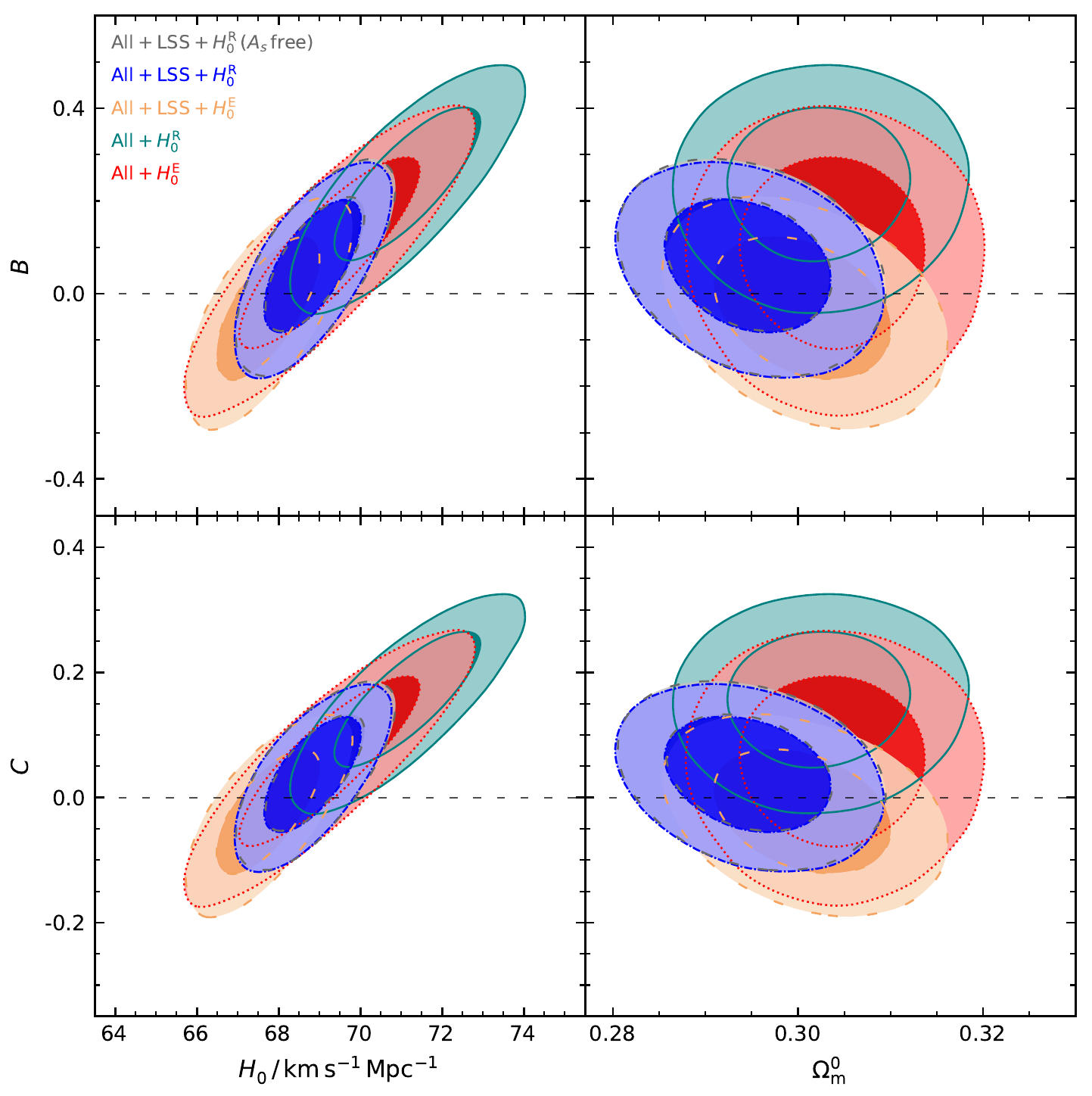}
\caption{\label{GRVMf}\emph{(top panel)} Marginalized posterior probability distributions for the GRVM parameter $B$ vs \emph{(left)} $H_0$, \emph{(right)} $\Omega_\text{m}^0$. The bottom panel shows analogous plots for the second GRVM parameter, $C$. Darker (lighter) shades denote $1\sigma$ ($2\sigma$) confidence regions (the first two data sets listed in the legend produce contours that overlap almost exactly). We assume a spatially flat space-time.}
\end{figure}

\begin{figure*}[ht!]
\hspace*{-0.5cm}
\includegraphics[width=0.7\textwidth]{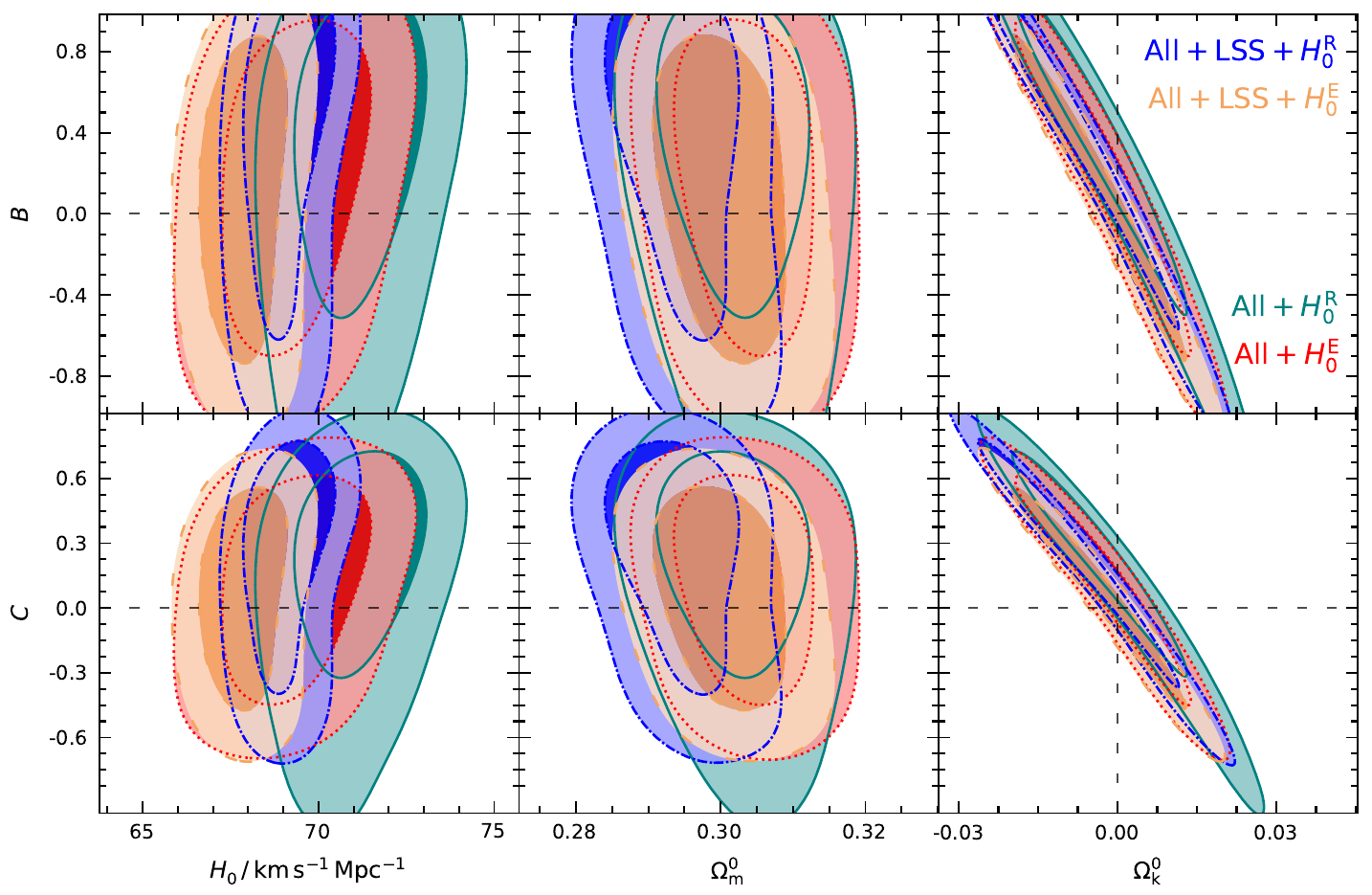}
\caption{\label{GRVMnf}\emph{(top panel)} Marginalized posterior probability distributions for the GRVM parameter $B$ vs \emph{(left)} $H_0$, \emph{(center)} $\Omega_\text{m}^0$, \emph{(right)} $\Omega_k^0$. The bottom panel shows analogous plots for the second GRVM parameter, $C$.}
\end{figure*}

The GRVM is characterized by two highly-correlated parameters, $B$ and $C$. The constraints we get in the flat case are nonetheless tight enough to be informative (Fig.~\ref{GRVMf}), but when $\Omega_k^0$ is allowed to vary, the data we use proves insufficient to break the degeneracy between $B$ and $C$ or between $\Omega_k^0$ and the model parameters (Fig.~\ref{GRVMnf}).

\renewcommand{\arraystretch}{1.5}
\setcounter{table}{6}
\begin{table}[ht!]
\caption{\label{TGRVMf-noLSS} Mean values and $1\sigma$ confidence limits obtained with each data set combination in the context of a flat GRVM scenario. LSS observations were excluded from the analysis. In the top block we present results for the baseline parameters, whereas in the last row we report constraints on the derived parameter $\Omega_{\Lambda(H)}^0$. $H_0$ is quoted in units of $\mathrm{km~s}^{-1}\,\mathrm{Mpc}^{-1}$.}
\vspace{2mm}
\begin{tabular}{l@{\hskip 0.3cm} A@{\hskip 0.3cm} A@{\hskip 0.3cm} A}  
\hline
\hline
\multicolumn{7}{l}{Parameter  ~~~~~~~~$\textsc{All}$ \qquad \quad~~\,$\textsc{All}+H_0^\text{R}$ \quad\quad\,~$\textsc{All}+H_0^\text{E}$}\\ \hline  
$H_0$ & 68.8330&_{-1.6725}^{+1.6845} & 71.1120&_{-1.1912}^{+1.2073} & 69.1850&_{-1.4954}^{+1.4937}\\
$10^3\,\Omega_{\mathrm{b}}^0 h^2$ & 22.4090&_{-0.1744}^{+0.1755} & 22.4190&_{-0.1726}^{+0.1747} & 22.4110&_{-0.1732}^{+0.1735}\\ 
$\Omega_{\mathrm{cdm}}^0 h^2$ & 0.1217&_{-0.0073}^{+0.0070} & 0.1305&_{-0.0059}^{+0.0056} & 0.1230&_{-0.0066}^{+0.0064}\\
$B$ & \phantom{-}0.0555&_{-0.1430}^{+0.1660} & \phantom{-}0.2341&_{-0.1069}^{+0.1186} & \phantom{-}0.0845&_{-0.1306}^{+0.1491}\\ 
$C$ & 0.0365&_{-0.0940}^{+0.1096} & 0.1553&_{-0.0701}^{+0.0776} & 0.0558&_{-0.0858}^{+0.0980}\\ \hline
$\Omega_{\Lambda(H)}^0$ & 0.6958&_{-0.0067}^{+0.0070} & 0.6976&_{-0.0065}^{+0.0069} & 0.6961&_{-0.0066}^{+0.0069}\\
\hline
\hline
\end{tabular}
\end{table} 

\renewcommand{\arraystretch}{1.5}
\begin{table*}[ht!]
\caption{\label{TGRVMf} Mean values and $1\sigma$ confidence limits obtained with each data set combination in the context of a flat GRVM scenario. In the top block we present constraints on the baseline parameters, whereas the last row features the derived parameter $\Omega_{\Lambda(H)}^0$. $H_0$ is quoted in units of $\mathrm{km~s}^{-1}\,\mathrm{Mpc}^{-1}$, and double dashes indicate cases in which $\ln(10^{10}A_s)$ was fixed at $3.062$ \cite{Planck2015}.}
\vspace{2mm}
\begin{tabular}{l@{\hskip 0.5cm} A@{\hskip 0.4cm} A@{\hskip 0.4cm} A@{\hskip 0.4cm} A@{\hskip 0.4cm} A@{\hskip 0.4cm} A}  
\hline
\hline
%\multicolumn{13}{l}{Parameter  ~~~~~~~~~~~~~$\textsc{All}$ \qquad \quad~~\,$\textsc{All}+H_0^\text{R}$ \quad\quad\,~\,$\textsc{All}+H_0^\text{E}$ \quad~~~\qquad\textsc{All} \quad\qquad\quad\,$\textsc{All}+H_0^\text{R}$ \quad\quad~\,\,$\textsc{All}+H_0^\text{E}$}\\ \hline 
Parameter & \multicolumn{2}{c}{\textsc{All}+LSS~~\,} & \multicolumn{2}{c}{\textsc{All}+LSS+$H_0^\text{R}$~~} & \multicolumn{2}{c}{\textsc{All}+LSS+$H_0^\text{E}$~~} & \multicolumn{2}{c}{\textsc{All}+LSS~~~} & \multicolumn{2}{c}{\textsc{All}+LSS+$H_0^\text{R}$~~} & \multicolumn{2}{c}{\textsc{All}+LSS+$H_0^\text{E}$~~}\\ \hline
$H_0$&67.5240&_{-0.8880}^{+0.8710}&68.8020&_{-0.8142}^{+0.7935}&67.7140&_{-0.8635}^{+0.8367}&67.5440&_{-0.8936}^{+0.8705}&68.8580&_{-0.8326}^{+0.8092}&67.7460&_{-0.8706}^{+0.8515}\\ 
$10^3\Omega_{\mathrm{b}}^0 h^2$&22.4140&_{-0.1732}^{+0.1750}&22.4470&_{-0.1742}^{+0.1729}&22.4210&_{-0.1743}^{+0.1737}&22.4160&_{-0.1734}^{+0.1732}&22.4460&_{-0.1740}^{+0.1754}&22.4200&_{-0.1724}^{+0.1728}\\ 
$\Omega_{\mathrm{cdm}}^0 h^2$&0.1151&_{-0.0025}^{+0.0024}&0.1170&_{-0.0024}^{+0.0024}&0.1154&_{-0.0025}^{+0.0024}&0.1152&_{-0.0026}^{+0.0025}&0.1173&_{-0.0026}^{+0.0025}&0.1155&_{-0.0026}^{+0.0025}\\ 
$B$&-0.0491&_{-0.1014}^{+0.1118}&0.0571&_{-0.0933}^{+0.1022}&-0.0338&_{-0.0996}^{+0.1088}&-0.0483&_{-0.1027}^{+0.1114}&0.0619&_{-0.0937}^{+0.1026}&-0.0301&_{-0.1000}^{+0.1093}\\ 
$C$&-0.0340&_{-0.0655}^{+0.0719}&0.0353&_{-0.0602}^{+0.0657}&-0.0240&_{-0.0644}^{+0.0700}&-0.0334&_{-0.0662}^{+0.0718}&0.0385&_{-0.0604}^{+0.0661}&-0.0215&_{-0.0644}^{+0.0704}\\ 
$n_s$&0.9683&_{-0.0052}^{+0.0052}&0.9688&_{-0.0052}^{+0.0053}&0.9684&_{-0.0052}^{+0.0053}&0.9683&_{-0.0052}^{+0.0052}&0.9688&_{-0.0052}^{+0.0052}&0.9683&_{-0.0052}^{+0.0052}\\ 
$\ln{\left(10^{10}A_s\right)}$& -&- & -&- & -&- &3.0580&_{-0.0294}^{+0.0310}&3.0503&_{-0.0298}^{+0.0314}&3.0571&_{-0.0296}^{+0.0309}\\ \hline
$\Omega_{\Lambda(H)}^0$&0.6983&_{-0.0065}^{+0.0067}&0.7052&_{-0.0059}^{+0.0062}&0.6993&_{-0.0064}^{+0.0066}&0.6982&_{-0.0064}^{+0.0068}&0.7051&_{-0.0060}^{+0.0062}&0.6993&_{-0.0063}^{+0.0066}\\
%$\sigma_{8,0}$ & \multicolumn{2}{c}{0.7660} & \multicolumn{2}{c}{0.8010} & \multicolumn{2}{c}{0.7714} & \multicolumn{2}{c}{0.7663} & \multicolumn{2}{c}{0.8000} & \multicolumn{2}{c}{0.7721}\\
\hline
\hline
\end{tabular}
\end{table*} 

\renewcommand{\arraystretch}{1.5}
\begin{table}[ht!]
\caption{\label{TGRVMnf-noLSS} Mean values and $1\sigma$ confidence limits obtained in the context of a GRVM scenario. LSS data was excluded from the analysis, and the condition of spatial flatness was not imposed. More details may be found in the caption of Table \ref{TGRVMf-noLSS}~.}
\vspace{2mm}
\begin{tabular}{l@{\hskip 0.3cm} A@{\hskip 0.3cm} A@{\hskip 0.3cm} A}  
\hline
\hline
\multicolumn{7}{l}{Parameter  ~~~~~~~~$\textsc{All}$ \qquad \quad~~\,$\textsc{All}+H_0^\text{R}$ \quad\quad\,~$\textsc{All}+H_0^\text{E}$}\\ \hline
$H_0$ & 68.8780&_{-1.6844}^{+1.6279} & 71.0800&_{-1.1827}^{+1.1760} & 69.2380&_{-1.5051}^{+1.4922}\\
$10^3\,\Omega_{\mathrm{b}}^0 h^2$ & 22.4050&_{-0.1715}^{+0.1754} & 22.4180&_{-0.1739}^{+0.1698} & 22.4060&_{-0.1725}^{+0.1719}\\  
$\Omega_{\mathrm{cdm}}^0 h^2$ & 0.1219&_{-0.0072}^{+0.0069} & 0.1303&_{-0.0058}^{+0.0055} & 0.1232&_{-0.0067}^{+0.0063}\\ 
$\Omega_k^0$ & -0.0019&_{-0.0110}^{+0.0121} & -0.0020&_{-0.0150}^{+0.0067} & -0.0023&_{-0.0124}^{+0.0109}\\ 
$B$ & 0.1068&_{-0.3660}^{+0.7911} & 0.2685&_{-0.1610}^{+0.7313} & 0.1587&_{-0.2599}^{+0.8389}\\ 
$C$ & 0.0688&_{-0.2378}^{+0.5045} & 0.1765&_{-0.1036}^{+0.4745} & 0.1027&_{-0.1861}^{+0.5202}\\ \hline
$\Omega_{\Lambda(H)}^0$ & 0.6978&_{-0.0144}^{+0.0143} & 0.6998&_{-0.0110}^{+0.0173} & 0.6986&_{-0.0140}^{+0.0145}\\ 
\hline
\hline
\end{tabular}
\end{table}

\renewcommand{\arraystretch}{1.5}
\begin{table*}[ht!]
\caption{\label{TGRVMnf} Mean values and $1\sigma$ confidence limits obtained in the context of a GRVM scenario. $\Omega_k^0$ was treated as a free parameter. More details may be found in the caption of Table \ref{TGRVMf}~.}
%\vspace{2mm}
\begin{tabular}{l@{\hskip 0.5cm} A@{\hskip 0.4cm} A@{\hskip 0.4cm} A@{\hskip 0.4cm} A@{\hskip 0.4cm} A@{\hskip 0.4cm} A}  
\hline
\hline
%\multicolumn{13}{l}{Parameter  ~~~~~~~~~~~~~$\textsc{All}$ \qquad \quad~~\,$\textsc{All}+H_0^\text{R}$ \quad\quad\,~\,~$\textsc{All}+H_0^\text{E}$ \quad~~~\qquad\textsc{All} \quad\qquad\quad\,\,$\textsc{All}+H_0^\text{R}$ \quad\quad~\,~$\textsc{All}+H_0^\text{E}$}\\ \hline
Parameter & \multicolumn{2}{c}{\textsc{All}+LSS~~~} & \multicolumn{2}{c}{\textsc{All}+LSS+$H_0^\text{R}$~~} & \multicolumn{2}{c}{\textsc{All}+LSS+$H_0^\text{E}$~~} & \multicolumn{2}{c}{\textsc{All}+LSS~~~} & \multicolumn{2}{c}{\textsc{All}+LSS+$H_0^\text{R}$~~} & \multicolumn{2}{c}{\textsc{All}+LSS+$H_0^\text{E}$}\\ \hline
$H_0$&67.7460&_{-0.9245}^{+0.8958}&69.0740&_{-0.8411}^{+0.7971}&67.8870&_{-0.8824}^{+0.8656}&67.7420&_{-0.9459}^{+0.9102}&69.1410&_{-0.8330}^{+0.8361}&67.9370&_{-0.8964}^{+0.8750}\\ 
$10^3\Omega_{\mathrm{b}}^0 h^2$&22.4070&_{-0.1729}^{+0.1716}&22.4430&_{-0.1756}^{+0.1723}&22.4140&_{-0.1733}^{+0.1743}&22.4090&_{-0.1730}^{+0.1716}&22.4400&_{-0.1741}^{+0.1664}&22.4180&_{-0.1709}^{+0.1739}\\ 
$\Omega_{\mathrm{cdm}}^0 h^2$&0.1157&_{-0.0028}^{+0.0026}&0.1180&_{-0.0027}^{+0.0026}&0.1160&_{-0.0028}^{+0.0026}&0.1159&_{-0.0028}^{+0.0027}&0.1183&_{-0.0026}^{+0.0027}&0.1162&_{-0.0028}^{+0.0027}\\ 
$\Omega_k^0$&-0.0050&_{-0.0141}^{+0.0106}&-0.0073&_{-0.0155}^{+0.0064}&-0.0029&_{-0.0124}^{+0.0105}&-0.0035&_{-0.0129}^{+0.0111}&-0.0062&_{-0.0160}^{+0.0071}&-0.0040&_{-0.0149}^{+0.0108}\\ 
$B$&0.1637&_{-0.2790}^{+0.8362}&0.3591&_{-0.2050}^{+0.6489}&0.0834&_{-0.4039}^{+0.6882}&0.0979&_{-0.4212}^{+0.7169}&0.3148&_{-0.3449}^{+0.6953}&0.1248&_{-0.3556}^{+0.8194}\\ 
$C$&0.1019&_{-0.1816}^{+0.5339}&0.2282&_{-0.1312}^{+0.4148}&0.0508&_{-0.2577}^{+0.4411}&0.0599&_{-0.2687}^{+0.4592}&0.2000&_{-0.2230}^{+0.4444}&0.0773&_{-0.2270}^{+0.5241}\\ 
$n_s$&0.9680&_{-0.0051}^{+0.0052}&0.9687&_{-0.0052}^{+0.0050}&0.9682&_{-0.0052}^{+0.0053}&0.9681&_{-0.0052}^{+0.0051}&0.9686&_{-0.0052}^{+0.0051}&0.9682&_{-0.0053}^{+0.0049}\\ 
$\ln{\left(10^{10}A_s\right)}$&-&-&-&-&-&-&3.0584&_{-0.0293}^{+0.0311}&3.0514&_{-0.0275}^{+0.0309}&3.0578&_{-0.0293}^{+0.0302}\\ \hline
$\Omega_{\Lambda(H)}^0$&0.7038&_{-0.0132}^{+0.0170}&0.7128&_{-0.0091}^{+0.0193}&0.7024&_{-0.0130}^{+0.0154}&0.7021&_{-0.0144}^{+0.0152}&0.7116&_{-0.0102}^{+0.0198}&0.7034&_{-0.0141}^{+0.0176}\\
%$\sigma_{8,0}$&\multicolumn{2}{c}{0.7751}&\multicolumn{2}{c}{0.8141}&\multicolumn{2}{c}{0.7772}&\multicolumn{2}{c}{0.7732}&\multicolumn{2}{c}{0.8118}&\multicolumn{2}{c}{0.7787} \\
\hline
\hline
\end{tabular}
\end{table*}
%Nonetheless, one notes that the $\textsc{All} + \textsc{LSS} + H_0^\text{R}$ data set produces tapered contour plots and thus goes some way towards achieving this. 

The challenges posed by the fact that $B$ is correlated with $C$ are also highlighted in Ref.~\cite{Valent2015}. In this work, the authors find a way around the problem by defining a particular combination of $\nu\,(\,=B/3)$ and $\alpha\,(\,=C/2)$ as another effective parameter -- labeled $\nu_\text{eff}$ -- that is then constrained instead of the original two. They do this by making the approximation
\begin{equation}
\xi=\frac{1-\nu}{1-\alpha}\sim1-(\nu-\alpha)\equiv 1-\nu_\text{eff}~,
\end{equation}
which is justified on the basis that $|\nu|$ and $|\alpha|$ must both be much smaller than unity if the deviation from $\Lambda$CDM is to be mild. The parameter $\xi$ controls the way the matter energy density ($\rho_\text{m}$) scales with $a$, and for the purposes of data fitting the authors assume that $\rho_\text{r}$ evolves as in the standard model. 

There are several reasons, however, why the approach outlined in Ref.~\cite{Valent2015} cannot be taken here. To begin with, the authors determine $\rho_\text{m}$ and $\rho_\text{r}$ in terms of $a$ by considering the cosmic fluid to have only two components at any given time -- dynamical dark energy and either matter or radiation, depending on which of the two dominates. The expressions thus obtained are then used to formulate $\rho_{\Lambda(H)}$ as a function of $\rho_\text{m}$ and $\rho_\text{r}$. The fact that we do not simplify our analysis likewise introduces more terms into the relevant equations, as does our decision to treat $\Omega_k^0$ as a free parameter for part of the study. In conclusion, the relations we get for $\rho_\text{cdm}$ and $\rho_{\Lambda(H)}$ -- Eqs.~($\ref{rhocdm}$) and ($\ref{rhoLambdaFinal}$), respectively -- include several different combinations of $B$ and $C$, so that it is not possible to reduce the number of degrees of freedom as detailed in Ref.~\cite{Valent2015}. 

\renewcommand{\arraystretch}{1.5}
\begin{table}[ht!]
\caption{\label{TRVMf-noLSS} Mean values and $1\sigma$ confidence limits obtained in the context of a flat RVM scenario. LSS data was excluded from the analysis. More details may be found in the caption of Table \ref{TGRVMf-noLSS}~.} 
%In the top block we present results for the baseline parameters, whereas in the last row we report constraints on the derived parameter $\Omega_\Lambda^0$. $H_0$ is quoted in units of $\mathrm{km~s}^{-1}\,\mathrm{Mpc}^{-1}$.}
\vspace{2mm}
\begin{tabular}{l@{\hskip 0.3cm} A@{\hskip 0.3cm} A@{\hskip 0.3cm} A}  
\hline
\hline
\multicolumn{7}{l}{Parameter  ~~~~~~~\,$\textsc{All}$ \qquad \quad~~\,$\textsc{All}+H_0^\text{R}$ \quad\quad\,~$\textsc{All}+H_0^\text{E}$}\\ \hline
$H_0$ & 67.4240&_{-0.4944}^{+0.4759} & 67.8950&_{-0.4939}^{+0.4785} & 67.4900&_{-0.4939}^{+0.4723}\\ 
${10^3\,\Omega_\text{b}^0 h^2 }$ & 22.3030&_{-0.1387}^{+0.1380} & 22.3830&_{-0.1373}^{+0.1378} & 22.3140&_{-0.1383}^{+0.1393}\\
${\Omega_\text{cdm}^0 h^2 }$ & \phantom{-}0.1175&_{-0.0035}^{+0.0032} & \phantom{-}0.1190&_{-0.0036}^{+0.0033} & \phantom{-}0.1177&_{-0.0035}^{+0.0032}\\
$10^3B$ & 3.0279&_{-3.1698}^{+3.1358} & 1.0225&_{-3.1745}^{+3.1603} & 2.7498&_{-3.1691}^{+3.1121}\\ \hline
$\Omega_{\Lambda(H)}^0$ & 0.6925&_{-0.0064}^{+0.0066} & 0.6932&_{-0.0064}^{+0.0067} & 0.6926&_{-0.0064}^{+0.0066}\\ 
\hline
\hline
\end{tabular}
\end{table}

The constraints we obtain in the context of a flat geometry are nonetheless instructive. The most prominent feature of Fig.~\ref{GRVMf} is the shift in the marginalized 2D posteriors that is brought about by the addition of LSS data. Table \ref{TGRVMf} shows that (in the absence of the $H_0^\text{R}$ likelihood) this shift results in negative mean values for $B$ and $C$ -- rather than the positive ones we get otherwise (Table \ref{TGRVMf-noLSS}). A second characteristic which emerges in Fig.~\ref{GRVMf} is the correlation between $B$ (or $C$) and $H_0$. In the case of $B$, this behavior is in stark contrast with the negative correlation observed in the RVM scenario (Figs.~\ref{RVMf} and \ref{RVMnf}). The fact that a larger value of $H_0$ favors a larger $B$ explains why, in the top panel of Fig.~\ref{GRVMf}\,, the contours obtained with the $\textsc{All} + H_0^\text{R}$ and $\textsc{All} + \text{LSS} + H_0^\text{R}$ data sets have a marked shift in the direction of increasing $B$ relative to their $H_0^\text{E}$ counterparts. The same holds true for $C$ (Fig.~\ref{GRVMf}, bottom panel). Consequently, in the context of a flat geometry, the $\textsc{All} + H_0^\text{R}$ mean values of $B$ and $C$ are inconsistent with zero within a 1$\sigma$ confidence interval. However, Fig.~\ref{GRVMf} plainly demonstrates that the introduction of growth data causes the contours to close around the $\Lambda$CDM limit. Additionally, the 2D posteriors for $B$ (or $C$) vs $H_0$ make it clear that LSS data lends support to the Hubble constant as established by \emph{Planck} $\left(H_0=67.27\pm0.60\,\text{km\,s}^{-1}\text{Mpc}^{-1}\right)$ \cite{Planck2018}, rather than to $H_0^\text{R}$. This may be observed in both the flat and non-flat cases (results for the latter are shown in Tables \ref{TGRVMnf-noLSS} and \ref{TGRVMnf}). We find that even the $\textsc{All}+H_0^\text{R}$ mean values for $H_0$ become more compatible with the \emph{Planck} constraints when we add the LSS likelihood. Moreover, the Hubble constant from \emph{Planck} is endorsed irrespectively of whether $A_s$ is allowed to vary, which makes it less likely that this is an indirect consequence of using the $\Lambda$CDM value for $A_s$. Before the possibility can be ruled out, however, one would need to repeat the procedure with a wider Gaussian likelihood for $A_s$.\footnote{If we opt for a flat prior instead, the LSS likelihood attempts to make model predictions compatible with data by `picking' values of $A_s$ well outside the established range.} 

%Interestingly, the $2\sigma$ confidence interval that the \textsc{All} dataset yields for the Hubble constant (in the flat case) has an upper bound of $72.19\,\text{km\,s}^{-1}\text{Mpc}^{-1}$, meaning that it intersects the $1\sigma$ confidence interval of the Riess value for $H_0$. This is noteworthy, especially since the \textsc{All} dataset contains CMB data. One recalls that the CMB constraint on $H_0$ obtained within the $\Lambda$CDM framework is in notable tension with the Riess et al. measurement \cite{Planck2018}.

\subsection{The RVM}

\renewcommand{\arraystretch}{1.5}
\begin{table*}[ht!]
\caption{\label{TRVMf} Mean values and $1\sigma$ confidence limits obtained in the context of a flat RVM scenario. More details may be found in the caption of Table \ref{TGRVMf}~.}
\vspace{2mm}
\begin{tabular}{l@{\hskip 0.5cm} A@{\hskip 0.4cm} A@{\hskip 0.4cm} A@{\hskip 0.4cm} A@{\hskip 0.4cm} A@{\hskip 0.4cm} A}  
\hline
\hline
%\multicolumn{13}{l}{Parameter  ~~~~~~~~~~~~$\textsc{All}$ \qquad \quad~~\,$\textsc{All}+H_0^\text{R}$ \quad\quad\,~$\textsc{All}+H_0^\text{E}$ \quad~~\,\qquad\textsc{All} \quad\qquad\quad$\textsc{All}+H_0^\text{R}$ \quad\quad~~$\textsc{All}+H_0^\text{E}$}\\ \hline 
Parameter & \multicolumn{2}{c}{\textsc{All}+LSS~~~\,} & \multicolumn{2}{c}{\textsc{All}+LSS+$H_0^\text{R}$~~} & \multicolumn{2}{c}{\textsc{All}+LSS+$H_0^\text{E}$~~} & \multicolumn{2}{c}{\textsc{All}+LSS~~~\,} & \multicolumn{2}{c}{\textsc{All}+LSS+$H_0^\text{R}$~~} & \multicolumn{2}{c}{\textsc{All}+LSS+$H_0^\text{E}$~~}\\ \hline
$H_0$&67.8130&_{-0.6275}^{+0.6130}&68.4770&_{-0.5896}^{+0.5816}&67.9030&_{-0.6135}^{+0.6059}&67.8270&_{-0.6319}^{+0.6179}&68.5050&_{-0.6036}^{+0.5812}&67.9160&_{-0.6155}^{+0.6059}\\ 
$10^3\Omega_{\mathrm{b}}^0 h^2$&22.3780&_{-0.1561}^{+0.1568}&22.4970&_{-0.1497}^{+0.1517}&22.3940&_{-0.1551}^{+0.1551}&22.3800&_{-0.1560}^{+0.1550}&22.4990&_{-0.1509}^{+0.1512}&22.3960&_{-0.1547}^{+0.1548}\\ 
$\Omega_{\mathrm{cdm}}^0 h^2$&0.1158&_{-0.0019}^{+0.0019}&0.1161&_{-0.0019}^{+0.0019}&0.1159&_{-0.0019}^{+0.0019}&0.1159&_{-0.0020}^{+0.0019}&0.1163&_{-0.0020}^{+0.0019}&0.1160&_{-0.0020}^{+0.0019}\\ 
$10^3 B$&3.6665&_{-2.0391}^{+1.9661}&2.3309&_{-1.9689}^{+1.8986}&3.4801&_{-2.0277}^{+1.9351}&3.5647&_{-2.1165}^{+2.0408}&2.1456&_{-2.0357}^{+1.9740}&3.3725&_{-2.0964}^{+2.0146}\\ 
$n_s$&0.9670&_{-0.0046}^{+0.0045}&0.9705&_{-0.0044}^{+0.0044}&0.9675&_{-0.0045}^{+0.0045}&0.9671&_{-0.0045}^{+0.0046}&0.9706&_{-0.0044}^{+0.0044}&0.9675&_{-0.0045}^{+0.0045}\\ 
$\ln{\left(10^{10}A_s\right)}$&-&-&-&-&-&-&3.0564&_{-0.0294}^{+0.0307}&3.0525&_{-0.0294}^{+0.0311}&3.0559&_{-0.0293}^{+0.0308}\\ \hline
$\Omega_{\Lambda(H)}^0$&0.6994&_{-0.0061}^{+0.0063}&0.7043&_{-0.0058}^{+0.0060}&0.7000&_{-0.0060}^{+0.0062}&0.6993&_{-0.0060}^{+0.0063}&0.7042&_{-0.0058}^{+0.0060}&0.6999&_{-0.0060}^{+0.0062}\\
%$\sigma_{8,0}$&\multicolumn{2}{c}{0.7748}&\multicolumn{2}{c}{0.7909}&\multicolumn{2}{c}{0.7773}&\multicolumn{2}{c}{0.7741}&\multicolumn{2}{c}{0.7900}&\multicolumn{2}{c}{0.7764}\\
\hline
\hline
\end{tabular}
\end{table*}

\begin{figure*}[ht!]
\hspace*{-1.3cm}
\includegraphics[width=0.93\textwidth]{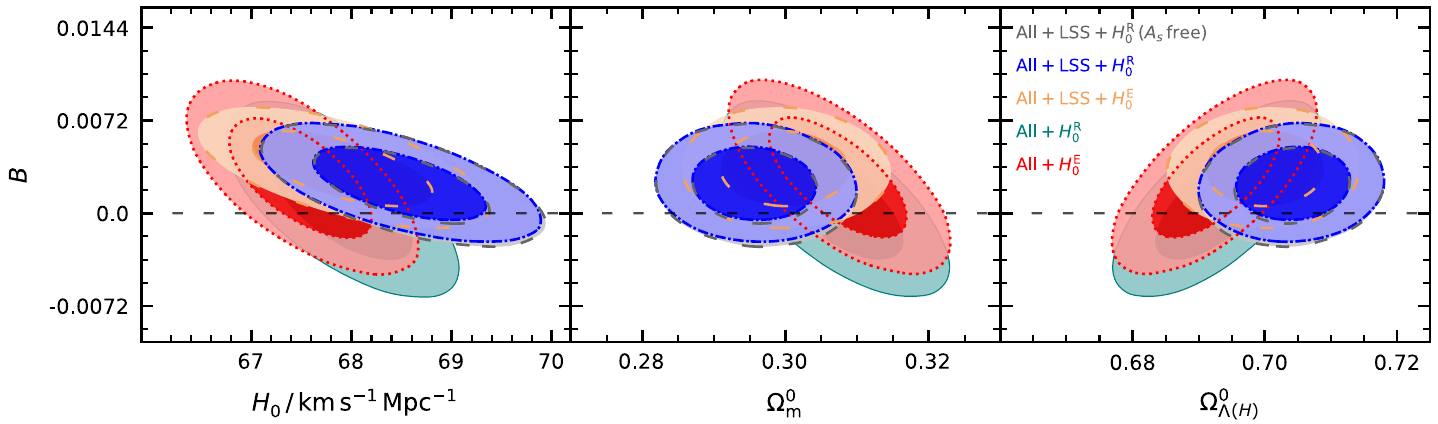}
\caption{\label{RVMf} Marginalized posterior probability distributions for the RVM parameter $B$ vs \emph{(left)} $H_0$, \emph{(center)} $\Omega_\text{m}^0$, \emph{(right)} $\Omega_{\Lambda(H)}^0$. We assume a spatially flat space-time.}
\end{figure*}

\begin{figure*}[ht!]
\hspace*{-1.3cm}
\includegraphics[width=0.93\textwidth]{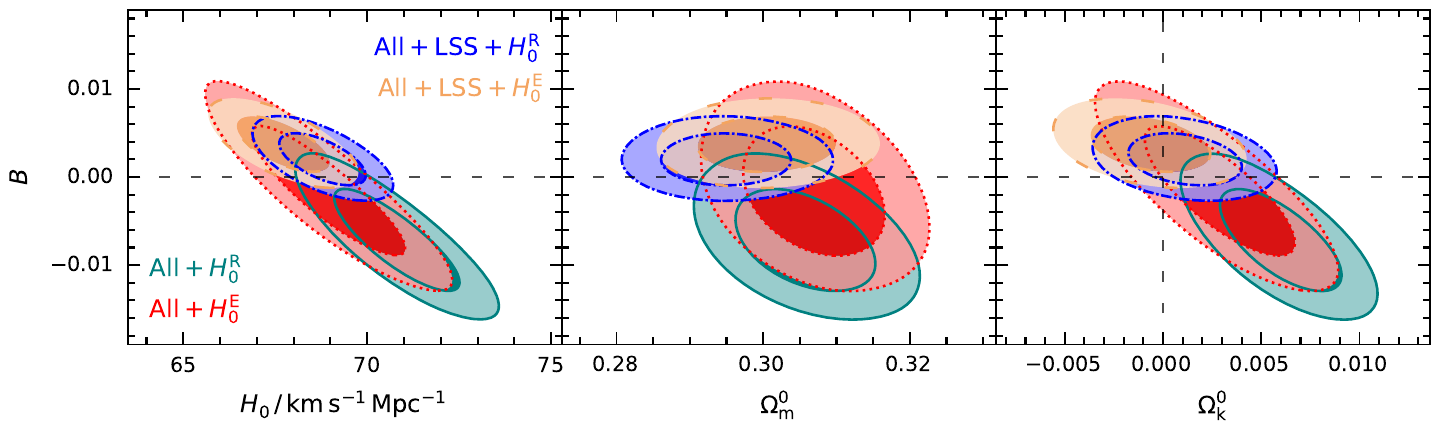}
\caption{\label{RVMnf} Marginalized posterior probability distributions for the RVM parameter $B$ vs \emph{(left)} $H_0$, \emph{(center)} $\Omega_\text{m}^0$, \emph{(right)} $\Omega_k^0$.}
\end{figure*}

The results for the RVM are summarized in Tables \ref{TRVMf-noLSS}--\ref{TRVMnf}. As can be deduced from Figs.~$\ref{RVMf}$ and $\ref{RVMnf}$ (left panel), there is significant negative correlation between the model parameter $B$ and the Hubble constant $H_0$, although in the flat case the use of LSS data makes this much less pronounced. The said correlation explains why including $H_0^\text{R}$ with the observational data -- rather than the lower value of $H_0^\text{E}$ -- shifts the corresponding contours in all the plots of Figs.~$\ref{RVMf}$ and $\ref{RVMnf}$ downwards, in the direction of decreasing $B$.

The introduction of LSS data is a game-changer. In the flat scenario, it reduces or even neutralizes the positive/negative correlation between $B$ and the parameters $H_0$, $\Omega_\text{m}^0$ and $\Omega_{\Lambda(H)}^0$ (Fig.~\ref{RVMf}). This makes the constraints on $B$ less compatible with the $\Lambda$CDM limit, and indeed the new 1D posteriors for $B$ exclude a null value at a little over $1\sigma$. Similar behavior is noted when $\Omega_k^0$ is allowed to vary. A few differences are worth mentioning, however: in the presence of spatial curvature, the effect of the LSS likelihood on the negative correlation between $B$ and $H_0$ is less significant (Fig.~\ref{RVMnf}). Furthermore, comparison of the average values of $B$ in Tables \ref{TRVMnf-noLSS} and \ref{TRVMnf} reveals that the addition of growth data changes their sign from negative to positive (which results from the tightening of contours around positive values of $B$; see Fig.~\ref{RVMnf}). Another point of interest is the fact that while the $\textsc{All}+H_0^\text{R}$ posteriors favor an open universe at more than $2\sigma$, the inclusion of LSS data causes them to close up around $\Omega_k=0$. Moreover, we note that the tendency of the LSS likelihood to select smaller values for the Hubble constant emerges again in the non-flat case.  It may easily be deduced that the resulting mean values of $H_0$ resonate with the \emph{Planck} constraint rather than with $H_0^\text{R}$.

\renewcommand{\arraystretch}{1.5}
\begin{table}[ht!]
\caption{\label{TRVMnf-noLSS} Mean values and $1\sigma$ confidence limits obtained in the context of an RVM scenario. LSS data was excluded from the analysis, and spatial flatness was not imposed. More details may be found in the caption of Table \ref{TGRVMf-noLSS}~.}
\vspace{2mm}
\begin{tabular}{l@{\hskip 0.3cm} A@{\hskip 0.3cm} A@{\hskip 0.3cm} A}  
\hline
\hline
\multicolumn{7}{l}{Parameter  ~~~~~~~~$\textsc{All}$ \qquad \quad~~$\textsc{All}+H_0^\text{R}$ \quad\quad\,~$\textsc{All}+H_0^\text{E}$}\\ \hline   
$H_0$ & 68.5350&_{-1.5236}^{+1.5435} & 70.7750&_{-1.1488}^{+1.1456} & 68.8920&_{-1.3974}^{+1.4166}\\
$10^3\,\Omega_{\mathrm{b}}^0 h^2$ & 22.2870&_{-0.1399}^{+0.1420} & 22.2940&_{-0.1396}^{+0.1409} & 22.2870&_{-0.1404}^{+0.1394}\\
$\Omega_{\mathrm{cdm}}^0 h^2$ & 0.1220&_{-0.0069}^{+0.0066} & 0.1309&_{-0.0057}^{+0.0055} & 0.1234&_{-0.0065}^{+0.0062}\\ 
$10^3\Omega_k^0$ & 2.1474&_{-2.6744}^{+2.9676} & 5.9320&_{-2.0502}^{+2.1120} & 2.7793&_{-2.5027}^{+2.6796}\\ 
$10^3B$ & -0.3265&_{-5.8219}^{+4.9193} & -7.1221&_{-4.1096}^{+3.6466} &-1.5054&_{-5.2781}^{+4.6128}\\ \hline
$\Omega_{\Lambda(H)}^0$ & 0.6908&_{-0.0068}^{+0.0069} & 0.6883&_{-0.0065}^{+0.0067} & 0.6903&_{-0.0067}^{+0.0068}\\
\hline
\hline
\end{tabular}
\end{table}

\renewcommand{\arraystretch}{1.5}
\begin{table*}[ht!]
\caption{\label{TRVMnf} Mean values and $1\sigma$ confidence limits obtained in the context of an RVM scenario. $\Omega_k^0$ was treated as a free parameter. More details may be found in the caption of Table \ref{TGRVMf}~.}
\vspace{2mm}
\begin{tabular}{l@{\hskip 0.5cm} A@{\hskip 0.4cm} A@{\hskip 0.4cm} A@{\hskip 0.4cm} A@{\hskip 0.4cm} A@{\hskip 0.4cm} A}  
\hline
\hline
%\multicolumn{13}{l}{Parameter  ~~~~~~~~~~~~$\textsc{All}$ \qquad \quad~~\,~$\textsc{All}+H_0^\text{R}$ \quad\quad\,~\,$\textsc{All}+H_0^\text{E}$ \quad~~~\qquad\textsc{All} \quad\qquad\quad$\textsc{All}+H_0^\text{R}$ \quad\quad~~\,$\textsc{All}+H_0^\text{E}$}\\ \hline 
Parameter & \multicolumn{2}{c}{\textsc{All}+LSS~~~} & \multicolumn{2}{c}{\textsc{All}+LSS+$H_0^\text{R}$~~} & \multicolumn{2}{c}{\textsc{All}+LSS+$H_0^\text{E}$~~} & \multicolumn{2}{c}{\textsc{All}+LSS~~~} & \multicolumn{2}{c}{\textsc{All}+LSS+$H_0^\text{R}$~~} & \multicolumn{2}{c}{\textsc{All}+LSS+$H_0^\text{E}$~~}\\ \hline
$H_0$&67.5010&_{-0.8693}^{+0.8679}&68.7760&_{-0.7956}^{+0.7858}&67.7000&_{-0.8546}^{+0.8428}&67.5280&_{-0.8918}^{+0.8876}&68.8250&_{-0.8075}^{+0.8011}&67.7310&_{-0.8716}^{+0.8496}\\ 
$10^3\Omega_{\mathrm{b}}^0 h^2$&22.4130&_{-0.1696}^{+0.1730}&22.4500&_{-0.1714}^{+0.1730}&22.4180&_{-0.1723}^{+0.1728}&22.4120&_{-0.1725}^{+0.1723}&22.4490&_{-0.1725}^{+0.1724}&22.4190&_{-0.1733}^{+0.1717}\\ 
$\Omega_{\mathrm{cdm}}^0 h^2$&0.1150&_{-0.0026}^{+0.0026}&0.1170&_{-0.0025}^{+0.0025}&0.1153&_{-0.0026}^{+0.0025}&0.1151&_{-0.0027}^{+0.0026}&0.1173&_{-0.0027}^{+0.0026}&0.1154&_{-0.0027}^{+0.0026}\\ 
$10^3\Omega_k^0$&-1.0358&_{-2.0708}^{+2.1119}&1.0969&_{-1.9154}^{+1.9576}&-0.6868&_{-2.0362}^{+2.0555}&-0.9842&_{-2.0898}^{+2.1179}&1.1876&_{-1.9402}^{+1.9680}&-0.6480&_{-2.0397}^{+2.0761}\\ 
$10^3 B$&4.0567&_{-2.2043}^{+2.1064}&2.0203&_{-2.0197}^{+1.9293}&3.7450&_{-2.1945}^{+2.0332}&3.9637&_{-2.3301}^{+2.1855}&1.7766&_{-2.1403}^{+2.0335}&3.6112&_{-2.2666}^{+2.1647}\\ 
$n_s$&0.9682&_{-0.0051}^{+0.0052}&0.9689&_{-0.0051}^{+0.0051}&0.9683&_{-0.0052}^{+0.0052}&0.9682&_{-0.0052}^{+0.0051}&0.9689&_{-0.0052}^{+0.0052}&0.9683&_{-0.0052}^{+0.0052}\\ 
$\ln{\left(10^{10}A_s\right)}$&-&-&-&-&-&-&3.0581&_{-0.0292}^{+0.0311}&3.0504&_{-0.0302}^{+0.0315}&3.0572&_{-0.0294}^{+0.0312}\\ \hline
$\Omega_{\Lambda(H)}^0$&0.6994&_{-0.0059}^{+0.0063}&0.7039&_{-0.0057}^{+0.0060}&0.7001&_{-0.0060}^{+0.0062}&0.6993&_{-0.0061}^{+0.0063}&0.7036&_{-0.0058}^{+0.0060}&0.7000&_{-0.0060}^{+0.0063}\\
%$\sigma_{8,0}$&\multicolumn{2}{c}{0.7676}&\multicolumn{2}{c}{0.7977}&\multicolumn{2}{c}{0.7722}&\multicolumn{2}{c}{0.7675}&\multicolumn{2}{c}{0.7969}&\multicolumn{2}{c}{0.7721} \\
\hline
\hline
\end{tabular}
\end{table*}

Let us now take a look at the literature and see how our findings for the flat scenario fare in comparison (the studies we consider are based on the assumption of spatial flatness). The authors of Ref.~\cite{JSola2017} report that the RVM appears to be more consistent with observations than $\Lambda$CDM, and furthermore remark that the inclusion of an LSS likelihood tips the balance in favor of the \emph{Planck} value for $H_0$. Our results paint a somewhat different picture. We find no statistically significant evidence that the RVM is preferred over $\Lambda$CDM (more details are provided in subsection \ref{subsec:LCDM}). Secondly, the addition of LSS data increases the mean values of $H_0$ slightly when $\Omega_k^0=0$, although we cannot say that it spoils the consistency with the \emph{Planck} constraints. We nonetheless note that once the conversion from $\nu$ to $B~(\,=3\nu)$ is made, the mean values of $B$ and the corresponding uncertainties (Tables \ref{TRVMf-noLSS} and \ref{TRVMf}) are found to be of the same order of magnitude\footnote{Two values ($a\times10^p$ and $b\times10^p$, where $1\leq a,b<10$ and $p$ is an integer) shall be deemed to have the same magnitude if $|a-b|<5$.\\ Given the quantity $M^{+n}_{-\ell}$, we shall refer to $n+\ell$ as the \emph{uncertainty} in $M$.} as those obtained in Ref.~\cite{JSola2017} with the full data set.

The tendency of the RVM to lend support to the \emph{Planck} bounds for $H_0$ is also pointed out in Ref.~\cite{PeracaulaJ}. The authors find that $\nu=0$ is excluded at more than $3\sigma$ when they include LSS data, and although our results do not corroborate this conclusion, the mean and uncertainty for $\nu$ again translate into values for $B$ that match ours in order of magnitude. The same can be said of the constraints placed on $\nu$ in Ref.~\cite{Sola2017} by means of a fit to SNeIa+BAO+$H(z)$+LSS+BBN+CMB data. This despite the fact that the study in question considers radiation to interact with dark energy, which is not the case here. We note that our mean values for $B$ (excluding the ones obtained with the $\textsc{All}+H_0^\text{R}$ and $\textsc{All}+\text{LSS}+H_0^\text{R}$ data sets) have $\nu$ equivalents that lie within $1\sigma$ of the value found in Ref.~\cite{Sola2017} using the full data set.

\renewcommand{\arraystretch}{1.5}
\begin{table}[ht!]
\caption{\label{TGRVSf-noLSS} Mean values and $1\sigma$ confidence limits obtained in the context of a flat GRVS scenario. LSS data was excluded from the analysis. More details may be found in the caption of Table \ref{TGRVMf-noLSS}~.}
\vspace{2mm}
\begin{tabular}{l@{\hskip 0.3cm} A@{\hskip 0.3cm} A@{\hskip 0.3cm} A}  
\hline
\hline
\multicolumn{7}{l}{Parameter  ~~~~~~~\,$\textsc{All}$ \qquad \quad~~\,$\textsc{All}+H_0^\text{R}$ \quad\quad\,~$\textsc{All}+H_0^\text{E}$}\\ \hline       
$H_0$ & 68.2640&_{-0.8232}^{+0.7812} & 69.2520&_{-0.7772}^{+0.7381} & 68.3850&_{-0.8033}^{+0.7663}\\
$10^3\,\Omega_{\mathrm{b}}^0 h^2$ & 22.4300&_{-0.1640}^{+0.1647} & 22.5740&_{-0.1563}^{+0.1559} & 22.4490&_{-0.1617}^{+0.1627}\\
$\Omega_{\mathrm{cdm}}^0 h^2$ & 0.1193&_{-0.0040}^{+0.0036} & 0.1222&_{-0.0041}^{+0.0037} & 0.1197&_{-0.0040}^{+0.0036}\\
$10^3C$ & -0.1050&_{-2.4237}^{+2.4635} & \phantom{-}2.2458&_{-2.2670}^{+2.3401} & \phantom{-}0.1792&_{-2.3773}^{+2.4189}\\ \hline
$\Omega_{\Lambda(H)}^0$ & 0.6957&_{-0.0066}^{+0.0069} & 0.6980&_{-0.0066}^{+0.0069} & 0.6960&_{-0.0066}^{+0.0069}\\
\hline
\hline
\end{tabular} 
\end{table}

\renewcommand{\arraystretch}{1.5}
\begin{table*}[ht!]
\caption{\label{TGRVSf} Mean values and $1\sigma$ confidence limits obtained in the context of a flat GRVS scenario. More details may be found in the caption of Table \ref{TGRVMf}~.}
\vspace{2mm}
\begin{tabular}{l@{\hskip 0.5cm} A@{\hskip 0.4cm} A@{\hskip 0.4cm} A@{\hskip 0.4cm} A@{\hskip 0.4cm} A@{\hskip 0.4cm} A}  
\hline
\hline
%\multicolumn{13}{l}{Parameter  ~~~~~~~~~~~~$\textsc{All}$ \qquad \quad~~\,~\,$\textsc{All}+H_0^\text{R}$ \quad\quad\,~\,\,$\textsc{All}+H_0^\text{E}$ \quad~~~\qquad\textsc{All} \quad\qquad\quad~$\textsc{All}+H_0^\text{R}$ \quad\quad~~\,$\textsc{All}+H_0^\text{E}$}\\ \hline  
Parameter & \multicolumn{2}{c}{\textsc{All}+LSS~~~} & \multicolumn{2}{c}{\textsc{All}+LSS+$H_0^\text{R}$~~} & \multicolumn{2}{c}{\textsc{All}+LSS+$H_0^\text{E}$~~} & \multicolumn{2}{c}{\textsc{All}+LSS~~~} & \multicolumn{2}{c}{\textsc{All}+LSS+$H_0^\text{R}$~~} & \multicolumn{2}{c}{\textsc{All}+LSS+$H_0^\text{E}$~~}\\ \hline
$H_0$&67.7840&_{-0.6298}^{+0.6186}&68.4770&_{-0.5972}^{+0.5943}&67.8800&_{-0.6200}^{+0.6095}&67.8000&_{-0.6370}^{+0.6284}&68.4960&_{-0.5968}^{+0.5958}&67.8990&_{-0.6292}^{+0.6132}\\ 
$10^3\Omega_{\mathrm{b}}^0 h^2$&22.3790&_{-0.1565}^{+0.1543}&22.5010&_{-0.1500}^{+0.1519}&22.3960&_{-0.1539}^{+0.1540}&22.3820&_{-0.1567}^{+0.1549}&22.5020&_{-0.1493}^{+0.1501}&22.3980&_{-0.1540}^{+0.1547}\\ 
$\Omega_{\mathrm{cdm}}^0 h^2$&0.1157&_{-0.0019}^{+0.0019}&0.1161&_{-0.0019}^{+0.0019}&0.1158&_{-0.0019}^{+0.0019}&0.1158&_{-0.0020}^{+0.0019}&0.1163&_{-0.0020}^{+0.0019}&0.1159&_{-0.0020}^{+0.0020}\\ 
$10^3 C$&-2.3810&_{-1.2607}^{+1.3190}&-1.4761&_{-1.2102}^{+1.2662}&-2.2558&_{-1.2541}^{+1.2957}&-2.3120&_{-1.3179}^{+1.3602}&-1.3684&_{-1.2737}^{+1.3103}&-2.1795&_{-1.3042}^{+1.3530}\\ 
$n_s$&0.9671&_{-0.0045}^{+0.0045}&0.9706&_{-0.0044}^{+0.0044}&0.9675&_{-0.0044}^{+0.0044}&0.9671&_{-0.0045}^{+0.0045}&0.9707&_{-0.0044}^{+0.0044}&0.9676&_{-0.0045}^{+0.0045}\\ 
$\ln{\left(10^{10}A_s\right)}$&-&-&-&-&-&-&3.0565&_{-0.0295}^{+0.0309}&3.0522&_{-0.0294}^{+0.0309}&3.0560&_{-0.0292}^{+0.0310}\\ \hline
$\Omega_{\Lambda(H)}^0$&0.6993&_{-0.0061}^{+0.0063}&0.7043&_{-0.0057}^{+0.0060}&0.7000&_{-0.0060}^{+0.0062}&0.6992&_{-0.0061}^{+0.0063}&0.7041&_{-0.0057}^{+0.0060}&0.6999&_{-0.0060}^{+0.0062}\\
%$\sigma_{8,0}$&\multicolumn{2}{c}{0.7736}&\multicolumn{2}{c}{0.7911}&\multicolumn{2}{c}{0.7762}&\multicolumn{2}{c}{0.7730}&\multicolumn{2}{c}{0.7898}&\multicolumn{2}{c}{0.7756}\\
\hline
\hline
\end{tabular} 
\end{table*}

\begin{figure*}[ht!]
\hspace*{-1.3cm}
\includegraphics[width=0.93\textwidth]{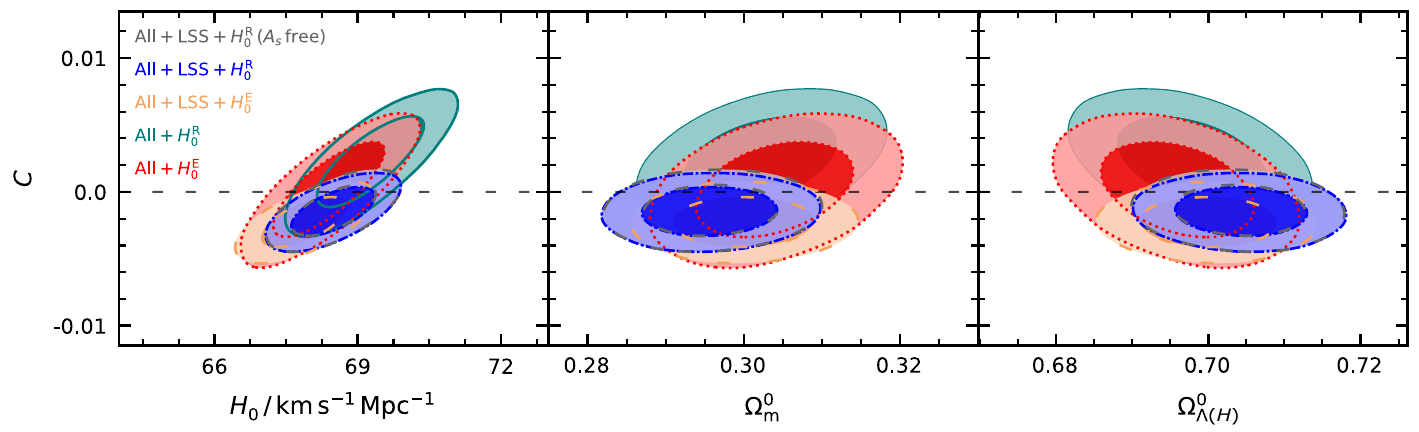}
\caption{\label{GRVSf} Marginalized posterior probability distributions for the GRVS parameter $C$ vs \emph{(left)} $H_0$, \emph{(center)} $\Omega_\text{m}^0$, \emph{(right)} $\Omega_{\Lambda(H)}^0$. We assume a spatially flat space-time.}
\end{figure*}

\renewcommand{\arraystretch}{1.5}
\begin{table}[ht!]
\caption{\label{TGRVSnf-noLSS} Mean values and $1\sigma$ confidence limits obtained in the context of a GRVS scenario. LSS data was excluded from the analysis, and the condition of spatial flatness was not imposed. More details may be found in the caption of Table \ref{TGRVMf-noLSS}~.}
\vspace{2mm}
\begin{tabular}{l@{\hskip 0.3cm} A@{\hskip 0.3cm} A@{\hskip 0.3cm} A}  
\hline
\hline
\multicolumn{7}{l}{Parameter  ~~~~~~~~$\textsc{All}$ \qquad \quad~~\,$\textsc{All}+H_0^\text{R}$ \quad\quad\,~$\textsc{All}+H_0^\text{E}$}\\ \hline       
$H_0$ & 68.7530&_{-1.5507}^{+1.5454} & 70.9130&_{-1.1578}^{+1.1628} & 69.0730&_{-1.4109}^{+1.4051}\\
$10^3\,\Omega_{\mathrm{b}}^0 h^2$ & 22.4090&_{-0.1718}^{+0.1726} & 22.4320&_{-0.1700}^{+0.1720} & 22.4120&_{-0.1715}^{+0.1716}\\
$\Omega_{\mathrm{cdm}}^0 h^2$ & 0.1214&_{-0.0070}^{+0.0066} & 0.1298&_{-0.0058}^{+0.0056} & 0.1227&_{-0.0064}^{+0.0062}\\
$10^3\,\Omega_k^0$ & 1.0452&_{-2.7865}^{+2.9292} & 4.3585&_{-2.2526}^{+2.3047} & 1.5638&_{-2.6167}^{+2.7015}\\ 
$10^3 C$ & \phantom{-}0.7443&_{-3.2210}^{+3.7198} & \phantom{-}4.9508&_{-2.3946}^{+2.6763} & \phantom{-}1.4163&_{-2.9272}^{+3.3940}\\ \hline
$\Omega_{\Lambda(H)}^0$ & 0.6946&_{-0.0073}^{+0.0074} & 0.6928&_{-0.0072}^{+0.0074} & 0.6943&_{-0.0072}^{+0.0074}\\
\hline
\hline
\end{tabular} 
\end{table}

\renewcommand{\arraystretch}{1.5}
\begin{table*}[ht!]
\caption{\label{TGRVSnf} Mean values and $1\sigma$ confidence limits obtained with each data set combination in the context of a GRVS scenario. $\Omega_k^0$ was treated as a free parameter. More details may be found in the caption of Table \ref{TGRVMf}~.}
\vspace{2mm}
\begin{tabular}{l@{\hskip 0.5cm} A@{\hskip 0.4cm} A@{\hskip 0.4cm} A@{\hskip 0.4cm} A@{\hskip 0.4cm} A@{\hskip 0.4cm} A}  
\hline
\hline
%\multicolumn{13}{l}{Parameter  ~~~~~~~~~~~~$\textsc{All}$ \qquad \quad~~\,~\,$\textsc{All}+H_0^\text{R}$ \quad\quad\,~\,\,$\textsc{All}+H_0^\text{E}$ \quad~~~\qquad\textsc{All} \quad\qquad\quad~$\textsc{All}+H_0^\text{R}$ \quad\quad~~\,$\textsc{All}+H_0^\text{E}$}\\ \hline   
Parameter & \multicolumn{2}{c}{\textsc{All}+LSS~~~} & \multicolumn{2}{c}{\textsc{All}+LSS+$H_0^\text{R}$~~} & \multicolumn{2}{c}{\textsc{All}+LSS+$H_0^\text{E}$~~} & \multicolumn{2}{c}{\textsc{All}+LSS~~~} & \multicolumn{2}{c}{\textsc{All}+LSS+$H_0^\text{R}$~~} & \multicolumn{2}{c}{\textsc{All}+LSS+$H_0^\text{E}$}\\ \hline
$H_0$&67.5080&_{-0.8940}^{+0.8646}&68.7710&_{-0.7891}^{+0.7891}&67.7040&_{-0.8632}^{+0.8326}&67.5270&_{-0.9011}^{+0.8799}&68.8250&_{-0.8066}^{+0.7975}&67.7270&_{-0.8650}^{+0.8537}\\ 
$10^3\Omega_{\mathrm{b}}^0 h^2$&22.4130&_{-0.1742}^{+0.1718}&22.4490&_{-0.1728}^{+0.1719}&22.4190&_{-0.1709}^{+0.1739}&22.4140&_{-0.1718}^{+0.1734}&22.4500&_{-0.1716}^{+0.1720}&22.4170&_{-0.1723}^{+0.1707}\\ 
$\Omega_{\mathrm{cdm}}^0 h^2$&0.1150&_{-0.0026}^{+0.0025}&0.1170&_{-0.0025}^{+0.0025}&0.1153&_{-0.0026}^{+0.0025}&0.1151&_{-0.0027}^{+0.0026}&0.1173&_{-0.0026}^{+0.0026}&0.1154&_{-0.0027}^{+0.0026}\\ 
$10^3\Omega_k^0$&-0.9464&_{-2.0889}^{+2.0824}&1.1408&_{-1.9133}^{+1.9543}&-0.6118&_{-2.0364}^{+2.0552}&-0.9229&_{-2.0864}^{+2.1097}&1.2022&_{-1.9261}^{+1.9458}&-0.5637&_{-2.0289}^{+2.0727}\\ 
$10^3 C$&-2.5839&_{-1.3420}^{+1.4209}&-1.2927&_{-1.2392}^{+1.2993}&-2.3797&_{-1.3091}^{+1.3910}&-2.5323&_{-1.4069}^{+1.4719}&-1.1341&_{-1.2930}^{+1.3627}&-2.3232&_{-1.3732}^{+1.4490}\\ 
$n_s$&0.9682&_{-0.0052}^{+0.0052}&0.9689&_{-0.0052}^{+0.0051}&0.9683&_{-0.0051}^{+0.0052}&0.9682&_{-0.0052}^{+0.0051}&0.9690&_{-0.0052}^{+0.0051}&0.9683&_{-0.0052}^{+0.0052}\\ 
$\ln{\left(10^{10}A_s\right)}$&-&-&-&-&-&-&3.0584&_{-0.0297}^{+0.0309}&3.0504&_{-0.0299}^{+0.0313}&3.0571&_{-0.0296}^{+0.0312}\\ \hline
$\Omega_{\Lambda(H)}^0$&0.6993&_{-0.0061}^{+0.0063}&0.7038&_{-0.0059}^{+0.0060}&0.7000&_{-0.0060}^{+0.0062}&0.6993&_{-0.0061}^{+0.0063}&0.7036&_{-0.0058}^{+0.0061}&0.6999&_{-0.0061}^{+0.0062}\\
%$\sigma_{8,0}$&\multicolumn{2}{c}{0.7676}&\multicolumn{2}{c}{0.7976}&\multicolumn{2}{c}{0.7722}&\multicolumn{2}{c}{0.7675}&\multicolumn{2}{c}{0.7969}&\multicolumn{2}{c}{0.7717} \\
\hline
\hline
\end{tabular} 
\end{table*}

On the contrary, our results are in some tension with that of Ref.~\cite{Geng2017} $\left[\nu=B/3=\left(1.37^{+0.72}_{-0.95}\right)\times 10^{-4}\right]$. The authors attribute their strong constraints on $\nu$ to the effectiveness of CMB temperature fluctuations as cosmological probes \cite{Geng2017}. One should keep in mind, however, that the approach taken in Ref.~\cite{Geng2017} differs from ours in a number of ways, the most prominent being the assumption that dark energy decays into both radiation and matter, and the incorporation of massive neutrinos into the model. 

We turn our attention to the study presented in Ref.~\cite{Peracaula2018} next. Here, the joint analysis is based on measurements of observables associated with SNeIa, BAOs, cosmic chronometers, LSS and the CMB, and again it transpires that the mean value of $\nu$ and associated standard deviation have the same order of magnitude as the ones we get (for $B/3$). The authors also investigate the impact of the individual likelihoods on the results, and observe that using both LSS and CMB data tightens constraints on $\nu$, consequently endowing it with a definite sign. They go on to show that the absence of either makes $\nu$ compatible with the $\Lambda$CDM limit ($\nu=0$). The authors conclude that the BAO+LSS+CMB combination excludes the standard model at more than $3\sigma$. 
%We note with satisfaction that the constraints the authors place on $\nu$ (when LSS data is excluded) have order-of-magnitude equivalence with the \textsc{All}, $\textsc{All}+H_0^\text{R}$ and $\textsc{All}+H_0^\text{E}$ results for $3B$.

%Finally, we consider Ref.~\cite{Perico2017}. The approach taken here is curious and insightful: photons, massless and massive neutrinos, cold dark matter and baryons are each paired with a portion of the vacuum, and interactions on both background and perturbation levels are only allowed between members of the same pair. The energy densities of the (independent) vacuum components are then summed up to get the total energy density of the running vacuum. It is in this context that the authors of Ref.~\cite{Perico2017} constrain the RVM parameter $\nu$ (although they call it $\alpha$), and find that $\nu=(-4.7\pm6.5)\times 10^{-4}$. The mean values and $1\sigma$ uncertainties we get for $3B$ in the absence of LSS data agree with theirs in order of magnitude. However, the LSS likelihood effectively rules out negative values for $\nu$. We note that the numerical analysis presented in Ref.~\cite{Perico2017} only makes use of \emph{Planck} data.

Although some of the above studies allow radiation to couple with vacuum energy, we decide not to do likewise. There is an important reason for this: namely, any interaction between radiation and dark energy would cause the CMB temperature $(T_\text{CMB})$ to scale differently with redshift than it does in $\Lambda$CDM. Additionally, any net changes in photon number would alter the relation between the angular diameter and luminosity distances \cite{Avg2012}. The literature contains many examples of studies that have constrained departures from the standard-model prediction for $T_\text{CMB}$ $[T_\text{CMB}\propto (1+z)]$ \cite{Avg2012,Luz2015, Noterdaeme2011, Avgoustidis2016}, or placed bounds on violations of the distance-duality relation \cite{Lin2018, Debernardis2006, Holanda2016, Ma2018}. As yet, however, no compelling evidence of deviations from $\Lambda$CDM has been found. In other words, there is currently little observational justification for energy exchange between radiation and the vacuum to be incorporated into a cosmological model.

In the same vein, since $\Omega_\text{b}^0$ is subject to very tight constraints, we refrain from coupling the baryon component with dark energy, as this would alter the way in which $\rho_\text{b}$ scales with redshift. The reader is referred to Refs.~\cite{McGaugh2004} and \cite{McGaugh2015} (and the works cited therein) for a review of the said constraints.

\subsection{The GRVS}
\begin{figure*}[ht!]
\hspace*{-1.3cm}
\includegraphics[width=0.95\textwidth]{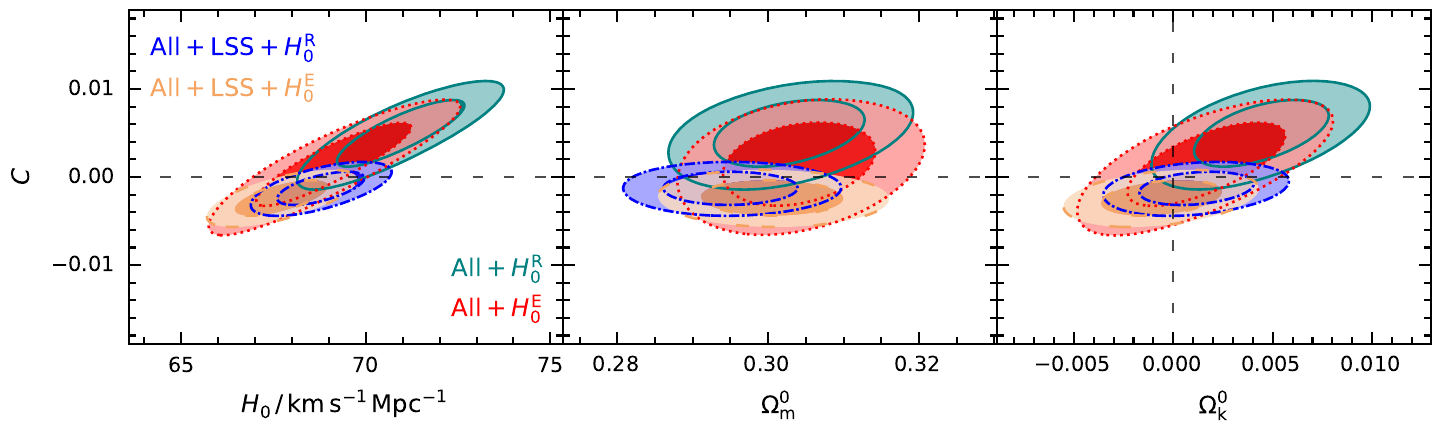}
\caption{\label{GRVSnf} Marginalized posterior probability distributions for the GRVS parameter $C$ vs \emph{(left)} $H_0$, \emph{(center)} $\Omega_\text{m}^0$, \emph{(right)} $\Omega_k^0$.}
\end{figure*}

\begin{figure}[ht!]
%\hspace*{-1.3cm}
\includegraphics[width=0.5\textwidth]{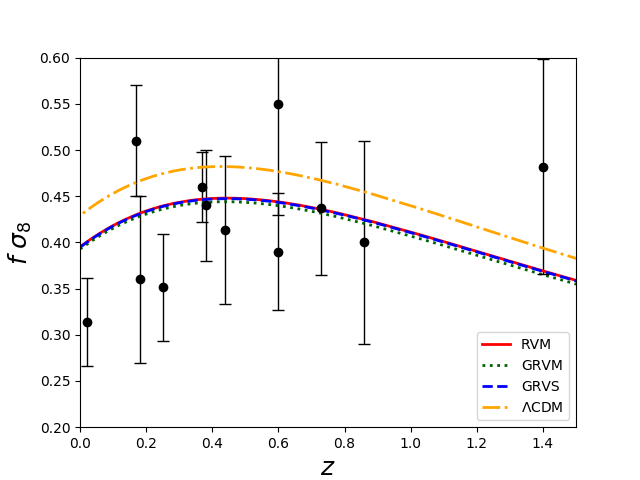}
\caption{The variation of $f\sigma_8$ with $z$ in a flat scenario. The data points of Table~\ref{TLSS} are shown as black circles with $1\sigma$ error bars, and it can be seen that they mainly probe redshifts less than unity, at which different models are less likely to be degenerate \cite{Lavrentios}. The dynamical-$\Lambda$ curves are based on the \textsc{All}+LSS (+ fixed $A_s$) mean values, and have $\sigma_{8,0}$ equal to \num{0.7748} (RVM), \num{0.7660} (GRVM) and \num{0.7736} (GRVS). The $\Lambda$CDM curve was obtained using the TT+lowP+lensing results from Ref.~\cite{Planck2015}.}
\label{fsigma8_plot1}
\end{figure}

\begin{figure}[ht!]
%\hspace*{-1.3cm}
\includegraphics[width=0.5\textwidth]{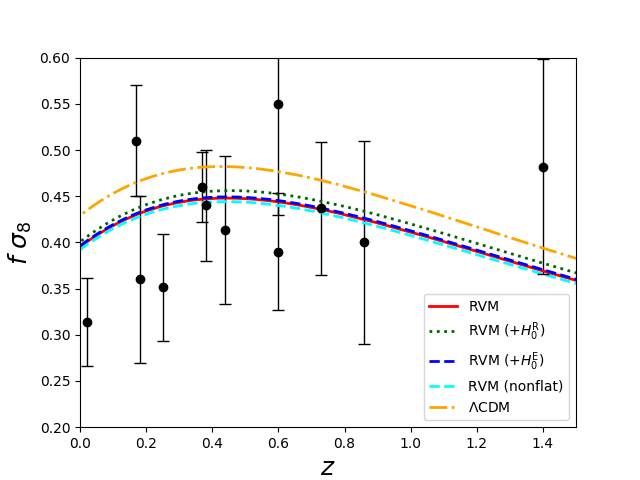}
\caption{The variation of $f\sigma_8$ with $z$. The dynamical-$\Lambda$ curves are based on the mean values obtained using the \textsc{All}+LSS data set (or, where indicated in the legend, \textsc{All}+LSS+$H_0^\text{R}/$+$H_0^\text{E}$) with $\ln{(10^{10}A_s)}$ fixed at 3.062; $\sigma_{8,0}$ equates to \num{0.7748} (RVM), \num{0.7909} (RVM+$H_0^\text{R}$), \num{0.7773} (RVM+$H_0^\text{E}$) and \num{0.7676} (RVM-nonflat). We assume a flat geometry in all cases but one. The sole exception is labeled accordingly.}
\label{fsigma8_plot2}
\end{figure}

Results are presented in Tables \ref{TGRVSf-noLSS}--\ref{TGRVSnf}. The inclusion of LSS data again proves to be important. In the flat case, it tightens constraints on $C$ and endows it with a definite (negative) sign, while also reducing (or even neutralizing) the correlation between $C$ and the parameters $H_0$, $\Omega_\text{m}^0$ and $\Omega_{\Lambda(H)}^0$ (Fig.~\ref{GRVSf}). As a result, the 1D posteriors for $C$ exclude the $\Lambda$CDM limit at a little over $1\sigma$. The situation is in many ways analogous to the RVM scenario. When the assumption of spatial flatness is relaxed, we again find that LSS data shows mild preference for a closed (rather than open) geometry, and tends to decrease the mean values of $H_0$ (Fig.~\ref{GRVSnf}). Contrary to what was observed for the RVM, the latter effect is also noted in the flat case. 

The correlation between $C$ and $H_0$ explains why the $H_0^\text{R}$ likelihood shifts the contours in Figs.~\ref{GRVSf} and \ref{GRVSnf} in the direction of increasing $C$. The introduction of growth data makes this displacement much less pronounced.

Before we move on, let us consider how well the GRVM, RVM and GRVS account for RSD measurements. The values of $f\sigma_8(z)$ inferred from CMB data (for a $\Lambda$CDM cosmology) seem to be in excess of what observations related to structure growth suggest. This is a result of the fact that the constraints on $\Omega_\text{m}^0$ and $\sigma_{8,0}$ obtained from weak lensing, Sunyaev-Zel'dovich cluster counts, and RSDs appear to be in some tension with the
Planck analysis of primary fluctuations \cite{Lavrentios}. However, the cause of the discrepancy is as yet a subject of debate. According to a recent study, the lower value for the reionization optical depth reported in more recent \emph{Planck} papers has partially solved the problem \cite{Douspis}. Whether or not any tension is detected also depends on the choice of data set. In particular, RSD measurements published in the last few years tend to probe higher redshifts, at which degeneracies can set in between different models. Such measurements are therefore more likely to be consistent with the $\Lambda$CDM values for $f\sigma_8$ \cite{Lavrentios}.

Figs.~\ref{fsigma8_plot1} and \ref{fsigma8_plot2} show the variation of $f\sigma_8$ with $z$ for the dynamical-$\Lambda$ models and $\Lambda$CDM. Most of the data points are located below the $\Lambda$CDM curve, so the fact that the GRVM, RVM and GRVS yield smaller values for $f\sigma_8(z)$ augurs well, and indeed one notes that -- in the majority of cases -- the dynamical-$\Lambda$ curves are closer to the mean values of the observations. By comparing the results for $\sigma_{8,0}$ (provided in the captions of Figs.~\ref{fsigma8_plot1} and \ref{fsigma8_plot2}\,) with the \emph{Planck} value of $\sim 0.8$ \cite{Planck2018}, we may additionally deduce that the lower $f\sigma_8$ curves are mainly a consequence of a smaller $\sigma_{8,0}$. Finally, it appears that the addition of $H_0^\text{R}$ to the data set yields a slightly higher value of $f\sigma_8$ at a given $z$. This observation is perfectly in accord with our conclusion that LSS data lends support to the \emph{Planck} constraints on $H_0$.

\subsection{Comparison with $\Lambda$CDM \label{subsec:LCDM}}

In this subsection, we consider the cosmological parameter constraints obtained by using the $\textsc{All} + \text{LSS}$, $\textsc{All} + \text{LSS} + H_0^\text{R}$ and $\textsc{All} + \text{LSS} + H_0^\text{E}$ data sets in the framework of a $\Lambda$CDM cosmology with freely-varying $\Omega_k^0$. The mean values and $1\sigma$ confidence limits are presented in Table \ref{TLCDMnf}.

How may we compare the results obtained for $\Lambda$CDM with those for the GRVM, RVM and GRVS? The number of baseline parameters differs from model to model (11 for the GRVM, 10 for the RVM and GRVS, and 9 in the case of $\Lambda$CDM), so one cannot simply assess performance by looking at the minimum $\chi^2$. Instead, we employ the Akaike Information Criterion (AIC) \cite{Akaike1974}. This takes into account both the number of free parameters ($p$) and the value of the maximum likelihood ($\mathcal{L}_\text{max}$):  
\begin{equation}
\text{AIC} = 2p - 2\ln{\left(\mathcal{L}_\text{max}\right)}~.
\label{AIC}
\end{equation}

\renewcommand{\arraystretch}{1.5}
\begin{table*}
\caption{\label{TLCDMnf} Mean values and $1\sigma$ confidence limits for the baseline cosmological parameters \emph{(top block)} and the derived quantity $\Omega_\Lambda^0$ \emph{(last row)} in the context of a $\Lambda$CDM cosmology. The condition $\Omega_k^0=0$ was not imposed.}
\vspace{2mm}
\begin{tabular}{l@{\hskip 0.5cm} A@{\hskip 0.4cm} A@{\hskip 0.4cm} A@{\hskip 0.4cm} A@{\hskip 0.4cm} A@{\hskip 0.4cm} A}  
\hline
\hline
%\multicolumn{13}{l}{Parameter  ~~~~~~~~~~~~$\textsc{All}$ \qquad \quad~~\,\,$\textsc{All}+H_0^\text{R}$ \quad\quad\,\,\,$\textsc{All}+H_0^\text{E}$ \quad~~\,\qquad\textsc{All} \quad\qquad\quad$\textsc{All}+H_0^\text{R}$ \quad\quad~\,$\textsc{All}+H_0^\text{E}$}\\ \hline  
Parameter & \multicolumn{2}{c}{\textsc{All}+LSS~~~~} & \multicolumn{2}{c}{\textsc{All}+LSS+$H_0^\text{R}$~~} & \multicolumn{2}{c}{\textsc{All}+LSS+$H_0^\text{E}$~~} & \multicolumn{2}{c}{\textsc{All}+LSS~~~~} & \multicolumn{2}{c}{\textsc{All}+LSS+$H_0^\text{R}$~~} & \multicolumn{2}{c}{\textsc{All}+LSS+$H_0^\text{E}$~~}\\ \hline       
$H_0$&68.6250&_{-0.6720}^{+0.6557}&69.2690&_{-0.6256}^{+0.6243}&68.7000&_{-0.6569}^{+0.6456}&68.6080&_{-0.6631}^{+0.6579}&69.2560&_{-0.6348}^{+0.6229}&68.6790&_{-0.6644}^{+0.6499}\\ 
$10^3\Omega_{\mathrm{b}}^0 h^2$&22.4650&_{-0.1697}^{+0.1700}&22.4740&_{-0.1707}^{+0.1697}&22.4640&_{-0.1705}^{+0.1711}&22.4570&_{-0.1696}^{+0.1707}&22.4680&_{-0.1694}^{+0.1706}&22.4600&_{-0.1711}^{+0.1703}\\ 
$\Omega_{\mathrm{cdm}}^0 h^2$&0.1190&_{-0.0015}^{+0.0015}&0.1191&_{-0.0015}^{+0.0015}&0.1190&_{-0.0015}^{+0.0015}&0.1191&_{-0.0015}^{+0.0015}&0.1191&_{-0.0015}^{+0.0015}&0.1191&_{-0.0015}^{+0.0015}\\ 
$10^3\Omega_k^0$&0.3456&_{-1.9447}^{+1.9551}&1.6406&_{-1.8476}^{+1.8794}&0.5135&_{-1.9091}^{+1.9412}&0.4033&_{-1.9491}^{+1.9647}&1.6730&_{-1.8582}^{+1.8754}&0.5333&_{-1.9232}^{+1.9657}\\ 
$n_s$&0.9698&_{-0.0051}^{+0.0051}&0.9697&_{-0.0051}^{+0.0051}&0.9698&_{-0.0051}^{+0.0051}&0.9696&_{-0.0051}^{+0.0051}&0.9696&_{-0.0051}^{+0.0051}&0.9696&_{-0.0051}^{+0.0051}\\ 
$\ln{\left(10^{10}A_s\right)}$&-&-&-&-&-&-&3.0417&_{-0.0287}^{+0.0300}&3.0425&_{-0.0286}^{+0.0300}&3.0418&_{-0.0282}^{+0.0303}\\ \hline
$\Omega_{\Lambda}^0$&0.6991&_{-0.0060}^{+0.0063}&0.7032&_{-0.0057}^{+0.0060}&0.6995&_{-0.0060}^{+0.0063}&0.6987&_{-0.0060}^{+0.0063}&0.7029&_{-0.0058}^{+0.0060}&0.6992&_{-0.0060}^{+0.0062}\\
%$\sigma_{8,0}$&\multicolumn{2}{c}{0.8273}&\multicolumn{2}{c}{0.8282}&\multicolumn{2}{c}{0.8274}&\multicolumn{2}{c}{0.8193}&\multicolumn{2}{c}{0.8202}&\multicolumn{2}{c}{0.8194} \\
\hline
\hline
\end{tabular}
\end{table*}

Additionally, since the theoretical quantities associated with each likelihood are treated as samples from a multivariate Gaussian distribution, the AIC may equivalently be expressed in terms of the minimum $\chi^2$:
\begin{equation}
\text{AIC} = 2p +\chi^2_\text{min}~.
\end{equation}
Another tool for model selection is the Bayesian Information Criterion (BIC) \cite{Schwarz1978}:
\begin{equation}
\text{BIC}=p\ln{N}-2\ln\left(\mathcal{L}_\text{max}\right)~.
\label{BIC}
\end{equation}
Here, $N$ is the number of observations, which in this work amounts to 798 or 799,\footnote{JLA: 740, $H(z)$: 30, CMB: 4, BAO: 12, LSS: 12~.} depending on whether $H_0^\text{R}$ or $H_0^\text{E}$ are included. Provided the assumption of sampling from a multivariate Gaussian distribution holds, Eq.~($\ref{BIC}$) may alternatively take the form 
\begin{equation}
\text{BIC}=p\ln{N}+\chi^2_\text{min}~.
\end{equation}
The AIC and BIC for the RVM and GRVS can be found in Tables \ref{TComparison} and \ref{TComparison2}. We do not include the GRVM, since some of its parameters are not well-constrained in the non-flat case.

\begin{table}
\caption{\label{TComparison} The AIC and BIC statistics for the RVM, GRVS and $\Lambda$CDM. The values in the top, middle and bottom sections were obtained using the $\textsc{All}+\text{LSS}$, $\textsc{All}+\text{LSS}+H_0^\text{R}$ and $\textsc{All}+\text{LSS}+H_0^\text{E}$ data sets, respectively. $\Omega_k^0$ was treated as a free parameter in all cases, while $A_s$ was set to a fixed value.}
\vspace{1.5mm}
\begin{tabular}{c S[table-format=+3.1] S[table-format=+3.1] S[table-format=+1.1] S[table-format=+3.1] S[table-format=+1.1]} 
\hline
\hline
Model & {~~\,\,$\chi^2_\text{min.}$} & ~\,\,AIC & {~\,\,$\Delta$AIC} & ~\,\,BIC & {~\,$\Delta$BIC}\\ 
\hline
RVM & 722.9 & 742.9 & -1.8 & 789.7 & 2.9\\ 
GRVS & 722.9 & 742.9 & -1.8 & 789.7 & 2.9\\
$\Lambda$CDM & 726.7 & 744.7 & 00.0 & 786.8 & 00.0\\ \hline
RVM ($+H_0^\text{R}$) & 733.2 & 753.2 & 1.1 & 800.0 & 5.8\\
GRVS ($+H_0^\text{R}$) & 733.1 & 753.1 & 1.0 & 799.9 & 5.7\\
$\Lambda$CDM ($+H_0^\text{R}$) & 734.1 & 752.1 & 0.0 & 794.3 & 0.0\\ \hline
RVM ($+H_0^\text{E}$) & 723.7 & 743.7 & -1.4 & 790.5 & 3.3\\
GRVS ($+H_0^\text{E}$) & 723.7 & 743.7 & -1.4 & 790.5 & 3.3\\
$\Lambda$CDM ($+H_0^\text{E}$) & 727.1 & 745.1 & 0.0 & 787.3 & 0.0\\ 
\hline
\hline
\end{tabular}
\end{table}

\begin{table}
\caption{\label{TComparison2} The AIC and BIC statistics for the RVM, GRVS and $\Lambda$CDM. $\Omega_k^0$ and $A_s$ were treated as free parameters in all cases.}
\vspace{1.5mm}
\begin{tabular}{c S[table-format=+3.1] S[table-format=+3.1] S[table-format=+1.1] S[table-format=+3.1] S[table-format=+1.1]} 
\hline
\hline
Model & {~~\,\,$\chi^2_\text{min.}$} & ~\,\,AIC & {~\,\,$\Delta$AIC} & ~\,\,BIC & {~\,$\Delta$BIC}\\ 
\hline
RVM & 723.1 & 745.1 & -1.2 & 796.6 & 3.5\\ 
GRVS & 722.7 & 744.7 & -1.6 & 796.2 & 3.1\\
$\Lambda$CDM & 726.3 & 746.3 & 00.0 & 793.1 & 00.0\\ \hline
RVM ($+H_0^\text{R}$) & 733.2 & 755.2 & 1.5 & 806.7 & 6.2\\
GRVS ($+H_0^\text{R}$) & 733.1 & 755.1 & 1.4 & 806.6 & 6.1\\
$\Lambda$CDM ($+H_0^\text{R}$) & 733.7 & 753.7 & 0.0 & 800.5 & 0.0\\ \hline
RVM ($+H_0^\text{E}$) & 724.0 & 746.0 & -0.6 & 797.5 & 4.1\\
GRVS ($+H_0^\text{E}$) & 723.7 & 745.7 & -0.9 & 797.2 & 3.8\\
$\Lambda$CDM ($+H_0^\text{E}$) & 726.6 & 746.6 & 0.0 & 793.4 & 0.0\\ 
\hline
\hline
\end{tabular}
\end{table}
Tables \ref{TComparison} and \ref{TComparison2} demonstrate that the minimum $\chi^2$ for the RVM and GRVS is smaller than its $\Lambda$CDM counterpart. What is more, this holds for all three data set combinations. One must however determine whether the difference in $\chi^2_\text{min}$ is enough to justify the extra free parameter of the RVM and GRVS. The reason is that although the addition of parameters introduces more degrees of freedom -- and hence allows the model to better approximate the data -- it does not necessarily yield a model of greater merit, because when the information supplied by the data has to be `shared' among more parameters the resulting estimates tend to be less precise \cite{Schwarz2011}. In such cases, information criteria like the AIC and BIC become indispensable to find a trade-off.\footnote{There is nonetheless a caveat: the AIC and BIC should, strictly speaking, only be applied if certain conditions are satisfied \cite{Link2010, Kuha2004}. For instance, they are both meant to be used with independent observations \cite{Liddle2007, Jordanger2014}.} As can be deduced from Eqs.~($\ref{AIC}$) and ($\ref{BIC}$), the AIC and BIC statistics do not only penalize for a smaller value of $\mathcal{L}_\text{max}$, but also for a larger number of free parameters. In general, a smaller AIC/BIC indicates better performance.

Let us consider this in more detail. We start by noting that $\Delta$AIC indicates the level of support the data provides for the model with the smaller AIC. An absolute value between 0 and 2 is usually not deemed enough to draw conclusions. If $|\Delta\text{AIC}|$ lies in the range from 2 to 4, the model with the larger AIC is considerably disfavored, while a value of $|\Delta\text{AIC}|>10$ renders it practically irrelevant. In the same vein, a difference of magnitude 2 in the BIC is considered as evidence against the model with the larger BIC, while a difference of magnitude 6 or more constitutes strong evidence \cite{Nesseris2010}. We may thus conclude that the models perform similarly when assessed by the AIC. However, the BIC penalizes for extra parameters more harshly than the AIC \cite{Liddle2004}, and consequently provides a considerable level of support for $\Lambda$CDM.
 
\section{Conclusion}
\label{sec:conclusion}
Many of the studies that investigate the nature of dark energy are based on the premise of a spatially flat Universe. However, it has been shown that if the true geometry is not exactly flat, this practice could critically distort the conclusions reached about the dynamics of dark energy \cite{Clarkson2007, Virey2008}. It is therefore important to ask what implications a non-zero $\Omega_k^0$ would have for dark energy scenarios. To this end, we consider three models from the literature that feature a dynamical $\Lambda$: the GRVM, whose characteristic $\Lambda(H)$ takes the form $A+BH^2+C \dot{H}$ \cite{Basilakos2012}, and two sub-cases: the RVM, obtained by setting $C$ to zero \cite{Shapiro2003}, and the model we call the GRVS, which has a null value for $B$ instead \cite{Karimkhani2015}. We assume that the vacuum only exchanges energy with cold dark matter as it decays. The parameters $B$ and/or $C$ are constrained by means of an MCMC analysis, initially using data for the observables associated with SNeIa, cosmic chronometers, the CMB, and BAOs. Each model is first investigated in the context of a flat space-time. Then we allow $\Omega_k^0$ to vary and look for any differences. We also analyze the effects of introducing a measurement of the Hubble constant as a fifth likelihood, and consider two different values for $H_0$: the one reported by Riess et al.~$\left(H_0^\text{R}=73.48\pm1.66~\si{km.s^{-1}.Mpc^{-1}}\right)$ \cite{Riess2018}, and that obtained by Efstathiou $\left(H_0^\text{E}\right)$ \cite{Efstathiou2014}, equal to $70.6\pm3.3~\si{km.s^{-1}.Mpc^{-1}}$. Finally, we include LSS observations in the collection of data sets and repeat the whole procedure. The amplitude of the primordial scalar power spectrum, $A_s$, is either assigned a fixed value or treated as a freely-varying parameter. 

In the case of the GRVM, the data we use is insufficient to break parameter degeneracies when the assumption of flatness is relaxed. The constraints we get in the flat scenario are, however, informative: we find that the addition of the LSS likelihood makes the posterior distributions for $B$ and $C$ close around the $\Lambda$CDM limit. It also changes the mean values of $B$ and $C$ from positive to negative -- although only if $H_0^\text{R}$ is absent from the combination of data sets, because the correlation between $B$ (or $C$) and the Hubble constant causes the $H_0^\text{R}$ likelihood to shift the posteriors in the direction of increasing $B$ (or $C$). We furthermore note that the inclusion of growth data lowers the averages for $H_0$, enhancing compatibility with the range of values established by \emph{Planck} \cite{Planck2018}.

Next, we turn our attention to the RVM. This time, the use of LSS data excludes the $\Lambda$CDM limit at a little over $1\sigma$ (in both the flat and non-flat cases, with one exception in the latter), while also serving to tighten the 2D posteriors in Fig.~\ref{RVMnf}. When $\Omega_k^0\neq0$, we note a change in the sign of the mean values of $B$ from negative to positive, and again find that growth data lends support to values of $H_0$ which resonate with the \emph{Planck} result. Of particular relevance is the fact that all the constraints on $\Omega_k^0$ become consistent with a flat geometry (within a $1\sigma$ confidence interval) once LSS observations are taken into account. %Furthermore, Fig. \ref{RVMnf} reveals a curious attribute: the $H_0^\text{R}$ likelihood appears to lend support to an open universe, and in fact shifts the curvature constraints in the direction of increasing $\Omega_k^0$. Consequently, the $1\sigma$ confidence interval for $\Omega_k^0$ is displaced away from zero when the values it encloses are mostly positive, but \emph{towards} zero if it mainly contains negative values instead.

The GRVS parallels the RVM in many ways. Here, too, growth data is responsible for a dynamical $\Lambda$ being preferred to a cosmological constant at a little over $1\sigma$, and once more this turns out to hold (for the most part) whether or not we assume that the Universe is spatially flat. The LSS likelihood establishes a definite (negative) sign for $C$ and increases compatibility between the average values of $H_0$ and the \emph{Planck} constraints. Its effect on the 1D posteriors for $\Omega_k^0$ is similar to what was noted for the RVM. In all cases, the addition of LSS observations puts $\Omega_k^0=0$ at less than $1\sigma$ from the resulting mean, but in its absence both the RVM and GRVS show some preference for an open universe, and indeed the level of support provided by the RVM for a positive $\Omega_k^0$ is over $2\sigma$ when $H_0^\text{R}$ is included with the data.

%The reason why the constraints placed on the model parameters and on $Omega_k^0$ by the $H_0^\text{R}$ data sets stand out from the rest is the significant level of correlation between $B$ (or $C$) and the Hubble constant. The appreciable difference between $H_0^\text{R}$ and $H_0^\text{E}$ can thus explain the tension in the results obtained with the respective data sets. 

In conclusion, our study indicates that a mildly-evolving $\Lambda$ (modeled as in the RVM or GRVS) is only marginally favored over the cosmological constant. Moreover, this comes at the cost of an extra parameter which -- while not given much weight by the AIC -- has a negative impact on the BIC score. Another point of interest is the fact that merging the RVM and GRVS expressions for $\Lambda$ into a two-parameter combination (the GRVM) appears to weaken the support for dynamical vacuum energy. We also investigate what happens when the assumption of spatial flatness is relaxed, and find that the addition of LSS data reduces the ability of the models to accommodate a non-zero $\Omega_k^0$.

\begin{acknowledgments}
This research has been carried out using computational facilities procured through the European Regional Development Fund, Project ERDF-080 \emph{``A super-computing laboratory for the University of Malta''}. The research work of C. F. disclosed in this publication is partially funded by the Endeavour Scholarship Scheme (Malta).  Scholarships are part-financed by the European Union -- European Social Fund (ESF) -- Operational Programme II\nopagebreak -- Cohesion Policy 2014-2020 \emph{``Investing in human capital to create more opportunities and promote the well-being of society''}.

The authors would like to thank the anonymous referee for his/her invaluable comments and suggestions. 
\end{acknowledgments}

\bibliographystyle{apsrev4-2}
\bibliography{Dynamical_Lambda_plus_curvature}

\begin{thebibliography}{164}%
\makeatletter
\providecommand \@ifxundefined [1]{%
 \@ifx{#1\undefined}
}%
\providecommand \@ifnum [1]{%
 \ifnum #1\expandafter \@firstoftwo
 \else \expandafter \@secondoftwo
 \fi
}%
\providecommand \@ifx [1]{%
 \ifx #1\expandafter \@firstoftwo
 \else \expandafter \@secondoftwo
 \fi
}%
\providecommand \natexlab [1]{#1}%
\providecommand \enquote  [1]{``#1''}%
\providecommand \bibnamefont  [1]{#1}%
\providecommand \bibfnamefont [1]{#1}%
\providecommand \citenamefont [1]{#1}%
\providecommand \href@noop [0]{\@secondoftwo}%
\providecommand \href [0]{\begingroup \@sanitize@url \@href}%
\providecommand \@href[1]{\@@startlink{#1}\@@href}%
\providecommand \@@href[1]{\endgroup#1\@@endlink}%
\providecommand \@sanitize@url [0]{\catcode `\\12\catcode `\$12\catcode
  `\&12\catcode `\#12\catcode `\^12\catcode `\_12\catcode `\%12\relax}%
\providecommand \@@startlink[1]{}%
\providecommand \@@endlink[0]{}%
\providecommand \url  [0]{\begingroup\@sanitize@url \@url }%
\providecommand \@url [1]{\endgroup\@href {#1}{\urlprefix }}%
\providecommand \urlprefix  [0]{URL }%
\providecommand \Eprint [0]{\href }%
\providecommand \doibase [0]{https://doi.org/}%
\providecommand \selectlanguage [0]{\@gobble}%
\providecommand \bibinfo  [0]{\@secondoftwo}%
\providecommand \bibfield  [0]{\@secondoftwo}%
\providecommand \translation [1]{[#1]}%
\providecommand \BibitemOpen [0]{}%
\providecommand \bibitemStop [0]{}%
\providecommand \bibitemNoStop [0]{.\EOS\space}%
\providecommand \EOS [0]{\spacefactor3000\relax}%
\providecommand \BibitemShut  [1]{\csname bibitem#1\endcsname}%
\let\auto@bib@innerbib\@empty
%</preamble>
\bibitem [{\citenamefont {Miao}\ \emph {et~al.}(2011)\citenamefont {Miao},
  \citenamefont {Xiao-Dong}, \citenamefont {Shuang},\ and\ \citenamefont
  {Yi}}]{Miao2011}%
  \BibitemOpen
  \bibfield  {author} {\bibinfo {author} {\bibfnamefont {L.}~\bibnamefont
  {Miao}}, \bibinfo {author} {\bibfnamefont {L.}~\bibnamefont {Xiao-Dong}},
  \bibinfo {author} {\bibfnamefont {W.}~\bibnamefont {Shuang}},\ and\ \bibinfo
  {author} {\bibfnamefont {W.}~\bibnamefont {Yi}},\ }\href
  {https://doi.org/10.1088/0253-6102/56/3/24} {\bibfield  {journal} {\bibinfo
  {journal} {Commun. Theor. Phys.}\ }\textbf {\bibinfo {volume} {56}},\
  \bibinfo {pages} {525} (\bibinfo {year} {2011})}\BibitemShut {NoStop}%
\bibitem [{\citenamefont {Yoo}\ and\ \citenamefont {Watanabe}(2012)}]{Yoo2012}%
  \BibitemOpen
  \bibfield  {author} {\bibinfo {author} {\bibfnamefont {J.}~\bibnamefont
  {Yoo}}\ and\ \bibinfo {author} {\bibfnamefont {Y.}~\bibnamefont {Watanabe}},\
  }\href {https://doi.org/10.1142/S0218271812300029} {\bibfield  {journal}
  {\bibinfo  {journal} {Int. J. Mod. Phys. D}\ }\textbf {\bibinfo {volume}
  {21}},\ \bibinfo {pages} {1230002} (\bibinfo {year} {2012})}\BibitemShut
  {NoStop}%
\bibitem [{\citenamefont {{A. G. Riess \emph{et al.}}}(1998)}]{Riess1998}%
  \BibitemOpen
  \bibfield  {author} {\bibinfo {author} {\bibnamefont {{A. G. Riess \emph{et
  al.}}}},\ }\href {https://doi.org/10.1086/300499} {\bibfield  {journal}
  {\bibinfo  {journal} {Astron. J.}\ }\textbf {\bibinfo {volume} {116}},\
  \bibinfo {pages} {1009} (\bibinfo {year} {1998})}\BibitemShut {NoStop}%
\bibitem [{\citenamefont {{S. Perlmutter \emph{et al.} (The Supernova Cosmology
  Project)}}(1999)}]{Perlmutter1999}%
  \BibitemOpen
  \bibfield  {author} {\bibinfo {author} {\bibnamefont {{S. Perlmutter \emph{et
  al.} (The Supernova Cosmology Project)}}},\ }\href
  {https://doi.org/10.1086/307221} {\bibfield  {journal} {\bibinfo  {journal}
  {Astrophys. J.}\ }\textbf {\bibinfo {volume} {517}},\ \bibinfo {pages} {565}
  (\bibinfo {year} {1999})}\BibitemShut {NoStop}%
\bibitem [{\citenamefont {{B. P. Schmidt \emph{et al.}}}(1998)}]{Schmidt1998}%
  \BibitemOpen
  \bibfield  {author} {\bibinfo {author} {\bibnamefont {{B. P. Schmidt \emph{et
  al.}}}},\ }\href {https://doi.org/10.1086/306308} {\bibfield  {journal}
  {\bibinfo  {journal} {Astrophys. J.}\ }\textbf {\bibinfo {volume} {507}},\
  \bibinfo {pages} {46} (\bibinfo {year} {1998})}\BibitemShut {NoStop}%
\bibitem [{\citenamefont {Einstein}(1917)}]{Einstein1917}%
  \BibitemOpen
  \bibfield  {author} {\bibinfo {author} {\bibfnamefont {A.}~\bibnamefont
  {Einstein}},\ }\href
  {http://articles.adsabs.harvard.edu/cgi-bin/get_file?pdfs/SPAW./1917/1917SPAW.......142E.pdf}
  {\bibfield  {journal} {\bibinfo  {journal} {Sitzungsber. Preuss. Akad. Wiss.
  Berlin (Math. Phys.)}\ }\textbf {\bibinfo {volume} {1917}},\ \bibinfo {pages}
  {142} (\bibinfo {year} {1917})},\ \bibinfo {note} {{\emph{translated into
  English by} W. Perrett and G. B. Jeffery, in \emph{The Principle of
  Relativity}, by H. A. Lorentz, A. Einstein, H. Minkowski, and H. Weyl (Dover,
  New York, 1952) pp.~175--188. With notes by A. Sommerfeld.}}\BibitemShut
  {Stop}%
\bibitem [{\citenamefont {Tegmark}\ \emph {et~al.}()\citenamefont {Tegmark},
  \citenamefont {Eisenstein}, \citenamefont {Hu},\ and\ \citenamefont
  {Kron}}]{Tegmark1998}%
  \BibitemOpen
  \bibfield  {author} {\bibinfo {author} {\bibfnamefont {M.}~\bibnamefont
  {Tegmark}}, \bibinfo {author} {\bibfnamefont {D.~J.}\ \bibnamefont
  {Eisenstein}}, \bibinfo {author} {\bibfnamefont {W.}~\bibnamefont {Hu}},\
  and\ \bibinfo {author} {\bibfnamefont {R.}~\bibnamefont {Kron}},\ }\href
  {http://arxiv.org/abs/astro-ph/9805117} {\ }\Eprint
  {https://arxiv.org/abs/astro-ph/9805117} {arXiv:astro-ph/9805117}
  \BibitemShut {NoStop}%
\bibitem [{\citenamefont {{P. A. R. Ade \emph{et
  al.}}}(2016{\natexlab{a}})}]{Ade2016}%
  \BibitemOpen
  \bibfield  {author} {\bibinfo {author} {\bibnamefont {{P. A. R. Ade \emph{et
  al.}}}} (\bibinfo {collaboration} {Planck Collaboration}),\ }\href
  {https://doi.org/10.1051/0004-6361/201525814} {\bibfield  {journal} {\bibinfo
   {journal} {Astron. \& Astrophys.}\ }\textbf {\bibinfo {volume} {594}},\
  \bibinfo {pages} {A14} (\bibinfo {year} {2016}{\natexlab{a}})}\BibitemShut
  {NoStop}%
\bibitem [{\citenamefont {{S. Joudaki \emph{et al.}}}(2017)}]{Joudaki2017}%
  \BibitemOpen
  \bibfield  {author} {\bibinfo {author} {\bibnamefont {{S. Joudaki \emph{et
  al.}}}},\ }\href {https://doi.org/10.1093/mnras/stx998} {\bibfield  {journal}
  {\bibinfo  {journal} {Mon. Not. R. Astron. Soc.}\ }\textbf {\bibinfo {volume}
  {471}},\ \bibinfo {pages} {1259} (\bibinfo {year} {2017})}\BibitemShut
  {NoStop}%
\bibitem [{\citenamefont {Einstein}()}]{Einstein1915}%
  \BibitemOpen
  \bibfield  {author} {\bibinfo {author} {\bibfnamefont {A.}~\bibnamefont
  {Einstein}},\ }\href {http://inspirehep.net/record/42609/} {\bibfield
  {journal} {\bibinfo  {journal} {Sitzungsber. Preuss. Akad. Wiss. Berlin}\
  }\textbf {\bibinfo {volume} {1915}},\ \bibinfo {pages} {831}},\ \bibinfo
  {note} {{\emph{translated into English by} A. Engel, in \emph{The Collected
  Papers of Albert Einstein, Volume 6: The Berlin Years: Writings, 1914--1917},
  (Princeton University Press, Princeton, New Jersey, 1997) pp.~112--116. With
  E. Schucking as consultant.}}\BibitemShut {Stop}%
\bibitem [{\citenamefont {Shapiro}(1990)}]{Shapiro1990}%
  \BibitemOpen
  \bibfield  {author} {\bibinfo {author} {\bibfnamefont {I.~i.}\ \bibnamefont
  {Shapiro}},\ }in\ \href
  {https://inis.iaea.org/search/search.aspx?orig_q=RN:22036209} {\emph
  {\bibinfo {booktitle} {General Relativity and Gravitation, 1989: Proceedings
  of the 12th International Conference on General Relativity and
  Gravitation}}},\ \bibinfo {editor} {edited by\ \bibinfo {editor}
  {\bibfnamefont {N.}~\bibnamefont {Ashby}}, \bibinfo {editor} {\bibfnamefont
  {D.~F.}\ \bibnamefont {Bartlett}},\ and\ \bibinfo {editor} {\bibfnamefont
  {W.}~\bibnamefont {Wyss}}}\ (\bibinfo  {publisher} {Cambridge University
  Press},\ \bibinfo {address} {Cambridge, New York},\ \bibinfo {year} {1990})\
  pp.\ \bibinfo {pages} {313--330}\BibitemShut {NoStop}%
\bibitem [{\citenamefont {{D. J. Eisenstein \emph{et
  al.}}}(2005)}]{Eisenstein2005}%
  \BibitemOpen
  \bibfield  {author} {\bibinfo {author} {\bibnamefont {{D. J. Eisenstein
  \emph{et al.}}}},\ }\href {https://doi.org/10.1086/466512} {\bibfield
  {journal} {\bibinfo  {journal} {Astrophys. J.}\ }\textbf {\bibinfo {volume}
  {633}},\ \bibinfo {pages} {560} (\bibinfo {year} {2005})}\BibitemShut
  {NoStop}%
\bibitem [{\citenamefont {{B. P. Abbott \emph{et al.} (LIGO Scientific
  Collaboration and \textsc{Virgo} Collaboration)}}(2016)}]{Abbott2016}%
  \BibitemOpen
  \bibfield  {author} {\bibinfo {author} {\bibnamefont {{B. P. Abbott \emph{et
  al.} (LIGO Scientific Collaboration and \textsc{Virgo} Collaboration)}}},\
  }\href {https://doi.org/10.1103/PhysRevLett.116.061102} {\bibfield  {journal}
  {\bibinfo  {journal} {Phys. Rev. Lett.}\ }\textbf {\bibinfo {volume} {116}},\
  \bibinfo {pages} {061102} (\bibinfo {year} {2016})}\BibitemShut {NoStop}%
\bibitem [{\citenamefont {{N. Aghanim \emph{et al.}}}()}]{Planck2018}%
  \BibitemOpen
  \bibfield  {author} {\bibinfo {author} {\bibnamefont {{N. Aghanim \emph{et
  al.}}}} (\bibinfo {collaboration} {Planck Collaboration}),\ }\href
  {http://arxiv.org/abs/1807.06209} {\ }\Eprint
  {https://arxiv.org/abs/1807.06209} {arXiv:1807.06209} \BibitemShut {NoStop}%
\bibitem [{\citenamefont {{P. A. R. Ade \emph{et
  al.}}}(2016{\natexlab{b}})}]{Planck2015}%
  \BibitemOpen
  \bibfield  {author} {\bibinfo {author} {\bibnamefont {{P. A. R. Ade \emph{et
  al.}}}} (\bibinfo {collaboration} {Planck Collaboration}),\ }\href
  {https://doi.org/10.1051/0004-6361/201525830} {\bibfield  {journal} {\bibinfo
   {journal} {Astron. \& Astrophys.}\ }\textbf {\bibinfo {volume} {594}},\
  \bibinfo {pages} {A13} (\bibinfo {year} {2016}{\natexlab{b}})}\BibitemShut
  {NoStop}%
\bibitem [{\citenamefont {{C. L. Bennett \emph{et al.}}}(2013)}]{Bennett2013}%
  \BibitemOpen
  \bibfield  {author} {\bibinfo {author} {\bibnamefont {{C. L. Bennett \emph{et
  al.}}}},\ }\href {https://doi.org/10.1088/0067-0049/208/2/20} {\bibfield
  {journal} {\bibinfo  {journal} {Astrophys. J. Suppl. Ser.}\ }\textbf
  {\bibinfo {volume} {208}},\ \bibinfo {pages} {20} (\bibinfo {year}
  {2013})}\BibitemShut {NoStop}%
\bibitem [{\citenamefont {Weinberg}(2001)}]{Weinberg2001}%
  \BibitemOpen
  \bibfield  {author} {\bibinfo {author} {\bibfnamefont {S.}~\bibnamefont
  {Weinberg}},\ }in\ \href {https://doi.org/10.1007/978-3-662-04587-9_2} {\emph
  {\bibinfo {booktitle} {Sources and Detection of Dark Matter and Dark Energy
  in the Universe}}},\ \bibinfo {editor} {edited by\ \bibinfo {editor}
  {\bibfnamefont {D.~B.}\ \bibnamefont {Cline}}}\ (\bibinfo  {publisher}
  {Springer-Verlag},\ \bibinfo {address} {Berlin \& Heidelberg},\ \bibinfo
  {year} {2001})\ pp.\ \bibinfo {pages} {18--26}\BibitemShut {NoStop}%
\bibitem [{\citenamefont {{A.~G. Riess \emph{et al.}}}(2018)}]{Riess2018}%
  \BibitemOpen
  \bibfield  {author} {\bibinfo {author} {\bibnamefont {{A.~G. Riess \emph{et
  al.}}}},\ }\href {https://doi.org/10.3847/1538-4357/aaadb7} {\bibfield
  {journal} {\bibinfo  {journal} {Astrophys. J.}\ }\textbf {\bibinfo {volume}
  {855}},\ \bibinfo {pages} {136} (\bibinfo {year} {2018})}\BibitemShut
  {NoStop}%
\bibitem [{\citenamefont {Nakama}\ \emph {et~al.}(2017)\citenamefont {Nakama},
  \citenamefont {Chluba},\ and\ \citenamefont {Kamionkowski}}]{Nakama2017}%
  \BibitemOpen
  \bibfield  {author} {\bibinfo {author} {\bibfnamefont {T.}~\bibnamefont
  {Nakama}}, \bibinfo {author} {\bibfnamefont {J.}~\bibnamefont {Chluba}},\
  and\ \bibinfo {author} {\bibfnamefont {M.}~\bibnamefont {Kamionkowski}},\
  }\href {https://doi.org/10.1103/PhysRevD.95.121302} {\bibfield  {journal}
  {\bibinfo  {journal} {Phys. Rev. D}\ }\textbf {\bibinfo {volume} {95}},\
  \bibinfo {pages} {121302} (\bibinfo {year} {2017})}\BibitemShut {NoStop}%
\bibitem [{\citenamefont {Hu}\ and\ \citenamefont {Sawicki}(2007)}]{Hu2007}%
  \BibitemOpen
  \bibfield  {author} {\bibinfo {author} {\bibfnamefont {W.}~\bibnamefont
  {Hu}}\ and\ \bibinfo {author} {\bibfnamefont {I.}~\bibnamefont {Sawicki}},\
  }\href {https://doi.org/10.1103/PhysRevD.76.064004} {\bibfield  {journal}
  {\bibinfo  {journal} {Phys. Rev. D}\ }\textbf {\bibinfo {volume} {76}},\
  \bibinfo {pages} {064004} (\bibinfo {year} {2007})}\BibitemShut {NoStop}%
\bibitem [{\citenamefont {Will}(2006)}]{Will2006}%
  \BibitemOpen
  \bibfield  {author} {\bibinfo {author} {\bibfnamefont {C.~M.}\ \bibnamefont
  {Will}},\ }\href {https://doi.org/10.12942/lrr-2006-3} {\bibfield  {journal}
  {\bibinfo  {journal} {Living Rev. Relativ.}\ }\textbf {\bibinfo {volume}
  {9}},\ \bibinfo {pages} {3} (\bibinfo {year} {2006})}\BibitemShut {NoStop}%
\bibitem [{\citenamefont {Khoury}(2013)}]{Khoury2013}%
  \BibitemOpen
  \bibfield  {author} {\bibinfo {author} {\bibfnamefont {J.}~\bibnamefont
  {Khoury}},\ }\href {https://doi.org/10.1088/0264-9381/30/21/214004}
  {\bibfield  {journal} {\bibinfo  {journal} {Class. Quant. Grav.}\ }\textbf
  {\bibinfo {volume} {30}},\ \bibinfo {pages} {214004} (\bibinfo {year}
  {2013})}\BibitemShut {NoStop}%
\bibitem [{\citenamefont {Basilakos}\ \emph {et~al.}(2012)\citenamefont
  {Basilakos}, \citenamefont {Polarski},\ and\ \citenamefont
  {Sol\`{a}}}]{Basilakos2012}%
  \BibitemOpen
  \bibfield  {author} {\bibinfo {author} {\bibfnamefont {S.}~\bibnamefont
  {Basilakos}}, \bibinfo {author} {\bibfnamefont {D.}~\bibnamefont
  {Polarski}},\ and\ \bibinfo {author} {\bibfnamefont {J.}~\bibnamefont
  {Sol\`{a}}},\ }\href {https://doi.org/10.1103/PhysRevD.86.043010} {\bibfield
  {journal} {\bibinfo  {journal} {Phys. Rev. D}\ }\textbf {\bibinfo {volume}
  {86}},\ \bibinfo {pages} {043010} (\bibinfo {year} {2012})}\BibitemShut
  {NoStop}%
\bibitem [{\citenamefont {Shapiro}\ \emph {et~al.}(2003)\citenamefont
  {Shapiro}, \citenamefont {Sol\`{a}}, \citenamefont {{Espa\~{n}a-Bonet}},\
  and\ \citenamefont {Ruiz-Lapuente}}]{Shapiro2003}%
  \BibitemOpen
  \bibfield  {author} {\bibinfo {author} {\bibfnamefont {I.~L.}\ \bibnamefont
  {Shapiro}}, \bibinfo {author} {\bibfnamefont {J.}~\bibnamefont {Sol\`{a}}},
  \bibinfo {author} {\bibfnamefont {C.}~\bibnamefont {{Espa\~{n}a-Bonet}}},\
  and\ \bibinfo {author} {\bibfnamefont {P.}~\bibnamefont {Ruiz-Lapuente}},\
  }\href {https://doi.org/10.1016/j.physletb.2003.09.016} {\bibfield  {journal}
  {\bibinfo  {journal} {Phys. Lett. B}\ }\textbf {\bibinfo {volume} {574}},\
  \bibinfo {pages} {149} (\bibinfo {year} {2003})}\BibitemShut {NoStop}%
\bibitem [{\citenamefont {G\'{o}mez-Valent}\ \emph
  {et~al.}({\natexlab{a}})\citenamefont {G\'{o}mez-Valent}, \citenamefont
  {Sol\`{a}},\ and\ \citenamefont {Basilakos}}]{Valent2015}%
  \BibitemOpen
  \bibfield  {author} {\bibinfo {author} {\bibfnamefont {A.}~\bibnamefont
  {G\'{o}mez-Valent}}, \bibinfo {author} {\bibfnamefont {J.}~\bibnamefont
  {Sol\`{a}}},\ and\ \bibinfo {author} {\bibfnamefont {S.}~\bibnamefont
  {Basilakos}},\ }\href {https://doi.org/10.1088/1475-7516/2015/01/004}
  {\bibfield  {journal} {\bibinfo  {journal} {J. Cosmol. Astropart. Phys.}\
  }\textbf {\bibinfo {volume} {2015}}\bibinfo  {number} { (01)},\ \bibinfo
  {pages} {004}}\BibitemShut {NoStop}%
\bibitem [{\citenamefont {{Sol\`{a} Peracaula}}\ \emph
  {et~al.}(2018{\natexlab{a}})\citenamefont {{Sol\`{a} Peracaula}},
  \citenamefont {{de Cruz P\'{e}rez}},\ and\ \citenamefont
  {G\'{o}mez-Valent}}]{Peracaula2018}%
  \BibitemOpen
\bibfield  {number} {  }\bibfield  {author} {\bibinfo {author} {\bibfnamefont
  {J.}~\bibnamefont {{Sol\`{a} Peracaula}}}, \bibinfo {author} {\bibfnamefont
  {J.}~\bibnamefont {{de Cruz P\'{e}rez}}},\ and\ \bibinfo {author}
  {\bibfnamefont {A.}~\bibnamefont {G\'{o}mez-Valent}},\ }\href
  {https://doi.org/10.1093/mnras/sty1253} {\bibfield  {journal} {\bibinfo
  {journal} {Mon. Not. R. Astron. Soc.}\ }\textbf {\bibinfo {volume} {478}},\
  \bibinfo {pages} {4357} (\bibinfo {year} {2018}{\natexlab{a}})}\BibitemShut
  {NoStop}%
\bibitem [{\citenamefont {Basilakos}\ \emph {et~al.}(2009)\citenamefont
  {Basilakos}, \citenamefont {Plionis},\ and\ \citenamefont
  {Sol\`{a}}}]{Basilakos2009}%
  \BibitemOpen
  \bibfield  {author} {\bibinfo {author} {\bibfnamefont {S.}~\bibnamefont
  {Basilakos}}, \bibinfo {author} {\bibfnamefont {M.}~\bibnamefont {Plionis}},\
  and\ \bibinfo {author} {\bibfnamefont {J.}~\bibnamefont {Sol\`{a}}},\
  }\href@noop {} {\bibfield  {journal} {\bibinfo  {journal} {Phys. Rev. D}\
  }\textbf {\bibinfo {volume} {80}},\ \bibinfo {pages} {083511} (\bibinfo
  {year} {2009})}\BibitemShut {NoStop}%
\bibitem [{\citenamefont {Grande}\ \emph {et~al.}()\citenamefont {Grande},
  \citenamefont {Sol\`{a}}, \citenamefont {Basilakos},\ and\ \citenamefont
  {Plionis}}]{Grande}%
  \BibitemOpen
  \bibfield  {author} {\bibinfo {author} {\bibfnamefont {J.}~\bibnamefont
  {Grande}}, \bibinfo {author} {\bibfnamefont {J.}~\bibnamefont {Sol\`{a}}},
  \bibinfo {author} {\bibfnamefont {S.}~\bibnamefont {Basilakos}},\ and\
  \bibinfo {author} {\bibfnamefont {M.}~\bibnamefont {Plionis}},\ }\href@noop
  {} {\bibfield  {journal} {\bibinfo  {journal} {J. Cosmol. Astropart. Phys.}\
  }\textbf {\bibinfo {volume} {2011}}\bibinfo  {number} { (08)},\ \bibinfo
  {pages} {007}}\BibitemShut {NoStop}%
\bibitem [{\citenamefont {Geng}\ \emph {et~al.}()\citenamefont {Geng},
  \citenamefont {Lee},\ and\ \citenamefont {Yin}}]{Geng2017}%
  \BibitemOpen
\bibfield  {number} {  }\bibfield  {author} {\bibinfo {author} {\bibfnamefont
  {C.-Q.}\ \bibnamefont {Geng}}, \bibinfo {author} {\bibfnamefont {C.-C.}\
  \bibnamefont {Lee}},\ and\ \bibinfo {author} {\bibfnamefont {L.}~\bibnamefont
  {Yin}},\ }\href {https://doi.org/10.1088/1475-7516/2017/08/032} {\bibfield
  {journal} {\bibinfo  {journal} {J. Cosmol. Astropart. Phys.}\ }\textbf
  {\bibinfo {volume} {2017}}\bibinfo  {number} { (08)},\ \bibinfo {pages}
  {032}}\BibitemShut {NoStop}%
\bibitem [{\citenamefont {G\'{o}mez-Valent}\ \emph
  {et~al.}({\natexlab{b}})\citenamefont {G\'{o}mez-Valent}, \citenamefont
  {Karimkhani},\ and\ \citenamefont {Sol\`{a}}}]{Karimkhani2015}%
  \BibitemOpen
\bibfield  {number} {  }\bibfield  {author} {\bibinfo {author} {\bibfnamefont
  {A.}~\bibnamefont {G\'{o}mez-Valent}}, \bibinfo {author} {\bibfnamefont
  {E.}~\bibnamefont {Karimkhani}},\ and\ \bibinfo {author} {\bibfnamefont
  {J.}~\bibnamefont {Sol\`{a}}},\ }\href
  {https://doi.org/10.1088/1475-7516/2015/12/048} {\bibfield  {journal}
  {\bibinfo  {journal} {J. Cosmol. Astropart. Phys.}\ }\textbf {\bibinfo
  {volume} {2015}}\bibinfo  {number} { (12)},\ \bibinfo {pages}
  {048}}\BibitemShut {NoStop}%
\bibitem [{\citenamefont {Shapiro}\ and\ \citenamefont
  {Sol\`{a}}(2000)}]{Shapiro2000}%
  \BibitemOpen
\bibfield  {number} {  }\bibfield  {author} {\bibinfo {author} {\bibfnamefont
  {I.~L.}\ \bibnamefont {Shapiro}}\ and\ \bibinfo {author} {\bibfnamefont
  {J.}~\bibnamefont {Sol\`{a}}},\ }\href
  {https://doi.org/10.1016/S0370-2693(00)00090-3} {\bibfield  {journal}
  {\bibinfo  {journal} {Phys. Lett. B}\ }\textbf {\bibinfo {volume} {475}},\
  \bibinfo {pages} {236} (\bibinfo {year} {2000})}\BibitemShut {NoStop}%
\bibitem [{\citenamefont {Shapiro}\ and\ \citenamefont
  {Sol\`{a}}(2002)}]{Shapiro2002}%
  \BibitemOpen
  \bibfield  {author} {\bibinfo {author} {\bibfnamefont {I.~L.}\ \bibnamefont
  {Shapiro}}\ and\ \bibinfo {author} {\bibfnamefont {J.}~\bibnamefont
  {Sol\`{a}}},\ }\href
  {http://jhep.sissa.it/archive/papers/jhep022002006/jhep022002006.pdf}
  {\bibfield  {journal} {\bibinfo  {journal} {J. High Energy Phys.}\ }\textbf
  {\bibinfo {volume} {2002}}\bibinfo  {number} { (02)},\ \bibinfo {pages}
  {006}}\BibitemShut {NoStop}%
\bibitem [{\citenamefont {{Shapiro}}\ and\ \citenamefont
  {Sol\`{a}}(2004)}]{Shapiro2004}%
  \BibitemOpen
\bibfield  {number} {  }\bibfield  {author} {\bibinfo {author} {\bibfnamefont
  {I.~L.}\ \bibnamefont {{Shapiro}}}\ and\ \bibinfo {author} {\bibfnamefont
  {J.}~\bibnamefont {Sol\`{a}}},\ }\href@noop {} {\bibfield  {journal}
  {\bibinfo  {journal} {Nucl. Phys. B}\ }\textbf {\bibinfo {volume} {127}},\
  \bibinfo {pages} {71} (\bibinfo {year} {2004})}\BibitemShut {NoStop}%
\bibitem [{\citenamefont {Sol\`{a}}(2008)}]{Sola2008}%
  \BibitemOpen
  \bibfield  {author} {\bibinfo {author} {\bibfnamefont {J.}~\bibnamefont
  {Sol\`{a}}},\ }\href@noop {} {\bibfield  {journal} {\bibinfo  {journal} {J.
  Phys. A -- Math. Theor.}\ }\textbf {\bibinfo {volume} {41}},\ \bibinfo
  {pages} {164066} (\bibinfo {year} {2008})}\BibitemShut {NoStop}%
\bibitem [{\citenamefont {Moreno-Pulido}\ and\ \citenamefont {{Sol\`{a}
  Peracaula}}()}]{MorenoPulido}%
  \BibitemOpen
  \bibfield  {author} {\bibinfo {author} {\bibfnamefont {C.}~\bibnamefont
  {Moreno-Pulido}}\ and\ \bibinfo {author} {\bibfnamefont {J.}~\bibnamefont
  {{Sol\`{a} Peracaula}}},\ }\href {https://arxiv.org/abs/2005.03164v2} {\
  }\Eprint {https://arxiv.org/abs/2005.03164v2} {arXiv:2005.03164v2}
  \BibitemShut {NoStop}%
\bibitem [{\citenamefont {Sol\`{a}}\ \emph {et~al.}(2017)\citenamefont
  {Sol\`{a}}, \citenamefont {G\'{o}mez-Valent},\ and\ \citenamefont {{de Cruz
  P\'{e}rez}}}]{JSola2017}%
  \BibitemOpen
  \bibfield  {author} {\bibinfo {author} {\bibfnamefont {J.}~\bibnamefont
  {Sol\`{a}}}, \bibinfo {author} {\bibfnamefont {A.}~\bibnamefont
  {G\'{o}mez-Valent}},\ and\ \bibinfo {author} {\bibfnamefont {J.}~\bibnamefont
  {{de Cruz P\'{e}rez}}},\ }\href@noop {} {\bibfield  {journal} {\bibinfo
  {journal} {Phys. Lett. B}\ }\textbf {\bibinfo {volume} {774}},\ \bibinfo
  {pages} {317} (\bibinfo {year} {2017})}\BibitemShut {NoStop}%
\bibitem [{\citenamefont {{Sol\`{a} Peracaula}}\ \emph
  {et~al.}(2018{\natexlab{b}})\citenamefont {{Sol\`{a} Peracaula}},
  \citenamefont {{de Cruz P\'{e}rez}},\ and\ \citenamefont
  {G\'{o}mez-Valent}}]{PeracaulaJ}%
  \BibitemOpen
  \bibfield  {author} {\bibinfo {author} {\bibfnamefont {J.}~\bibnamefont
  {{Sol\`{a} Peracaula}}}, \bibinfo {author} {\bibfnamefont {J.}~\bibnamefont
  {{de Cruz P\'{e}rez}}},\ and\ \bibinfo {author} {\bibfnamefont
  {A.}~\bibnamefont {G\'{o}mez-Valent}},\ }\href@noop {} {\bibfield  {journal}
  {\bibinfo  {journal} {Europhys. Lett.}\ }\textbf {\bibinfo {volume} {121}},\
  \bibinfo {pages} {39001} (\bibinfo {year} {2018}{\natexlab{b}})}\BibitemShut
  {NoStop}%
\bibitem [{\citenamefont {{Sol\`{a}}}\ \emph {et~al.}(2017)\citenamefont
  {{Sol\`{a}}}, \citenamefont {G\'{o}mez-Valent},\ and\ \citenamefont {{de Cruz
  P\'{e}rez}}}]{Sola2017}%
  \BibitemOpen
  \bibfield  {author} {\bibinfo {author} {\bibfnamefont {J.}~\bibnamefont
  {{Sol\`{a}}}}, \bibinfo {author} {\bibfnamefont {A.}~\bibnamefont
  {G\'{o}mez-Valent}},\ and\ \bibinfo {author} {\bibfnamefont {J.}~\bibnamefont
  {{de Cruz P\'{e}rez}}},\ }\href@noop {} {\bibfield  {journal} {\bibinfo
  {journal} {Astrophys. J.}\ }\textbf {\bibinfo {volume} {836}},\ \bibinfo
  {pages} {43} (\bibinfo {year} {2017})}\BibitemShut {NoStop}%
\bibitem [{\citenamefont {{S\'{a}nchez}}\ \emph {et~al.}(2009)\citenamefont
  {{S\'{a}nchez}}, \citenamefont {Crocce}, \citenamefont {Cabr\'{e}},
  \citenamefont {{Baugh}},\ and\ \citenamefont {Gazta\~{n}aga}}]{Sanchez}%
  \BibitemOpen
  \bibfield  {author} {\bibinfo {author} {\bibfnamefont {A.~G.}\ \bibnamefont
  {{S\'{a}nchez}}}, \bibinfo {author} {\bibfnamefont {M.}~\bibnamefont
  {Crocce}}, \bibinfo {author} {\bibfnamefont {A.}~\bibnamefont {Cabr\'{e}}},
  \bibinfo {author} {\bibfnamefont {C.~M.}\ \bibnamefont {{Baugh}}},\ and\
  \bibinfo {author} {\bibfnamefont {E.}~\bibnamefont {Gazta\~{n}aga}},\
  }\href@noop {} {\bibfield  {journal} {\bibinfo  {journal} {Mon. Not. R.
  Astron. Soc.}\ }\textbf {\bibinfo {volume} {400}},\ \bibinfo {pages} {1643}
  (\bibinfo {year} {2009})}\BibitemShut {NoStop}%
\bibitem [{\citenamefont {Moresco}\ \emph {et~al.}({\natexlab{a}})\citenamefont
  {Moresco}, \citenamefont {Jimenez}, \citenamefont {Verde}, \citenamefont
  {Cimatti}, \citenamefont {Pozzetti}, \citenamefont {Maraston},\ and\
  \citenamefont {Thomas}}]{Moresco}%
  \BibitemOpen
  \bibfield  {author} {\bibinfo {author} {\bibfnamefont {M.}~\bibnamefont
  {Moresco}}, \bibinfo {author} {\bibfnamefont {R.}~\bibnamefont {Jimenez}},
  \bibinfo {author} {\bibfnamefont {L.}~\bibnamefont {Verde}}, \bibinfo
  {author} {\bibfnamefont {A.}~\bibnamefont {Cimatti}}, \bibinfo {author}
  {\bibfnamefont {L.}~\bibnamefont {Pozzetti}}, \bibinfo {author}
  {\bibfnamefont {C.}~\bibnamefont {Maraston}},\ and\ \bibinfo {author}
  {\bibfnamefont {D.}~\bibnamefont {Thomas}},\ }\href@noop {} {\bibfield
  {journal} {\bibinfo  {journal} {J. Cosmol. Astropart. Phys.}\ }\textbf
  {\bibinfo {volume} {2016}}\bibinfo  {number} { (12)},\ \bibinfo {pages}
  {039}}\BibitemShut {NoStop}%
\bibitem [{\citenamefont {Berman}(1991)}]{Berman1991}%
  \BibitemOpen
\bibfield  {number} {  }\bibfield  {author} {\bibinfo {author} {\bibfnamefont
  {M.~S.}\ \bibnamefont {Berman}},\ }\href
  {https://doi.org/10.1103/PhysRevD.43.1075} {\bibfield  {journal} {\bibinfo
  {journal} {Phys. Rev. D}\ }\textbf {\bibinfo {volume} {43}},\ \bibinfo
  {pages} {1075} (\bibinfo {year} {1991})}\BibitemShut {NoStop}%
\bibitem [{\citenamefont {Beesham}(1993)}]{Beesham1993}%
  \BibitemOpen
  \bibfield  {author} {\bibinfo {author} {\bibfnamefont {A.}~\bibnamefont
  {Beesham}},\ }\href {https://doi.org/10.1103/PhysRevD.48.3539} {\bibfield
  {journal} {\bibinfo  {journal} {Phys. Rev. D}\ }\textbf {\bibinfo {volume}
  {48}},\ \bibinfo {pages} {3539} (\bibinfo {year} {1993})}\BibitemShut
  {NoStop}%
\bibitem [{\citenamefont {Perico}\ \emph {et~al.}(2013)\citenamefont {Perico},
  \citenamefont {Lima}, \citenamefont {Basilakos},\ and\ \citenamefont
  {Sol{\`{a}}}}]{Perico2013}%
  \BibitemOpen
  \bibfield  {author} {\bibinfo {author} {\bibfnamefont {E.~L.~D.}\
  \bibnamefont {Perico}}, \bibinfo {author} {\bibfnamefont {J.~A.~S.}\
  \bibnamefont {Lima}}, \bibinfo {author} {\bibfnamefont {S.}~\bibnamefont
  {Basilakos}},\ and\ \bibinfo {author} {\bibfnamefont {J.}~\bibnamefont
  {Sol{\`{a}}}},\ }\href {https://doi.org/10.1103/PhysRevD.88.063531}
  {\bibfield  {journal} {\bibinfo  {journal} {Phys. Rev. D}\ }\textbf {\bibinfo
  {volume} {88}},\ \bibinfo {pages} {063531} (\bibinfo {year}
  {2013})}\BibitemShut {NoStop}%
\bibitem [{\citenamefont {Bertolami}(1986)}]{Bertolami1986}%
  \BibitemOpen
  \bibfield  {author} {\bibinfo {author} {\bibfnamefont {O.}~\bibnamefont
  {Bertolami}},\ }\href {https://doi.org/10.1007/BF02728301} {\bibfield
  {journal} {\bibinfo  {journal} {Nuovo Cim. B}\ }\textbf {\bibinfo {volume}
  {93}},\ \bibinfo {pages} {36} (\bibinfo {year} {1986})}\BibitemShut {NoStop}%
\bibitem [{\citenamefont {Kalligas}\ \emph {et~al.}(1992)\citenamefont
  {Kalligas}, \citenamefont {Wesson},\ and\ \citenamefont
  {Everitt}}]{Kalligas1992}%
  \BibitemOpen
  \bibfield  {author} {\bibinfo {author} {\bibfnamefont {D.}~\bibnamefont
  {Kalligas}}, \bibinfo {author} {\bibfnamefont {P.}~\bibnamefont {Wesson}},\
  and\ \bibinfo {author} {\bibfnamefont {C.~W.~F.}\ \bibnamefont {Everitt}},\
  }\href {https://doi.org/10.1007/BF00760411} {\bibfield  {journal} {\bibinfo
  {journal} {Gen. Relativ. Gravit.}\ }\textbf {\bibinfo {volume} {24}},\
  \bibinfo {pages} {351} (\bibinfo {year} {1992})}\BibitemShut {NoStop}%
\bibitem [{\citenamefont {Lopez}\ and\ \citenamefont
  {Nanopoulos}(1996)}]{Lopez1996}%
  \BibitemOpen
  \bibfield  {author} {\bibinfo {author} {\bibfnamefont {J.~L.}\ \bibnamefont
  {Lopez}}\ and\ \bibinfo {author} {\bibfnamefont {D.}~\bibnamefont
  {Nanopoulos}},\ }\href {https://doi.org/10.1142/S0217732396000023} {\bibfield
   {journal} {\bibinfo  {journal} {Mod. Phys. Lett. A}\ }\textbf {\bibinfo
  {volume} {11}},\ \bibinfo {pages} {1} (\bibinfo {year} {1996})}\BibitemShut
  {NoStop}%
\bibitem [{\citenamefont {Arbab}(1997)}]{Arbab1997}%
  \BibitemOpen
  \bibfield  {author} {\bibinfo {author} {\bibfnamefont {A.~I.}\ \bibnamefont
  {Arbab}},\ }\href {https://doi.org/10.1023/A:1010252130608} {\bibfield
  {journal} {\bibinfo  {journal} {Gen. Relativ. Gravit.}\ }\textbf {\bibinfo
  {volume} {29}},\ \bibinfo {pages} {61} (\bibinfo {year} {1997})}\BibitemShut
  {NoStop}%
\bibitem [{\citenamefont {Fujii}\ and\ \citenamefont
  {Nishioka}(1990)}]{Fujii1990}%
  \BibitemOpen
  \bibfield  {author} {\bibinfo {author} {\bibfnamefont {Y.}~\bibnamefont
  {Fujii}}\ and\ \bibinfo {author} {\bibfnamefont {T.}~\bibnamefont
  {Nishioka}},\ }\href {https://doi.org/10.1103/PhysRevD.42.361} {\bibfield
  {journal} {\bibinfo  {journal} {Phys. Rev. D}\ }\textbf {\bibinfo {volume}
  {42}},\ \bibinfo {pages} {361} (\bibinfo {year} {1990})}\BibitemShut
  {NoStop}%
\bibitem [{\citenamefont {Beesham}(1994)}]{Beesham1994}%
  \BibitemOpen
  \bibfield  {author} {\bibinfo {author} {\bibfnamefont {A.}~\bibnamefont
  {Beesham}},\ }\href {https://doi.org/10.1007/BF02105151} {\bibfield
  {journal} {\bibinfo  {journal} {Gen. Relativ. Gravit.}\ }\textbf {\bibinfo
  {volume} {26}},\ \bibinfo {pages} {159} (\bibinfo {year} {1994})}\BibitemShut
  {NoStop}%
\bibitem [{\citenamefont {Berman}\ and\ \citenamefont
  {Som}(1990)}]{Berman1990}%
  \BibitemOpen
  \bibfield  {author} {\bibinfo {author} {\bibfnamefont {M.~S.}\ \bibnamefont
  {Berman}}\ and\ \bibinfo {author} {\bibfnamefont {M.~M.}\ \bibnamefont
  {Som}},\ }\href {https://doi.org/10.1007/BF00674120} {\bibfield  {journal}
  {\bibinfo  {journal} {Int. J. Theor. Phys.}\ }\textbf {\bibinfo {volume}
  {29}},\ \bibinfo {pages} {1411} (\bibinfo {year} {1990})}\BibitemShut
  {NoStop}%
\bibitem [{\citenamefont {End\={o}}\ and\ \citenamefont
  {Fukui}(1977)}]{Endo1977}%
  \BibitemOpen
  \bibfield  {author} {\bibinfo {author} {\bibfnamefont {M.}~\bibnamefont
  {End\={o}}}\ and\ \bibinfo {author} {\bibfnamefont {T.}~\bibnamefont
  {Fukui}},\ }\href {https://doi.org/10.1007/BF00759587} {\bibfield  {journal}
  {\bibinfo  {journal} {Gen. Relativ. Gravit.}\ }\textbf {\bibinfo {volume}
  {8}},\ \bibinfo {pages} {833} (\bibinfo {year} {1977})}\BibitemShut {NoStop}%
\bibitem [{\citenamefont {{Ph. Spindel}}\ and\ \citenamefont
  {Brout}(1994)}]{Spindel1994}%
  \BibitemOpen
  \bibfield  {author} {\bibinfo {author} {\bibnamefont {{Ph. Spindel}}}\ and\
  \bibinfo {author} {\bibfnamefont {R.}~\bibnamefont {Brout}},\ }\href
  {https://doi.org/10.1016/0370-2693(94)90651-3} {\bibfield  {journal}
  {\bibinfo  {journal} {Phys. Lett. B}\ }\textbf {\bibinfo {volume} {320}},\
  \bibinfo {pages} {241} (\bibinfo {year} {1994})}\BibitemShut {NoStop}%
\bibitem [{\citenamefont {Dussattar}\ and\ \citenamefont
  {Vishwakarma}(1996)}]{Dussattar1996}%
  \BibitemOpen
  \bibfield  {author} {\bibinfo {author} {\bibfnamefont {A.~B.}\ \bibnamefont
  {Dussattar}}\ and\ \bibinfo {author} {\bibfnamefont {R.~G.}\ \bibnamefont
  {Vishwakarma}},\ }\href {https://doi.org/10.1007/BF02847165} {\bibfield
  {journal} {\bibinfo  {journal} {Pramana}\ }\textbf {\bibinfo {volume} {47}},\
  \bibinfo {pages} {41} (\bibinfo {year} {1996})}\BibitemShut {NoStop}%
\bibitem [{\citenamefont {John}\ and\ \citenamefont {{Babu
  Joseph}}(1997)}]{John1997}%
  \BibitemOpen
  \bibfield  {author} {\bibinfo {author} {\bibfnamefont {M.~V.}\ \bibnamefont
  {John}}\ and\ \bibinfo {author} {\bibfnamefont {K.}~\bibnamefont {{Babu
  Joseph}}},\ }\href {https://doi.org/10.1088/0264-9381/14/5/016} {\bibfield
  {journal} {\bibinfo  {journal} {Class. Quant. Grav.}\ }\textbf {\bibinfo
  {volume} {14}},\ \bibinfo {pages} {1115} (\bibinfo {year}
  {1997})}\BibitemShut {NoStop}%
\bibitem [{\citenamefont {Carvalho}\ \emph {et~al.}(1992)\citenamefont
  {Carvalho}, \citenamefont {Lima},\ and\ \citenamefont {Waga}}]{Carvalho1992}%
  \BibitemOpen
  \bibfield  {author} {\bibinfo {author} {\bibfnamefont {J.~C.}\ \bibnamefont
  {Carvalho}}, \bibinfo {author} {\bibfnamefont {J.~A.~S.}\ \bibnamefont
  {Lima}},\ and\ \bibinfo {author} {\bibfnamefont {I.}~\bibnamefont {Waga}},\
  }\href {https://doi.org/10.1103/PhysRevD.46.2404} {\bibfield  {journal}
  {\bibinfo  {journal} {Phys. Rev. D}\ }\textbf {\bibinfo {volume} {46}},\
  \bibinfo {pages} {2404} (\bibinfo {year} {1992})}\BibitemShut {NoStop}%
\bibitem [{\citenamefont {Waga}(1993)}]{Waga1993}%
  \BibitemOpen
  \bibfield  {author} {\bibinfo {author} {\bibfnamefont {I.}~\bibnamefont
  {Waga}},\ }\href
  {http://articles.adsabs.harvard.edu/cgi-bin/nph-iarticle_query?1993ApJ...414..436W&data_type=PDF_HIGH&whole_paper=YES&type=PRINTER&filetype=.pdf}
  {\bibfield  {journal} {\bibinfo  {journal} {Astrophys. J.}\ }\textbf
  {\bibinfo {volume} {414}},\ \bibinfo {pages} {436} (\bibinfo {year}
  {1993})}\BibitemShut {NoStop}%
\bibitem [{\citenamefont {Sister\'{o}}(1991)}]{Sistero1991}%
  \BibitemOpen
  \bibfield  {author} {\bibinfo {author} {\bibfnamefont {R.~F.}\ \bibnamefont
  {Sister\'{o}}},\ }\href {https://doi.org/10.1007/BF00756848} {\bibfield
  {journal} {\bibinfo  {journal} {Gen. Relativ. Gravit.}\ }\textbf {\bibinfo
  {volume} {23}},\ \bibinfo {pages} {1265} (\bibinfo {year}
  {1991})}\BibitemShut {NoStop}%
\bibitem [{\citenamefont {Overduin}\ and\ \citenamefont
  {Cooperstock}(1998)}]{Overduin1998}%
  \BibitemOpen
  \bibfield  {author} {\bibinfo {author} {\bibfnamefont {J.~M.}\ \bibnamefont
  {Overduin}}\ and\ \bibinfo {author} {\bibfnamefont {F.~I.}\ \bibnamefont
  {Cooperstock}},\ }\href {https://doi.org/10.1103/PhysRevD.58.043506}
  {\bibfield  {journal} {\bibinfo  {journal} {Phys. Rev. D}\ }\textbf {\bibinfo
  {volume} {58}},\ \bibinfo {pages} {043506} (\bibinfo {year}
  {1998})}\BibitemShut {NoStop}%
\bibitem [{\citenamefont {Arbab}\ and\ \citenamefont
  {Abdel-Rahman}(1994)}]{Arbab1994}%
  \BibitemOpen
  \bibfield  {author} {\bibinfo {author} {\bibfnamefont {A.~I.}\ \bibnamefont
  {Arbab}}\ and\ \bibinfo {author} {\bibfnamefont {A.-M.~M.}\ \bibnamefont
  {Abdel-Rahman}},\ }\href {https://doi.org/10.1103/PhysRevD.50.7725}
  {\bibfield  {journal} {\bibinfo  {journal} {Phys. Rev. D}\ }\textbf {\bibinfo
  {volume} {50}},\ \bibinfo {pages} {7725} (\bibinfo {year}
  {1994})}\BibitemShut {NoStop}%
\bibitem [{\citenamefont {Abdel-Rahman}(1992)}]{AbdelRahman1992}%
  \BibitemOpen
  \bibfield  {author} {\bibinfo {author} {\bibfnamefont {A.-M.~M.}\
  \bibnamefont {Abdel-Rahman}},\ }\href
  {https://doi.org/10.1103/PhysRevD.45.3497} {\bibfield  {journal} {\bibinfo
  {journal} {Phys. Rev. D}\ }\textbf {\bibinfo {volume} {45}},\ \bibinfo
  {pages} {3497} (\bibinfo {year} {1992})}\BibitemShut {NoStop}%
\bibitem [{\citenamefont {M\'{e}ndez}\ and\ \citenamefont
  {Pav\'{o}n}(1996)}]{Mendez1996}%
  \BibitemOpen
  \bibfield  {author} {\bibinfo {author} {\bibfnamefont {V.}~\bibnamefont
  {M\'{e}ndez}}\ and\ \bibinfo {author} {\bibfnamefont {D.}~\bibnamefont
  {Pav\'{o}n}},\ }\href {https://doi.org/10.1007/BF02104834} {\bibfield
  {journal} {\bibinfo  {journal} {Gen. Relativ. Gravit.}\ }\textbf {\bibinfo
  {volume} {28}},\ \bibinfo {pages} {679} (\bibinfo {year} {1996})}\BibitemShut
  {NoStop}%
\bibitem [{\citenamefont {Calvao}\ \emph {et~al.}(1992)\citenamefont {Calvao},
  \citenamefont {de~Oliveira}, \citenamefont {Pav\'{o}n},\ and\ \citenamefont
  {Salim}}]{Calvao1992}%
  \BibitemOpen
  \bibfield  {author} {\bibinfo {author} {\bibfnamefont {M.~O.}\ \bibnamefont
  {Calvao}}, \bibinfo {author} {\bibfnamefont {H.~P.}\ \bibnamefont
  {de~Oliveira}}, \bibinfo {author} {\bibfnamefont {D.}~\bibnamefont
  {Pav\'{o}n}},\ and\ \bibinfo {author} {\bibfnamefont {J.~M.}\ \bibnamefont
  {Salim}},\ }\href {https://doi.org/10.1103/PhysRevD.45.3869} {\bibfield
  {journal} {\bibinfo  {journal} {Phys. Rev. D}\ }\textbf {\bibinfo {volume}
  {45}},\ \bibinfo {pages} {3869} (\bibinfo {year} {1992})}\BibitemShut
  {NoStop}%
\bibitem [{\citenamefont {Silveira}\ and\ \citenamefont
  {Waga}(1997)}]{Silveira1997}%
  \BibitemOpen
  \bibfield  {author} {\bibinfo {author} {\bibfnamefont {V.}~\bibnamefont
  {Silveira}}\ and\ \bibinfo {author} {\bibfnamefont {I.}~\bibnamefont
  {Waga}},\ }\href {https://doi.org/10.1103/PhysRevD.56.4625} {\bibfield
  {journal} {\bibinfo  {journal} {Phys. Rev. D}\ }\textbf {\bibinfo {volume}
  {56}},\ \bibinfo {pages} {4625} (\bibinfo {year} {1997})}\BibitemShut
  {NoStop}%
\bibitem [{\citenamefont {Chen}\ and\ \citenamefont {Wu}(1990)}]{Chen1990}%
  \BibitemOpen
  \bibfield  {author} {\bibinfo {author} {\bibfnamefont {W.}~\bibnamefont
  {Chen}}\ and\ \bibinfo {author} {\bibfnamefont {Y.-S.}\ \bibnamefont {Wu}},\
  }\href {https://doi.org/10.1103/PhysRevD.41.695} {\bibfield  {journal}
  {\bibinfo  {journal} {Phys. Rev. D}\ }\textbf {\bibinfo {volume} {41}},\
  \bibinfo {pages} {695} (\bibinfo {year} {1990})}\BibitemShut {NoStop}%
\bibitem [{\citenamefont {\"{O}zer}\ and\ \citenamefont
  {Taha}(1986)}]{Ozer1986}%
  \BibitemOpen
  \bibfield  {author} {\bibinfo {author} {\bibfnamefont {M.}~\bibnamefont
  {\"{O}zer}}\ and\ \bibinfo {author} {\bibfnamefont {M.}~\bibnamefont
  {Taha}},\ }\href {https://doi.org/10.1016/0370-2693(86)91421-8} {\bibfield
  {journal} {\bibinfo  {journal} {Phys. Lett. B}\ }\textbf {\bibinfo {volume}
  {171}},\ \bibinfo {pages} {363} (\bibinfo {year} {1986})}\BibitemShut
  {NoStop}%
\bibitem [{\citenamefont {Freese}\ \emph {et~al.}(1987)\citenamefont {Freese},
  \citenamefont {Adams}, \citenamefont {Frieman},\ and\ \citenamefont
  {Mottola}}]{Freese1987}%
  \BibitemOpen
  \bibfield  {author} {\bibinfo {author} {\bibfnamefont {K.}~\bibnamefont
  {Freese}}, \bibinfo {author} {\bibfnamefont {F.~C.}\ \bibnamefont {Adams}},
  \bibinfo {author} {\bibfnamefont {J.~A.}\ \bibnamefont {Frieman}},\ and\
  \bibinfo {author} {\bibfnamefont {E.}~\bibnamefont {Mottola}},\ }\href
  {https://doi.org/10.1016/0550-3213(87)90129-5} {\bibfield  {journal}
  {\bibinfo  {journal} {Nucl. Phys. B}\ }\textbf {\bibinfo {volume} {287}},\
  \bibinfo {pages} {797} (\bibinfo {year} {1987})}\BibitemShut {NoStop}%
\bibitem [{\citenamefont {Wang}\ and\ \citenamefont {Meng}(2005)}]{Wang2005}%
  \BibitemOpen
  \bibfield  {author} {\bibinfo {author} {\bibfnamefont {P.}~\bibnamefont
  {Wang}}\ and\ \bibinfo {author} {\bibfnamefont {X.-H.}\ \bibnamefont
  {Meng}},\ }\href {https://doi.org/10.1088/0264-9381/22/2/003} {\bibfield
  {journal} {\bibinfo  {journal} {Class. Quant. Grav.}\ }\textbf {\bibinfo
  {volume} {22}},\ \bibinfo {pages} {283} (\bibinfo {year} {2005})}\BibitemShut
  {NoStop}%
\bibitem [{\citenamefont {Lima}\ and\ \citenamefont
  {Carvalho}(1994)}]{Lima1994}%
  \BibitemOpen
  \bibfield  {author} {\bibinfo {author} {\bibfnamefont {J.~A.~S.}\
  \bibnamefont {Lima}}\ and\ \bibinfo {author} {\bibfnamefont {J.~C.}\
  \bibnamefont {Carvalho}},\ }\href {https://doi.org/10.1007/BF02107147}
  {\bibfield  {journal} {\bibinfo  {journal} {Gen. Relativ. Gravit.}\ }\textbf
  {\bibinfo {volume} {26}},\ \bibinfo {pages} {909} (\bibinfo {year}
  {1994})}\BibitemShut {NoStop}%
\bibitem [{\citenamefont {Wetterich}(1995)}]{Wetterich1995}%
  \BibitemOpen
  \bibfield  {author} {\bibinfo {author} {\bibfnamefont {C.}~\bibnamefont
  {Wetterich}},\ }\href
  {http://articles.adsabs.harvard.edu/cgi-bin/nph-iarticle_query?1995A\%26A...301..321W&data_type=PDF_HIGH&whole_paper=YES&type=PRINTER&filetype=.pdf}
  {\bibfield  {journal} {\bibinfo  {journal} {Astron. \& Astrophys.}\ }\textbf
  {\bibinfo {volume} {301}},\ \bibinfo {pages} {321} (\bibinfo {year}
  {1995})}\BibitemShut {NoStop}%
\bibitem [{\citenamefont {Ray}\ \emph {et~al.}(2007)\citenamefont {Ray},
  \citenamefont {Mukhopadhyay},\ and\ \citenamefont {Meng}}]{Ray2007}%
  \BibitemOpen
  \bibfield  {author} {\bibinfo {author} {\bibfnamefont {S.}~\bibnamefont
  {Ray}}, \bibinfo {author} {\bibfnamefont {U.}~\bibnamefont {Mukhopadhyay}},\
  and\ \bibinfo {author} {\bibfnamefont {X.-H.}\ \bibnamefont {Meng}},\
  }\href@noop {} {\bibfield  {journal} {\bibinfo  {journal} {Gravit. Cosmol.}\
  }\textbf {\bibinfo {volume} {13}},\ \bibinfo {pages} {142} (\bibinfo {year}
  {2007})}\BibitemShut {NoStop}%
\bibitem [{\citenamefont {Ray}\ \emph {et~al.}(2011)\citenamefont {Ray},
  \citenamefont {Rahaman}, \citenamefont {Mukhopadhyay},\ and\ \citenamefont
  {Sarkar}}]{Ray2011}%
  \BibitemOpen
  \bibfield  {author} {\bibinfo {author} {\bibfnamefont {S.}~\bibnamefont
  {Ray}}, \bibinfo {author} {\bibfnamefont {F.}~\bibnamefont {Rahaman}},
  \bibinfo {author} {\bibfnamefont {U.}~\bibnamefont {Mukhopadhyay}},\ and\
  \bibinfo {author} {\bibfnamefont {R.}~\bibnamefont {Sarkar}},\ }\href
  {https://doi.org/10.1007/s10773-011-0766-2} {\bibfield  {journal} {\bibinfo
  {journal} {Int. J. Theor. Phys.}\ }\textbf {\bibinfo {volume} {50}},\
  \bibinfo {pages} {2687} (\bibinfo {year} {2011})}\BibitemShut {NoStop}%
\bibitem [{\citenamefont {Lima}\ and\ \citenamefont {Maia}(1994)}]{Maia1994}%
  \BibitemOpen
  \bibfield  {author} {\bibinfo {author} {\bibfnamefont {J.~A.~S.}\
  \bibnamefont {Lima}}\ and\ \bibinfo {author} {\bibfnamefont {J.~M.~F.}\
  \bibnamefont {Maia}},\ }\href {https://doi.org/10.1103/PhysRevD.49.5597}
  {\bibfield  {journal} {\bibinfo  {journal} {Phys. Rev. D}\ }\textbf {\bibinfo
  {volume} {49}},\ \bibinfo {pages} {5597} (\bibinfo {year}
  {1994})}\BibitemShut {NoStop}%
\bibitem [{\citenamefont {Easson}\ \emph {et~al.}(2011)\citenamefont {Easson},
  \citenamefont {Frampton},\ and\ \citenamefont {Smoot}}]{Easson2011}%
  \BibitemOpen
  \bibfield  {author} {\bibinfo {author} {\bibfnamefont {D.~A.}\ \bibnamefont
  {Easson}}, \bibinfo {author} {\bibfnamefont {P.~H.}\ \bibnamefont
  {Frampton}},\ and\ \bibinfo {author} {\bibfnamefont {G.~F.}\ \bibnamefont
  {Smoot}},\ }\href {https://doi.org/10.1016/J.PHYSLETB.2010.12.025} {\bibfield
   {journal} {\bibinfo  {journal} {Phys. Lett. B}\ }\textbf {\bibinfo {volume}
  {696}},\ \bibinfo {pages} {273} (\bibinfo {year} {2011})}\BibitemShut
  {NoStop}%
\bibitem [{\citenamefont {Basilakos}\ and\ \citenamefont
  {Sol\`{a}}(2014)}]{Basilakos2014}%
  \BibitemOpen
  \bibfield  {author} {\bibinfo {author} {\bibfnamefont {S.}~\bibnamefont
  {Basilakos}}\ and\ \bibinfo {author} {\bibfnamefont {J.}~\bibnamefont
  {Sol\`{a}}},\ }\href {https://doi.org/10.1103/PhysRevD.90.023008} {\bibfield
  {journal} {\bibinfo  {journal} {Phys. Rev. D}\ }\textbf {\bibinfo {volume}
  {90}},\ \bibinfo {pages} {023008} (\bibinfo {year} {2014})}\BibitemShut
  {NoStop}%
\bibitem [{\citenamefont {Sol\`{a}}(2011)}]{Sola2011}%
  \BibitemOpen
  \bibfield  {author} {\bibinfo {author} {\bibfnamefont {J.}~\bibnamefont
  {Sol\`{a}}},\ }\href {https://doi.org/10.1088/1742-6596/283/1/012033}
  {\bibfield  {journal} {\bibinfo  {journal} {J. Phys. Conf. Ser.}\ }\textbf
  {\bibinfo {volume} {283}},\ \bibinfo {pages} {012033} (\bibinfo {year}
  {2011})}\BibitemShut {NoStop}%
\bibitem [{\citenamefont {{I.~L. Shapiro}}(2008)}]{Shapiro}%
  \BibitemOpen
  \bibfield  {author} {\bibinfo {author} {\bibnamefont {{I.~L. Shapiro}}},\
  }\href@noop {} {\bibfield  {journal} {\bibinfo  {journal} {Class. Quant.
  Grav.}\ }\textbf {\bibinfo {volume} {25}},\ \bibinfo {pages} {103001}
  (\bibinfo {year} {2008})}\BibitemShut {NoStop}%
\bibitem [{\citenamefont {{Shapiro}}\ \emph {et~al.}()\citenamefont
  {{Shapiro}}, \citenamefont {Sol\`{a}},\ and\ \citenamefont
  {\u{S}tefan\u{c}i\'{c}}}]{Shapiro2005}%
  \BibitemOpen
  \bibfield  {author} {\bibinfo {author} {\bibfnamefont {I.~L.}\ \bibnamefont
  {{Shapiro}}}, \bibinfo {author} {\bibfnamefont {J.}~\bibnamefont
  {Sol\`{a}}},\ and\ \bibinfo {author} {\bibfnamefont {H.}~\bibnamefont
  {\u{S}tefan\u{c}i\'{c}}},\ }\href@noop {} {\bibfield  {journal} {\bibinfo
  {journal} {J. Cosmol. Astropart. Phys.}\ }\textbf {\bibinfo {volume}
  {2005}}\bibinfo  {number} { (01)},\ \bibinfo {pages} {012}}\BibitemShut
  {NoStop}%
\bibitem [{\citenamefont {Sol\`{a}}\ \emph {et~al.}(2015)\citenamefont
  {Sol\`{a}}, \citenamefont {G\'{o}mez-Valent},\ and\ \citenamefont {{de Cruz
  P\'{e}rez}}}]{Sola2015}%
  \BibitemOpen
\bibfield  {number} {  }\bibfield  {author} {\bibinfo {author} {\bibfnamefont
  {J.}~\bibnamefont {Sol\`{a}}}, \bibinfo {author} {\bibfnamefont
  {A.}~\bibnamefont {G\'{o}mez-Valent}},\ and\ \bibinfo {author} {\bibfnamefont
  {J.}~\bibnamefont {{de Cruz P\'{e}rez}}},\ }\href@noop {} {\bibfield
  {journal} {\bibinfo  {journal} {Astrophys. J. Lett.}\ }\textbf {\bibinfo
  {volume} {811}},\ \bibinfo {pages} {L14} (\bibinfo {year}
  {2015})}\BibitemShut {NoStop}%
\bibitem [{\citenamefont {Carroll}(2004)}]{Carroll2004}%
  \BibitemOpen
  \bibfield  {author} {\bibinfo {author} {\bibfnamefont {S.}~\bibnamefont
  {Carroll}},\ }\href@noop {} {\emph {\bibinfo {title} {{Spacetime and
  Geometry: An Introduction to General Relativity}}}}\ (\bibinfo  {publisher}
  {Addison Wesley},\ \bibinfo {address} {San Francisco},\ \bibinfo {year}
  {2004})\BibitemShut {NoStop}%
\bibitem [{\citenamefont {Lima}\ \emph {et~al.}(2015)\citenamefont {Lima},
  \citenamefont {Perico},\ and\ \citenamefont {Zilioti}}]{Lima2015}%
  \BibitemOpen
  \bibfield  {author} {\bibinfo {author} {\bibfnamefont {J.~A.~S.}\
  \bibnamefont {Lima}}, \bibinfo {author} {\bibfnamefont {E.~L.~D.}\
  \bibnamefont {Perico}},\ and\ \bibinfo {author} {\bibfnamefont {G.~J.~M.}\
  \bibnamefont {Zilioti}},\ }\href {https://doi.org/10.1142/S0218271815410060}
  {\bibfield  {journal} {\bibinfo  {journal} {Int. J. Mod. Phys. D}\ }\textbf
  {\bibinfo {volume} {24}},\ \bibinfo {pages} {1541006} (\bibinfo {year}
  {2015})}\BibitemShut {NoStop}%
\bibitem [{\citenamefont {Blas}\ \emph {et~al.}()\citenamefont {Blas},
  \citenamefont {Lesgourgues},\ and\ \citenamefont {Tram}}]{Blas2011}%
  \BibitemOpen
  \bibfield  {author} {\bibinfo {author} {\bibfnamefont {D.}~\bibnamefont
  {Blas}}, \bibinfo {author} {\bibfnamefont {J.}~\bibnamefont {Lesgourgues}},\
  and\ \bibinfo {author} {\bibfnamefont {T.}~\bibnamefont {Tram}},\ }\href
  {https://doi.org/10.1088/1475-7516/2011/07/034} {\bibfield  {journal}
  {\bibinfo  {journal} {J. Cosmol. Astropart. Phys.}\ }\textbf {\bibinfo
  {volume} {2011}}\bibinfo  {number} { (07)},\ \bibinfo {pages}
  {034}}\BibitemShut {NoStop}%
\bibitem [{\citenamefont {Audren}\ \emph {et~al.}()\citenamefont {Audren},
  \citenamefont {Lesgourgues}, \citenamefont {Benabed},\ and\ \citenamefont
  {Prunet}}]{Audren2013}%
  \BibitemOpen
\bibfield  {number} {  }\bibfield  {author} {\bibinfo {author} {\bibfnamefont
  {B.}~\bibnamefont {Audren}}, \bibinfo {author} {\bibfnamefont
  {J.}~\bibnamefont {Lesgourgues}}, \bibinfo {author} {\bibfnamefont
  {K.}~\bibnamefont {Benabed}},\ and\ \bibinfo {author} {\bibfnamefont
  {S.}~\bibnamefont {Prunet}},\ }\href
  {https://doi.org/10.1088/1475-7516/2013/02/001} {\bibfield  {journal}
  {\bibinfo  {journal} {J. Cosmol. Astropart. Phys.}\ }\textbf {\bibinfo
  {volume} {2013}}\bibinfo  {number} { (02)},\ \bibinfo {pages}
  {001}}\BibitemShut {NoStop}%
\bibitem [{\citenamefont {Lewis}(2015)}]{Lewis2015}%
  \BibitemOpen
\bibfield  {number} {  }\bibfield  {author} {\bibinfo {author} {\bibfnamefont
  {A.}~\bibnamefont {Lewis}},\ }\href@noop {} {\bibinfo {title} {{GetDist}}},\
  \bibinfo {howpublished}
  {\url{https://getdist.readthedocs.io/en/latest/intro.html}} (\bibinfo {year}
  {2015}),\ \bibinfo {note} {{date accessed: 11 Oct. 2018}}\BibitemShut
  {NoStop}%
\bibitem [{\citenamefont {Katz}\ \emph {et~al.}()\citenamefont {Katz},
  \citenamefont {Dong},\ and\ \citenamefont {Kushnir}}]{Katz}%
  \BibitemOpen
  \bibfield  {author} {\bibinfo {author} {\bibfnamefont {B.}~\bibnamefont
  {Katz}}, \bibinfo {author} {\bibfnamefont {S.}~\bibnamefont {Dong}},\ and\
  \bibinfo {author} {\bibfnamefont {D.}~\bibnamefont {Kushnir}},\ }\href@noop
  {} {\bibinfo {title} {{What Causes Supernovae Explosions?}}},\ \bibinfo
  {howpublished}
  {\url{www.ias.edu/ideas/2013/katz-dong-kushnir-1a-supernovae}},\ \bibinfo
  {note} {{Published in \emph{The Institute Letter -- Summer 2013} by the
  Institute for Advanced Study}}\BibitemShut {NoStop}%
\bibitem [{\citenamefont {{Carroll}}(2001)}]{Carroll}%
  \BibitemOpen
  \bibfield  {author} {\bibinfo {author} {\bibfnamefont {S.~M.}\ \bibnamefont
  {{Carroll}}},\ }\href@noop {} {\bibfield  {journal} {\bibinfo  {journal}
  {Living Rev. Relativ.}\ }\textbf {\bibinfo {volume} {4}},\ \bibinfo {pages}
  {1} (\bibinfo {year} {2001})},\ \bibinfo {note}
  {\url{www.livingreviews.org/lrr-2001-1}}\BibitemShut {NoStop}%
\bibitem [{\citenamefont {{M. Betoule \emph{et al.}}}(2014)}]{Betoule2014}%
  \BibitemOpen
  \bibfield  {author} {\bibinfo {author} {\bibnamefont {{M. Betoule \emph{et
  al.}}}},\ }\href {https://doi.org/10.1051/0004-6361/201423413} {\bibfield
  {journal} {\bibinfo  {journal} {Astron. \& Astrophys.}\ }\textbf {\bibinfo
  {volume} {568}},\ \bibinfo {pages} {A22} (\bibinfo {year}
  {2014})}\BibitemShut {NoStop}%
\bibitem [{\citenamefont {Zou}\ \emph {et~al.}(2018)\citenamefont {Zou},
  \citenamefont {Deng}, \citenamefont {Yin},\ and\ \citenamefont
  {Wei}}]{Zou2018}%
  \BibitemOpen
  \bibfield  {author} {\bibinfo {author} {\bibfnamefont {X.-B.}\ \bibnamefont
  {Zou}}, \bibinfo {author} {\bibfnamefont {H.-K.}\ \bibnamefont {Deng}},
  \bibinfo {author} {\bibfnamefont {Z.-Y.}\ \bibnamefont {Yin}},\ and\ \bibinfo
  {author} {\bibfnamefont {H.}~\bibnamefont {Wei}},\ }\href
  {https://doi.org/10.1016/J.PHYSLETB.2017.11.053} {\bibfield  {journal}
  {\bibinfo  {journal} {Phys. Lett. B}\ }\textbf {\bibinfo {volume} {776}},\
  \bibinfo {pages} {284} (\bibinfo {year} {2018})}\BibitemShut {NoStop}%
\bibitem [{\citenamefont {Zhang}\ \emph {et~al.}(2014)\citenamefont {Zhang},
  \citenamefont {Zhang}, \citenamefont {Yuan}, \citenamefont {Liu},
  \citenamefont {Zhang},\ and\ \citenamefont {Sun}}]{Zhang2014}%
  \BibitemOpen
  \bibfield  {author} {\bibinfo {author} {\bibfnamefont {C.}~\bibnamefont
  {Zhang}}, \bibinfo {author} {\bibfnamefont {H.}~\bibnamefont {Zhang}},
  \bibinfo {author} {\bibfnamefont {S.}~\bibnamefont {Yuan}}, \bibinfo {author}
  {\bibfnamefont {S.}~\bibnamefont {Liu}}, \bibinfo {author} {\bibfnamefont
  {T.-J.}\ \bibnamefont {Zhang}},\ and\ \bibinfo {author} {\bibfnamefont
  {Y.-C.}\ \bibnamefont {Sun}},\ }\href
  {https://doi.org/10.1088/1674-4527/14/10/002} {\bibfield  {journal} {\bibinfo
   {journal} {Res. Astron. Astrophys.}\ }\textbf {\bibinfo {volume} {14}},\
  \bibinfo {pages} {1221} (\bibinfo {year} {2014})}\BibitemShut {NoStop}%
\bibitem [{\citenamefont {Simon}\ \emph {et~al.}(2005)\citenamefont {Simon},
  \citenamefont {Verde},\ and\ \citenamefont {Jimenez}}]{Simon2005}%
  \BibitemOpen
  \bibfield  {author} {\bibinfo {author} {\bibfnamefont {J.}~\bibnamefont
  {Simon}}, \bibinfo {author} {\bibfnamefont {L.}~\bibnamefont {Verde}},\ and\
  \bibinfo {author} {\bibfnamefont {R.}~\bibnamefont {Jimenez}},\ }\href
  {https://doi.org/10.1103/PhysRevD.71.123001} {\bibfield  {journal} {\bibinfo
  {journal} {Phys. Rev. D}\ }\textbf {\bibinfo {volume} {71}},\ \bibinfo
  {pages} {123001} (\bibinfo {year} {2005})}\BibitemShut {NoStop}%
\bibitem [{\citenamefont {{M. Moresco \emph{et al.}}}()}]{Moresco2012}%
  \BibitemOpen
  \bibfield  {author} {\bibinfo {author} {\bibnamefont {{M. Moresco \emph{et
  al.}}}},\ }\href {https://doi.org/10.1088/1475-7516/2012/08/006} {\bibfield
  {journal} {\bibinfo  {journal} {J. Cosmol. Astropart. Phys.}\ }\textbf
  {\bibinfo {volume} {2012}}\bibinfo  {number} { (08)},\ \bibinfo {pages}
  {006}}\BibitemShut {NoStop}%
\bibitem [{\citenamefont {Moresco}\ \emph {et~al.}({\natexlab{b}})\citenamefont
  {Moresco}, \citenamefont {Pozzetti}, \citenamefont {Cimatti}, \citenamefont
  {Jimenez}, \citenamefont {Maraston}, \citenamefont {Verde}, \citenamefont
  {Thomas}, \citenamefont {Citro}, \citenamefont {Tojeiro},\ and\ \citenamefont
  {Wilkinson}}]{Moresco2016}%
  \BibitemOpen
\bibfield  {number} {  }\bibfield  {author} {\bibinfo {author} {\bibfnamefont
  {M.}~\bibnamefont {Moresco}}, \bibinfo {author} {\bibfnamefont
  {L.}~\bibnamefont {Pozzetti}}, \bibinfo {author} {\bibfnamefont
  {A.}~\bibnamefont {Cimatti}}, \bibinfo {author} {\bibfnamefont
  {R.}~\bibnamefont {Jimenez}}, \bibinfo {author} {\bibfnamefont
  {C.}~\bibnamefont {Maraston}}, \bibinfo {author} {\bibfnamefont
  {L.}~\bibnamefont {Verde}}, \bibinfo {author} {\bibfnamefont
  {D.}~\bibnamefont {Thomas}}, \bibinfo {author} {\bibfnamefont
  {A.}~\bibnamefont {Citro}}, \bibinfo {author} {\bibfnamefont
  {R.}~\bibnamefont {Tojeiro}},\ and\ \bibinfo {author} {\bibfnamefont
  {D.}~\bibnamefont {Wilkinson}},\ }\href
  {https://doi.org/10.1088/1475-7516/2016/05/014} {\bibfield  {journal}
  {\bibinfo  {journal} {J. Cosmol. Astropart. Phys.}\ }\textbf {\bibinfo
  {volume} {2016}}\bibinfo  {number} { (05)},\ \bibinfo {pages}
  {014}}\BibitemShut {NoStop}%
\bibitem [{\citenamefont {Ratsimbazafy}\ \emph {et~al.}(2017)\citenamefont
  {Ratsimbazafy}, \citenamefont {Loubser}, \citenamefont {Crawford},
  \citenamefont {Cress}, \citenamefont {Bassett}, \citenamefont {Nichol},\ and\
  \citenamefont {V{\"{a}}is{\"{a}}nen}}]{Ratsimbazafy2017}%
  \BibitemOpen
\bibfield  {number} {  }\bibfield  {author} {\bibinfo {author} {\bibfnamefont
  {A.~L.}\ \bibnamefont {Ratsimbazafy}}, \bibinfo {author} {\bibfnamefont
  {S.~I.}\ \bibnamefont {Loubser}}, \bibinfo {author} {\bibfnamefont {S.~M.}\
  \bibnamefont {Crawford}}, \bibinfo {author} {\bibfnamefont {C.~M.}\
  \bibnamefont {Cress}}, \bibinfo {author} {\bibfnamefont {B.~A.}\ \bibnamefont
  {Bassett}}, \bibinfo {author} {\bibfnamefont {R.~C.}\ \bibnamefont
  {Nichol}},\ and\ \bibinfo {author} {\bibfnamefont {P.}~\bibnamefont
  {V{\"{a}}is{\"{a}}nen}},\ }\href {https://doi.org/10.1093/mnras/stx301}
  {\bibfield  {journal} {\bibinfo  {journal} {Mon. Not. R. Astron. Soc.}\
  }\textbf {\bibinfo {volume} {467}},\ \bibinfo {pages} {3239} (\bibinfo {year}
  {2017})}\BibitemShut {NoStop}%
\bibitem [{\citenamefont {Stern}\ \emph {et~al.}()\citenamefont {Stern},
  \citenamefont {Jimenez}, \citenamefont {Verde}, \citenamefont
  {Kamionkowski},\ and\ \citenamefont {Stanford}}]{Stern2010}%
  \BibitemOpen
  \bibfield  {author} {\bibinfo {author} {\bibfnamefont {D.}~\bibnamefont
  {Stern}}, \bibinfo {author} {\bibfnamefont {R.}~\bibnamefont {Jimenez}},
  \bibinfo {author} {\bibfnamefont {L.}~\bibnamefont {Verde}}, \bibinfo
  {author} {\bibfnamefont {M.}~\bibnamefont {Kamionkowski}},\ and\ \bibinfo
  {author} {\bibfnamefont {S.~A.}\ \bibnamefont {Stanford}},\ }\href
  {https://doi.org/10.1088/1475-7516/2010/02/008} {\bibfield  {journal}
  {\bibinfo  {journal} {J. Cosmol. Astropart. Phys.}\ }\textbf {\bibinfo
  {volume} {2010}}\bibinfo  {number} { (02)},\ \bibinfo {pages}
  {008}}\BibitemShut {NoStop}%
\bibitem [{\citenamefont {Moresco}(2015)}]{Moresco2015}%
  \BibitemOpen
\bibfield  {number} {  }\bibfield  {author} {\bibinfo {author} {\bibfnamefont
  {M.}~\bibnamefont {Moresco}},\ }\href {https://doi.org/10.1093/mnrasl/slv037}
  {\bibfield  {journal} {\bibinfo  {journal} {Mon. Not. R. Astron. Soc.:
  Lett.}\ }\textbf {\bibinfo {volume} {450}},\ \bibinfo {pages} {L16} (\bibinfo
  {year} {2015})}\BibitemShut {NoStop}%
\bibitem [{\citenamefont {Jimenez}\ and\ \citenamefont
  {Loeb}(2002)}]{Jimenez2002}%
  \BibitemOpen
  \bibfield  {author} {\bibinfo {author} {\bibfnamefont {R.}~\bibnamefont
  {Jimenez}}\ and\ \bibinfo {author} {\bibfnamefont {A.}~\bibnamefont {Loeb}},\
  }\href {https://doi.org/10.1086/340549} {\bibfield  {journal} {\bibinfo
  {journal} {Astrophys. J.}\ }\textbf {\bibinfo {volume} {573}},\ \bibinfo
  {pages} {37} (\bibinfo {year} {2002})}\BibitemShut {NoStop}%
\bibitem [{\citenamefont {Bruzual}\ and\ \citenamefont
  {Charlot}(2003)}]{Bruzual2003}%
  \BibitemOpen
  \bibfield  {author} {\bibinfo {author} {\bibfnamefont {G.}~\bibnamefont
  {Bruzual}}\ and\ \bibinfo {author} {\bibfnamefont {S.}~\bibnamefont
  {Charlot}},\ }\href {https://doi.org/10.1046/j.1365-8711.2003.06897.x}
  {\bibfield  {journal} {\bibinfo  {journal} {Mon. Not. R. Astron. Soc.}\
  }\textbf {\bibinfo {volume} {344}},\ \bibinfo {pages} {1000} (\bibinfo {year}
  {2003})}\BibitemShut {NoStop}%
\bibitem [{\citenamefont {{E. Komatsu \emph{et al.}}}(2009)}]{Komatsu2009}%
  \BibitemOpen
  \bibfield  {author} {\bibinfo {author} {\bibnamefont {{E. Komatsu \emph{et
  al.}}}},\ }\href {https://doi.org/10.1088/0067-0049/180/2/330} {\bibfield
  {journal} {\bibinfo  {journal} {Astrophys. J. Suppl. Ser.}\ }\textbf
  {\bibinfo {volume} {180}},\ \bibinfo {pages} {330} (\bibinfo {year}
  {2009})}\BibitemShut {NoStop}%
\bibitem [{\citenamefont {Huang}\ \emph {et~al.}()\citenamefont {Huang},
  \citenamefont {Wang},\ and\ \citenamefont {Wang}}]{Huang2015}%
  \BibitemOpen
  \bibfield  {author} {\bibinfo {author} {\bibfnamefont {Q.-G.}\ \bibnamefont
  {Huang}}, \bibinfo {author} {\bibfnamefont {K.}~\bibnamefont {Wang}},\ and\
  \bibinfo {author} {\bibfnamefont {S.}~\bibnamefont {Wang}},\ }\href
  {https://doi.org/10.1088/1475-7516/2015/12/022} {\bibfield  {journal}
  {\bibinfo  {journal} {J. Cosmol. Astropart. Phys.}\ }\textbf {\bibinfo
  {volume} {2015}}\bibinfo  {number} { (12)},\ \bibinfo {pages}
  {022}}\BibitemShut {NoStop}%
\bibitem [{\citenamefont {{\'{E}. Aubourg \emph{et al.}}}(2015)}]{Aubourg2015}%
  \BibitemOpen
\bibfield  {number} {  }\bibfield  {author} {\bibinfo {author} {\bibnamefont
  {{\'{E}. Aubourg \emph{et al.}}}} (\bibinfo {collaboration} {BOSS
  Collaboration}),\ }\href {https://doi.org/10.1103/PhysRevD.92.123516}
  {\bibfield  {journal} {\bibinfo  {journal} {Phys. Rev. D}\ }\textbf {\bibinfo
  {volume} {92}},\ \bibinfo {pages} {123516} (\bibinfo {year}
  {2015})}\BibitemShut {NoStop}%
\bibitem [{\citenamefont {Mukherjee}\ \emph {et~al.}(2008)\citenamefont
  {Mukherjee}, \citenamefont {Kunz}, \citenamefont {Parkinson},\ and\
  \citenamefont {Wang}}]{Mukherjee}%
  \BibitemOpen
  \bibfield  {author} {\bibinfo {author} {\bibfnamefont {P.}~\bibnamefont
  {Mukherjee}}, \bibinfo {author} {\bibfnamefont {M.}~\bibnamefont {Kunz}},
  \bibinfo {author} {\bibfnamefont {D.}~\bibnamefont {Parkinson}},\ and\
  \bibinfo {author} {\bibfnamefont {Y.}~\bibnamefont {Wang}},\ }\href@noop {}
  {\bibfield  {journal} {\bibinfo  {journal} {Phys. Rev. D}\ }\textbf {\bibinfo
  {volume} {78}},\ \bibinfo {pages} {083529} (\bibinfo {year}
  {2008})}\BibitemShut {NoStop}%
\bibitem [{\citenamefont {{S. Alam \emph{et al.}}}(2017)}]{Alam2017}%
  \BibitemOpen
  \bibfield  {author} {\bibinfo {author} {\bibnamefont {{S. Alam \emph{et
  al.}}}},\ }\href {https://doi.org/10.1093/mnras/stx721} {\bibfield  {journal}
  {\bibinfo  {journal} {Mon. Not. R. Astron. Soc.}\ }\textbf {\bibinfo {volume}
  {470}},\ \bibinfo {pages} {2617} (\bibinfo {year} {2017})}\BibitemShut
  {NoStop}%
\bibitem [{\citenamefont {Bonvin}(2014)}]{Bonvin2014}%
  \BibitemOpen
  \bibfield  {author} {\bibinfo {author} {\bibfnamefont {C.}~\bibnamefont
  {Bonvin}},\ }\href@noop {} {\bibinfo {title} {{Baryon acoustic
  oscillations}}},\ \bibinfo {howpublished}
  {\url{www.blue-shift.ch/wp-content/uploads/downloads/2014/08/bao.pdf}}
  (\bibinfo {year} {2014}),\ \bibinfo {note} {{date accessed: 29 Sep.
  2018}}\BibitemShut {NoStop}%
\bibitem [{\citenamefont {Beutler}\ \emph {et~al.}(2011)\citenamefont
  {Beutler}, \citenamefont {Blake}, \citenamefont {Colless}, \citenamefont
  {Jones}, \citenamefont {Staveley-Smith}, \citenamefont {Campbell},
  \citenamefont {Parker}, \citenamefont {Saunders},\ and\ \citenamefont
  {Watson}}]{Beutler2011}%
  \BibitemOpen
  \bibfield  {author} {\bibinfo {author} {\bibfnamefont {F.}~\bibnamefont
  {Beutler}}, \bibinfo {author} {\bibfnamefont {C.}~\bibnamefont {Blake}},
  \bibinfo {author} {\bibfnamefont {M.}~\bibnamefont {Colless}}, \bibinfo
  {author} {\bibfnamefont {D.~H.}\ \bibnamefont {Jones}}, \bibinfo {author}
  {\bibfnamefont {L.}~\bibnamefont {Staveley-Smith}}, \bibinfo {author}
  {\bibfnamefont {L.}~\bibnamefont {Campbell}}, \bibinfo {author}
  {\bibfnamefont {Q.}~\bibnamefont {Parker}}, \bibinfo {author} {\bibfnamefont
  {W.}~\bibnamefont {Saunders}},\ and\ \bibinfo {author} {\bibfnamefont
  {F.}~\bibnamefont {Watson}},\ }\href
  {https://doi.org/10.1111/j.1365-2966.2011.19250.x} {\bibfield  {journal}
  {\bibinfo  {journal} {Mon. Not. R. Astron. Soc.}\ }\textbf {\bibinfo {volume}
  {416}},\ \bibinfo {pages} {3017} (\bibinfo {year} {2011})}\BibitemShut
  {NoStop}%
\bibitem [{\citenamefont {Ross}\ \emph {et~al.}(2015)\citenamefont {Ross},
  \citenamefont {Samushia}, \citenamefont {Howlett}, \citenamefont {Percival},
  \citenamefont {Burden},\ and\ \citenamefont {Manera}}]{Ross2015}%
  \BibitemOpen
  \bibfield  {author} {\bibinfo {author} {\bibfnamefont {A.~J.}\ \bibnamefont
  {Ross}}, \bibinfo {author} {\bibfnamefont {L.}~\bibnamefont {Samushia}},
  \bibinfo {author} {\bibfnamefont {C.}~\bibnamefont {Howlett}}, \bibinfo
  {author} {\bibfnamefont {W.~J.}\ \bibnamefont {Percival}}, \bibinfo {author}
  {\bibfnamefont {A.}~\bibnamefont {Burden}},\ and\ \bibinfo {author}
  {\bibfnamefont {M.}~\bibnamefont {Manera}},\ }\href
  {https://doi.org/10.1093/mnras/stv154} {\bibfield  {journal} {\bibinfo
  {journal} {Mon. Not. R. Astron. Soc.}\ }\textbf {\bibinfo {volume} {449}},\
  \bibinfo {pages} {835} (\bibinfo {year} {2015})}\BibitemShut {NoStop}%
\bibitem [{\citenamefont {{M. Ata \emph{et al.}}}(2018)}]{Ata2018}%
  \BibitemOpen
  \bibfield  {author} {\bibinfo {author} {\bibnamefont {{M. Ata \emph{et
  al.}}}},\ }\href {https://doi.org/10.1093/mnras/stx2630} {\bibfield
  {journal} {\bibinfo  {journal} {Mon. Not. R. Astron. Soc.}\ }\textbf
  {\bibinfo {volume} {473}},\ \bibinfo {pages} {4773} (\bibinfo {year}
  {2018})}\BibitemShut {NoStop}%
\bibitem [{\citenamefont {{J. E. Bautista \emph{et
  al.}}}(2017)}]{Bautista2017}%
  \BibitemOpen
  \bibfield  {author} {\bibinfo {author} {\bibnamefont {{J. E. Bautista
  \emph{et al.}}}},\ }\href {https://doi.org/10.1051/0004-6361/201730533}
  {\bibfield  {journal} {\bibinfo  {journal} {Astron. \& Astrophys.}\ }\textbf
  {\bibinfo {volume} {603}},\ \bibinfo {pages} {A12} (\bibinfo {year}
  {2017})}\BibitemShut {NoStop}%
\bibitem [{\citenamefont {{H. du Mas des Bourboux \emph{et
  al.}}}(2017)}]{Bourboux2017}%
  \BibitemOpen
  \bibfield  {author} {\bibinfo {author} {\bibnamefont {{H. du Mas des Bourboux
  \emph{et al.}}}},\ }\href {https://doi.org/10.1051/0004-6361/201731731}
  {\bibfield  {journal} {\bibinfo  {journal} {Astron. \& Astrophys.}\ }\textbf
  {\bibinfo {volume} {608}},\ \bibinfo {pages} {A130} (\bibinfo {year}
  {2017})}\BibitemShut {NoStop}%
\bibitem [{\citenamefont {{Percival}}\ \emph {et~al.}(2011)\citenamefont
  {{Percival}}, \citenamefont {Samushia}, \citenamefont {{Ross}}, \citenamefont
  {Shapiro},\ and\ \citenamefont {Raccanelli}}]{Percival}%
  \BibitemOpen
  \bibfield  {author} {\bibinfo {author} {\bibfnamefont {W.~J.}\ \bibnamefont
  {{Percival}}}, \bibinfo {author} {\bibfnamefont {L.}~\bibnamefont
  {Samushia}}, \bibinfo {author} {\bibfnamefont {A.~J.}\ \bibnamefont
  {{Ross}}}, \bibinfo {author} {\bibfnamefont {C.}~\bibnamefont {Shapiro}},\
  and\ \bibinfo {author} {\bibfnamefont {A.}~\bibnamefont {Raccanelli}},\
  }\href@noop {} {\bibfield  {journal} {\bibinfo  {journal} {Phil. Trans. R.
  Soc. A}\ }\textbf {\bibinfo {volume} {369}},\ \bibinfo {pages} {5058}
  (\bibinfo {year} {2011})}\BibitemShut {NoStop}%
\bibitem [{\citenamefont {Kwan}\ \emph {et~al.}(2012)\citenamefont {Kwan},
  \citenamefont {{Lewis}},\ and\ \citenamefont {{Linder}}}]{Kwan}%
  \BibitemOpen
  \bibfield  {author} {\bibinfo {author} {\bibfnamefont {J.}~\bibnamefont
  {Kwan}}, \bibinfo {author} {\bibfnamefont {G.~F.}\ \bibnamefont {{Lewis}}},\
  and\ \bibinfo {author} {\bibfnamefont {E.~V.}\ \bibnamefont {{Linder}}},\
  }\href@noop {} {\bibfield  {journal} {\bibinfo  {journal} {Astrophys. J.}\
  }\textbf {\bibinfo {volume} {748}},\ \bibinfo {pages} {78} (\bibinfo {year}
  {2012})}\BibitemShut {NoStop}%
\bibitem [{\citenamefont {Sagredo}\ \emph {et~al.}(2018)\citenamefont
  {Sagredo}, \citenamefont {Nesseris},\ and\ \citenamefont {Sapone}}]{Sagredo}%
  \BibitemOpen
  \bibfield  {author} {\bibinfo {author} {\bibfnamefont {B.}~\bibnamefont
  {Sagredo}}, \bibinfo {author} {\bibfnamefont {S.}~\bibnamefont {Nesseris}},\
  and\ \bibinfo {author} {\bibfnamefont {D.}~\bibnamefont {Sapone}},\
  }\href@noop {} {\bibfield  {journal} {\bibinfo  {journal} {Phys. Rev. D}\
  }\textbf {\bibinfo {volume} {98}},\ \bibinfo {pages} {083543} (\bibinfo
  {year} {2018})}\BibitemShut {NoStop}%
\bibitem [{\citenamefont {Mo}\ \emph {et~al.}(2010)\citenamefont {Mo},
  \citenamefont {{van den Bosch}},\ and\ \citenamefont {White}}]{Houjun}%
  \BibitemOpen
  \bibfield  {author} {\bibinfo {author} {\bibfnamefont {H.}~\bibnamefont
  {Mo}}, \bibinfo {author} {\bibfnamefont {F.}~\bibnamefont {{van den
  Bosch}}},\ and\ \bibinfo {author} {\bibfnamefont {S.}~\bibnamefont {White}},\
  }\href@noop {} {\emph {\bibinfo {title} {{Galaxy Formation and Evolution}}}}\
  (\bibinfo  {publisher} {Cambridge University Press},\ \bibinfo {address} {New
  York},\ \bibinfo {year} {2010})\BibitemShut {NoStop}%
\bibitem [{\citenamefont {{G{\'{o}}mez-Valent}}\ and\ \citenamefont
  {{Sol{\`{a}} Peracaula}}(2018)}]{Valent2018}%
  \BibitemOpen
  \bibfield  {author} {\bibinfo {author} {\bibfnamefont {A.}~\bibnamefont
  {{G{\'{o}}mez-Valent}}}\ and\ \bibinfo {author} {\bibfnamefont
  {J.}~\bibnamefont {{Sol{\`{a}} Peracaula}}},\ }\href@noop {} {\bibfield
  {journal} {\bibinfo  {journal} {Mon. Not. R. Astron. Soc.}\ }\textbf
  {\bibinfo {volume} {478}},\ \bibinfo {pages} {126} (\bibinfo {year}
  {2018})}\BibitemShut {NoStop}%
\bibitem [{\citenamefont {Amendola}\ and\ \citenamefont
  {Tsujikawa}(2010)}]{Amendola2010}%
  \BibitemOpen
  \bibfield  {author} {\bibinfo {author} {\bibfnamefont {L.}~\bibnamefont
  {Amendola}}\ and\ \bibinfo {author} {\bibfnamefont {S.}~\bibnamefont
  {Tsujikawa}},\ }\href@noop {} {\emph {\bibinfo {title} {{Dark Energy: Theory
  and Observations}}}}\ (\bibinfo  {publisher} {Cambridge University Press},\
  \bibinfo {address} {Cambridge, UK},\ \bibinfo {year} {2010})\ p.\ \bibinfo
  {pages} {103}\BibitemShut {NoStop}%
\bibitem [{\citenamefont {Abbott}\ and\ \citenamefont
  {Schaefer}(1986)}]{Abbott1986}%
  \BibitemOpen
  \bibfield  {author} {\bibinfo {author} {\bibfnamefont {L.~F.}\ \bibnamefont
  {Abbott}}\ and\ \bibinfo {author} {\bibfnamefont {R.~K.}\ \bibnamefont
  {Schaefer}},\ }\href@noop {} {\bibfield  {journal} {\bibinfo  {journal}
  {Astrophys. J.}\ }\textbf {\bibinfo {volume} {308}},\ \bibinfo {pages} {546}
  (\bibinfo {year} {1986})}\BibitemShut {NoStop}%
\bibitem [{\citenamefont {Chiu}\ \emph {et~al.}(2015)\citenamefont {Chiu},
  \citenamefont {Taylor}, \citenamefont {Shu},\ and\ \citenamefont
  {Tu}}]{Chiu}%
  \BibitemOpen
  \bibfield  {author} {\bibinfo {author} {\bibfnamefont {M.-C.}\ \bibnamefont
  {Chiu}}, \bibinfo {author} {\bibfnamefont {A.}~\bibnamefont {Taylor}},
  \bibinfo {author} {\bibfnamefont {C.}~\bibnamefont {Shu}},\ and\ \bibinfo
  {author} {\bibfnamefont {H.}~\bibnamefont {Tu}},\ }\href@noop {} {\bibfield
  {journal} {\bibinfo  {journal} {Phys. Rev. D}\ }\textbf {\bibinfo {volume}
  {92}},\ \bibinfo {pages} {103514} (\bibinfo {year} {2015})}\BibitemShut
  {NoStop}%
\bibitem [{\citenamefont {Ma}\ and\ \citenamefont
  {Bertschinger}(1995)}]{Bertschinger}%
  \BibitemOpen
  \bibfield  {author} {\bibinfo {author} {\bibfnamefont {C.-P.}\ \bibnamefont
  {Ma}}\ and\ \bibinfo {author} {\bibfnamefont {E.}~\bibnamefont
  {Bertschinger}},\ }\href@noop {} {\bibfield  {journal} {\bibinfo  {journal}
  {Astrophys. J.}\ }\textbf {\bibinfo {volume} {455}},\ \bibinfo {pages} {7}
  (\bibinfo {year} {1995})}\BibitemShut {NoStop}%
\bibitem [{\citenamefont {Mukhanov}\ \emph {et~al.}(1992)\citenamefont
  {Mukhanov}, \citenamefont {Feldman},\ and\ \citenamefont
  {Brandenberger}}]{Mukhanov}%
  \BibitemOpen
  \bibfield  {author} {\bibinfo {author} {\bibfnamefont {V.~F.}\ \bibnamefont
  {Mukhanov}}, \bibinfo {author} {\bibfnamefont {H.~A.}\ \bibnamefont
  {Feldman}},\ and\ \bibinfo {author} {\bibfnamefont {R.~H.}\ \bibnamefont
  {Brandenberger}},\ }\href@noop {} {\bibfield  {journal} {\bibinfo  {journal}
  {Phys. Rep.}\ }\textbf {\bibinfo {volume} {215}},\ \bibinfo {pages} {203}
  (\bibinfo {year} {1992})}\BibitemShut {NoStop}%
\bibitem [{\citenamefont {Eisenstein}\ and\ \citenamefont
  {Hu}(1999)}]{Eisenstein1999}%
  \BibitemOpen
  \bibfield  {author} {\bibinfo {author} {\bibfnamefont {D.~J.}\ \bibnamefont
  {Eisenstein}}\ and\ \bibinfo {author} {\bibfnamefont {W.}~\bibnamefont
  {Hu}},\ }\href@noop {} {\bibfield  {journal} {\bibinfo  {journal} {Astrophys.
  J.}\ }\textbf {\bibinfo {volume} {511}},\ \bibinfo {pages} {5} (\bibinfo
  {year} {1999})}\BibitemShut {NoStop}%
\bibitem [{\citenamefont {{P.~A.~R. Ade \emph{et al.}}}(2016)}]{Aghanim2016}%
  \BibitemOpen
  \bibfield  {author} {\bibinfo {author} {\bibnamefont {{P.~A.~R. Ade \emph{et
  al.}}}} (\bibinfo {collaboration} {Planck Collaboration}),\ }\href@noop {}
  {\bibfield  {journal} {\bibinfo  {journal} {Astron. \& Astrophys.}\ }\textbf
  {\bibinfo {volume} {594}},\ \bibinfo {pages} {A20} (\bibinfo {year}
  {2016})}\BibitemShut {NoStop}%
\bibitem [{\citenamefont {Dodelson}(2003)}]{Dodelson}%
  \BibitemOpen
  \bibfield  {author} {\bibinfo {author} {\bibfnamefont {S.}~\bibnamefont
  {Dodelson}},\ }\href@noop {} {\emph {\bibinfo {title} {{Modern Cosmology}}}}\
  (\bibinfo  {publisher} {Academic Press},\ \bibinfo {address} {San Diego,
  California},\ \bibinfo {year} {2003})\BibitemShut {NoStop}%
\bibitem [{\citenamefont {Eisenstein}\ and\ \citenamefont
  {Hu}(1998)}]{Eisenstein1998}%
  \BibitemOpen
  \bibfield  {author} {\bibinfo {author} {\bibfnamefont {D.~J.}\ \bibnamefont
  {Eisenstein}}\ and\ \bibinfo {author} {\bibfnamefont {W.}~\bibnamefont
  {Hu}},\ }\href {https://doi.org/10.1086/305424} {\bibfield  {journal}
  {\bibinfo  {journal} {Astrophys. J.}\ }\textbf {\bibinfo {volume} {496}},\
  \bibinfo {pages} {605} (\bibinfo {year} {1998})}\BibitemShut {NoStop}%
\bibitem [{\citenamefont {Davis}\ \emph {et~al.}(2011)\citenamefont {Davis},
  \citenamefont {Nusser}, \citenamefont {Masters}, \citenamefont {Springob},
  \citenamefont {Huchra},\ and\ \citenamefont {Lemson}}]{Davis}%
  \BibitemOpen
  \bibfield  {author} {\bibinfo {author} {\bibfnamefont {M.}~\bibnamefont
  {Davis}}, \bibinfo {author} {\bibfnamefont {A.}~\bibnamefont {Nusser}},
  \bibinfo {author} {\bibfnamefont {K.~L.}\ \bibnamefont {Masters}}, \bibinfo
  {author} {\bibfnamefont {C.}~\bibnamefont {Springob}}, \bibinfo {author}
  {\bibfnamefont {J.~P.}\ \bibnamefont {Huchra}},\ and\ \bibinfo {author}
  {\bibfnamefont {G.}~\bibnamefont {Lemson}},\ }\href@noop {} {\bibfield
  {journal} {\bibinfo  {journal} {Mon. Not. R. Astron. Soc.}\ }\textbf
  {\bibinfo {volume} {413}},\ \bibinfo {pages} {2906} (\bibinfo {year}
  {2011})}\BibitemShut {NoStop}%
\bibitem [{\citenamefont {Hudson}\ and\ \citenamefont
  {Turnbull}(2012)}]{Hudson}%
  \BibitemOpen
  \bibfield  {author} {\bibinfo {author} {\bibfnamefont {M.~J.}\ \bibnamefont
  {Hudson}}\ and\ \bibinfo {author} {\bibfnamefont {S.~J.}\ \bibnamefont
  {Turnbull}},\ }\href@noop {} {\bibfield  {journal} {\bibinfo  {journal}
  {Astrophys. J. Lett.}\ }\textbf {\bibinfo {volume} {751}},\ \bibinfo {pages}
  {L30} (\bibinfo {year} {2012})}\BibitemShut {NoStop}%
\bibitem [{\citenamefont {Song}\ and\ \citenamefont {Percival}()}]{Song}%
  \BibitemOpen
  \bibfield  {author} {\bibinfo {author} {\bibfnamefont {Y.-S.}\ \bibnamefont
  {Song}}\ and\ \bibinfo {author} {\bibfnamefont {W.~J.}\ \bibnamefont
  {Percival}},\ }\href@noop {} {\bibfield  {journal} {\bibinfo  {journal} {J.
  Cosmol. Astropart. Phys.}\ }\textbf {\bibinfo {volume} {2009}}\bibinfo
  {number} { (10)},\ \bibinfo {pages} {004}}\BibitemShut {NoStop}%
\bibitem [{\citenamefont {{C. Blake \emph{et al.}}}(2013)}]{Blake}%
  \BibitemOpen
\bibfield  {number} {  }\bibfield  {author} {\bibinfo {author} {\bibnamefont
  {{C. Blake \emph{et al.}}}},\ }\href@noop {} {\bibfield  {journal} {\bibinfo
  {journal} {Mon. Not. R. Astron. Soc.}\ }\textbf {\bibinfo {volume} {436}},\
  \bibinfo {pages} {3089} (\bibinfo {year} {2013})}\BibitemShut {NoStop}%
\bibitem [{\citenamefont {Samushia}\ \emph {et~al.}(2012)\citenamefont
  {Samushia}, \citenamefont {Percival},\ and\ \citenamefont
  {Raccanelli}}]{Samushia}%
  \BibitemOpen
  \bibfield  {author} {\bibinfo {author} {\bibfnamefont {L.}~\bibnamefont
  {Samushia}}, \bibinfo {author} {\bibfnamefont {W.~J.}\ \bibnamefont
  {Percival}},\ and\ \bibinfo {author} {\bibfnamefont {A.}~\bibnamefont
  {Raccanelli}},\ }\href@noop {} {\bibfield  {journal} {\bibinfo  {journal}
  {Mon. Not. R. Astron. Soc.}\ }\textbf {\bibinfo {volume} {420}},\ \bibinfo
  {pages} {2102} (\bibinfo {year} {2012})}\BibitemShut {NoStop}%
\bibitem [{\citenamefont {{C. Blake \emph{et al.}}}(2012)}]{Blake2012}%
  \BibitemOpen
  \bibfield  {author} {\bibinfo {author} {\bibnamefont {{C. Blake \emph{et
  al.}}}},\ }\href@noop {} {\bibfield  {journal} {\bibinfo  {journal} {Mon.
  Not. R. Astron. Soc.}\ }\textbf {\bibinfo {volume} {425}},\ \bibinfo {pages}
  {405} (\bibinfo {year} {2012})}\BibitemShut {NoStop}%
\bibitem [{\citenamefont {{A. Pezzotta \emph{et al.}}}(2017)}]{Pezzotta}%
  \BibitemOpen
  \bibfield  {author} {\bibinfo {author} {\bibnamefont {{A. Pezzotta \emph{et
  al.}}}},\ }\href@noop {} {\bibfield  {journal} {\bibinfo  {journal} {Astron.
  \& Astrophys.}\ }\textbf {\bibinfo {volume} {604}},\ \bibinfo {pages} {A33}
  (\bibinfo {year} {2017})}\BibitemShut {NoStop}%
\bibitem [{\citenamefont {{T. Okumura \emph{et al.}}}(2016)}]{Okumura}%
  \BibitemOpen
  \bibfield  {author} {\bibinfo {author} {\bibnamefont {{T. Okumura \emph{et
  al.}}}},\ }\href@noop {} {\bibfield  {journal} {\bibinfo  {journal} {Publ.
  Astron. Soc. Japan}\ }\textbf {\bibinfo {volume} {68}},\ \bibinfo {pages}
  {38} (\bibinfo {year} {2016})}\BibitemShut {NoStop}%
\bibitem [{\citenamefont {Nesseris}\ \emph {et~al.}(2017)\citenamefont
  {Nesseris}, \citenamefont {Pantazis},\ and\ \citenamefont
  {Perivolaropoulos}}]{Nesseris}%
  \BibitemOpen
  \bibfield  {author} {\bibinfo {author} {\bibfnamefont {S.}~\bibnamefont
  {Nesseris}}, \bibinfo {author} {\bibfnamefont {G.}~\bibnamefont {Pantazis}},\
  and\ \bibinfo {author} {\bibfnamefont {L.}~\bibnamefont {Perivolaropoulos}},\
  }\href@noop {} {\bibfield  {journal} {\bibinfo  {journal} {Phys. Rev. D}\
  }\textbf {\bibinfo {volume} {96}},\ \bibinfo {pages} {023542} (\bibinfo
  {year} {2017})}\BibitemShut {NoStop}%
\bibitem [{\citenamefont {Kazantzidis}\ and\ \citenamefont
  {Perivolaropoulos}(2018)}]{Lavrentios}%
  \BibitemOpen
  \bibfield  {author} {\bibinfo {author} {\bibfnamefont {L.}~\bibnamefont
  {Kazantzidis}}\ and\ \bibinfo {author} {\bibfnamefont {L.}~\bibnamefont
  {Perivolaropoulos}},\ }\href@noop {} {\bibfield  {journal} {\bibinfo
  {journal} {Phys. Rev. D}\ }\textbf {\bibinfo {volume} {97}},\ \bibinfo
  {pages} {103503} (\bibinfo {year} {2018})}\BibitemShut {NoStop}%
\bibitem [{\citenamefont {Alcock}\ and\ \citenamefont
  {Paczy\'{n}ski}(1979)}]{Alcock}%
  \BibitemOpen
  \bibfield  {author} {\bibinfo {author} {\bibfnamefont {C.}~\bibnamefont
  {Alcock}}\ and\ \bibinfo {author} {\bibfnamefont {B.}~\bibnamefont
  {Paczy\'{n}ski}},\ }\href@noop {} {\bibfield  {journal} {\bibinfo  {journal}
  {Nature}\ }\textbf {\bibinfo {volume} {281}},\ \bibinfo {pages} {358}
  (\bibinfo {year} {1979})}\BibitemShut {NoStop}%
\bibitem [{\citenamefont {Macaulay}\ \emph {et~al.}(2013)\citenamefont
  {Macaulay}, \citenamefont {Wehus},\ and\ \citenamefont {Eriksen}}]{Macaulay}%
  \BibitemOpen
  \bibfield  {author} {\bibinfo {author} {\bibfnamefont {E.}~\bibnamefont
  {Macaulay}}, \bibinfo {author} {\bibfnamefont {I.~K.}\ \bibnamefont
  {Wehus}},\ and\ \bibinfo {author} {\bibfnamefont {H.~K.}\ \bibnamefont
  {Eriksen}},\ }\href@noop {} {\bibfield  {journal} {\bibinfo  {journal} {Phys.
  Rev. Lett.}\ }\textbf {\bibinfo {volume} {111}},\ \bibinfo {pages} {161301}
  (\bibinfo {year} {2013})}\BibitemShut {NoStop}%
\bibitem [{\citenamefont {Efstathiou}(2014)}]{Efstathiou2014}%
  \BibitemOpen
  \bibfield  {author} {\bibinfo {author} {\bibfnamefont {G.}~\bibnamefont
  {Efstathiou}},\ }\href {https://doi.org/10.1093/mnras/stu278} {\bibfield
  {journal} {\bibinfo  {journal} {Mon. Not. R. Astron. Soc.}\ }\textbf
  {\bibinfo {volume} {440}},\ \bibinfo {pages} {1138} (\bibinfo {year}
  {2014})}\BibitemShut {NoStop}%
\bibitem [{\citenamefont {M\"{o}rtsell}\ and\ \citenamefont
  {Dhawan}()}]{Mortsell}%
  \BibitemOpen
  \bibfield  {author} {\bibinfo {author} {\bibfnamefont {E.}~\bibnamefont
  {M\"{o}rtsell}}\ and\ \bibinfo {author} {\bibfnamefont {S.}~\bibnamefont
  {Dhawan}},\ }\href@noop {} {\bibfield  {journal} {\bibinfo  {journal} {J.
  Cosmol. Astropart. Phys.}\ }\textbf {\bibinfo {volume} {2018}}\bibinfo
  {number} { (09)},\ \bibinfo {pages} {025}}\BibitemShut {NoStop}%
\bibitem [{\citenamefont {{Di Valentino}}\ \emph {et~al.}(2018)\citenamefont
  {{Di Valentino}}, \citenamefont {{Linder}},\ and\ \citenamefont
  {Melchiorri}}]{DiValentino}%
  \BibitemOpen
\bibfield  {number} {  }\bibfield  {author} {\bibinfo {author} {\bibfnamefont
  {E.}~\bibnamefont {{Di Valentino}}}, \bibinfo {author} {\bibfnamefont
  {E.~V.}\ \bibnamefont {{Linder}}},\ and\ \bibinfo {author} {\bibfnamefont
  {A.}~\bibnamefont {Melchiorri}},\ }\href@noop {} {\bibfield  {journal}
  {\bibinfo  {journal} {Phys. Rev. D}\ }\textbf {\bibinfo {volume} {97}},\
  \bibinfo {pages} {043528} (\bibinfo {year} {2018})}\BibitemShut {NoStop}%
\bibitem [{\citenamefont {Poulin}\ \emph {et~al.}(2019)\citenamefont {Poulin},
  \citenamefont {{Smith}}, \citenamefont {Karwal},\ and\ \citenamefont
  {Kamionkowski}}]{Poulin}%
  \BibitemOpen
  \bibfield  {author} {\bibinfo {author} {\bibfnamefont {V.}~\bibnamefont
  {Poulin}}, \bibinfo {author} {\bibfnamefont {T.~L.}\ \bibnamefont {{Smith}}},
  \bibinfo {author} {\bibfnamefont {T.}~\bibnamefont {Karwal}},\ and\ \bibinfo
  {author} {\bibfnamefont {M.}~\bibnamefont {Kamionkowski}},\ }\href@noop {}
  {\bibfield  {journal} {\bibinfo  {journal} {Phys. Rev. Lett.}\ }\textbf
  {\bibinfo {volume} {122}},\ \bibinfo {pages} {221301} (\bibinfo {year}
  {2019})}\BibitemShut {NoStop}%
\bibitem [{\citenamefont {{Sol\`{a} Peracaula}}(2018)}]{SolaPeracaula}%
  \BibitemOpen
  \bibfield  {author} {\bibinfo {author} {\bibfnamefont {J.}~\bibnamefont
  {{Sol\`{a} Peracaula}}},\ }\href@noop {} {\bibfield  {journal} {\bibinfo
  {journal} {Int. J. Mod. Phys. D}\ }\textbf {\bibinfo {volume} {27}},\
  \bibinfo {pages} {1847029} (\bibinfo {year} {2018})}\BibitemShut {NoStop}%
\bibitem [{\citenamefont {{Sol\`{a} Peracaula}}\ \emph
  {et~al.}(2019)\citenamefont {{Sol\`{a} Peracaula}}, \citenamefont
  {G\'{o}mez-Valent}, \citenamefont {{de Cruz P\'{e}rez}},\ and\ \citenamefont
  {Moreno-Pulido}}]{Peracaula}%
  \BibitemOpen
  \bibfield  {author} {\bibinfo {author} {\bibfnamefont {J.}~\bibnamefont
  {{Sol\`{a} Peracaula}}}, \bibinfo {author} {\bibfnamefont {A.}~\bibnamefont
  {G\'{o}mez-Valent}}, \bibinfo {author} {\bibfnamefont {J.}~\bibnamefont {{de
  Cruz P\'{e}rez}}},\ and\ \bibinfo {author} {\bibfnamefont {C.}~\bibnamefont
  {Moreno-Pulido}},\ }\href@noop {} {\bibfield  {journal} {\bibinfo  {journal}
  {Astrophys. J. Lett.}\ }\textbf {\bibinfo {volume} {886}},\ \bibinfo {pages}
  {L6} (\bibinfo {year} {2019})}\BibitemShut {NoStop}%
\bibitem [{\citenamefont {{de Cruz P\'{e}rez}}\ and\ \citenamefont {{Sol\`{a}
  Peracaula}}(2018)}]{CruzPerez}%
  \BibitemOpen
  \bibfield  {author} {\bibinfo {author} {\bibfnamefont {J.}~\bibnamefont {{de
  Cruz P\'{e}rez}}}\ and\ \bibinfo {author} {\bibfnamefont {J.}~\bibnamefont
  {{Sol\`{a} Peracaula}}},\ }\href@noop {} {\bibfield  {journal} {\bibinfo
  {journal} {Mod. Phys. Lett. A}\ }\textbf {\bibinfo {volume} {33}},\ \bibinfo
  {pages} {1850228} (\bibinfo {year} {2018})}\BibitemShut {NoStop}%
\bibitem [{\citenamefont {Mangano}\ \emph {et~al.}(2005)\citenamefont
  {Mangano}, \citenamefont {Miele}, \citenamefont {Pastor}, \citenamefont
  {Pinto}, \citenamefont {Pisanti},\ and\ \citenamefont
  {Serpico}}]{Mangano2005}%
  \BibitemOpen
  \bibfield  {author} {\bibinfo {author} {\bibfnamefont {G.}~\bibnamefont
  {Mangano}}, \bibinfo {author} {\bibfnamefont {G.}~\bibnamefont {Miele}},
  \bibinfo {author} {\bibfnamefont {S.}~\bibnamefont {Pastor}}, \bibinfo
  {author} {\bibfnamefont {T.}~\bibnamefont {Pinto}}, \bibinfo {author}
  {\bibfnamefont {O.}~\bibnamefont {Pisanti}},\ and\ \bibinfo {author}
  {\bibfnamefont {P.~D.}\ \bibnamefont {Serpico}},\ }\href
  {https://doi.org/10.1016/j.nuclphysb.2005.09.041} {\bibfield  {journal}
  {\bibinfo  {journal} {Nucl. Phys. B}\ }\textbf {\bibinfo {volume} {729}},\
  \bibinfo {pages} {221} (\bibinfo {year} {2005})}\BibitemShut {NoStop}%
\bibitem [{\citenamefont {Fixsen}(2009)}]{Fixsen2009}%
  \BibitemOpen
  \bibfield  {author} {\bibinfo {author} {\bibfnamefont {D.~J.}\ \bibnamefont
  {Fixsen}},\ }\href {https://doi.org/10.1088/0004-637X/707/2/916} {\bibfield
  {journal} {\bibinfo  {journal} {Astrophys. J.}\ }\textbf {\bibinfo {volume}
  {707}},\ \bibinfo {pages} {916} (\bibinfo {year} {2009})}\BibitemShut
  {NoStop}%
\bibitem [{\citenamefont {Avgoustidis}\ \emph {et~al.}()\citenamefont
  {Avgoustidis}, \citenamefont {Luzzi}, \citenamefont {Martins},\ and\
  \citenamefont {Monteiro}}]{Avg2012}%
  \BibitemOpen
  \bibfield  {author} {\bibinfo {author} {\bibfnamefont {A.}~\bibnamefont
  {Avgoustidis}}, \bibinfo {author} {\bibfnamefont {G.}~\bibnamefont {Luzzi}},
  \bibinfo {author} {\bibfnamefont {C.}~\bibnamefont {Martins}},\ and\ \bibinfo
  {author} {\bibfnamefont {A.}~\bibnamefont {Monteiro}},\ }\href
  {https://doi.org/10.1088/1475-7516/2012/02/013} {\bibfield  {journal}
  {\bibinfo  {journal} {J. Cosmol. Astropart. Phys.}\ }\textbf {\bibinfo
  {volume} {2012}}\bibinfo  {number} { (02)},\ \bibinfo {pages}
  {013}}\BibitemShut {NoStop}%
\bibitem [{\citenamefont {Luzzi}\ \emph {et~al.}()\citenamefont {Luzzi},
  \citenamefont {G\'{e}nova-Santos}, \citenamefont {Martins}, \citenamefont
  {Petris},\ and\ \citenamefont {Lamagna}}]{Luz2015}%
  \BibitemOpen
\bibfield  {number} {  }\bibfield  {author} {\bibinfo {author} {\bibfnamefont
  {G.}~\bibnamefont {Luzzi}}, \bibinfo {author} {\bibfnamefont
  {R.}~\bibnamefont {G\'{e}nova-Santos}}, \bibinfo {author} {\bibfnamefont
  {C.}~\bibnamefont {Martins}}, \bibinfo {author} {\bibfnamefont {M.~D.}\
  \bibnamefont {Petris}},\ and\ \bibinfo {author} {\bibfnamefont
  {L.}~\bibnamefont {Lamagna}},\ }\href
  {https://doi.org/10.1088/1475-7516/2015/09/011} {\bibfield  {journal}
  {\bibinfo  {journal} {J. Cosmol. Astropart. Phys.}\ }\textbf {\bibinfo
  {volume} {2015}}\bibinfo  {number} { (09)},\ \bibinfo {pages}
  {011}}\BibitemShut {NoStop}%
\bibitem [{\citenamefont {Noterdaeme}\ \emph {et~al.}(2011)\citenamefont
  {Noterdaeme}, \citenamefont {Petitjean}, \citenamefont {Srianand},
  \citenamefont {Ledoux},\ and\ \citenamefont {L{\'{o}}pez}}]{Noterdaeme2011}%
  \BibitemOpen
\bibfield  {number} {  }\bibfield  {author} {\bibinfo {author} {\bibfnamefont
  {P.}~\bibnamefont {Noterdaeme}}, \bibinfo {author} {\bibfnamefont
  {P.}~\bibnamefont {Petitjean}}, \bibinfo {author} {\bibfnamefont
  {R.}~\bibnamefont {Srianand}}, \bibinfo {author} {\bibfnamefont
  {C.}~\bibnamefont {Ledoux}},\ and\ \bibinfo {author} {\bibfnamefont
  {S.}~\bibnamefont {L{\'{o}}pez}},\ }\href
  {https://doi.org/10.1051/0004-6361/201016140} {\bibfield  {journal} {\bibinfo
   {journal} {Astron. \& Astrophys.}\ }\textbf {\bibinfo {volume} {526}},\
  \bibinfo {pages} {L7} (\bibinfo {year} {2011})}\BibitemShut {NoStop}%
\bibitem [{\citenamefont {Avgoustidis}\ \emph {et~al.}(2016)\citenamefont
  {Avgoustidis}, \citenamefont {G\'{e}nova-Santos}, \citenamefont {Luzzi},\
  and\ \citenamefont {Martins}}]{Avgoustidis2016}%
  \BibitemOpen
  \bibfield  {author} {\bibinfo {author} {\bibfnamefont {A.}~\bibnamefont
  {Avgoustidis}}, \bibinfo {author} {\bibfnamefont {R.~T.}\ \bibnamefont
  {G\'{e}nova-Santos}}, \bibinfo {author} {\bibfnamefont {G.}~\bibnamefont
  {Luzzi}},\ and\ \bibinfo {author} {\bibfnamefont {C.~J. A.~P.}\ \bibnamefont
  {Martins}},\ }\href {https://doi.org/10.1103/PhysRevD.93.043521} {\bibfield
  {journal} {\bibinfo  {journal} {Phys. Rev. D}\ }\textbf {\bibinfo {volume}
  {93}},\ \bibinfo {pages} {043521} (\bibinfo {year} {2016})}\BibitemShut
  {NoStop}%
\bibitem [{\citenamefont {Lin}\ \emph {et~al.}(2018)\citenamefont {Lin},
  \citenamefont {Li},\ and\ \citenamefont {Li}}]{Lin2018}%
  \BibitemOpen
  \bibfield  {author} {\bibinfo {author} {\bibfnamefont {H.-N.}\ \bibnamefont
  {Lin}}, \bibinfo {author} {\bibfnamefont {M.-H.}\ \bibnamefont {Li}},\ and\
  \bibinfo {author} {\bibfnamefont {X.}~\bibnamefont {Li}},\ }\href
  {https://doi.org/10.1093/mnras/sty2062} {\bibfield  {journal} {\bibinfo
  {journal} {Mon. Not. R. Astron. Soc.}\ }\textbf {\bibinfo {volume} {480}},\
  \bibinfo {pages} {3117} (\bibinfo {year} {2018})}\BibitemShut {NoStop}%
\bibitem [{\citenamefont {{De Bernardis}}\ \emph {et~al.}(2006)\citenamefont
  {{De Bernardis}}, \citenamefont {Giusarma},\ and\ \citenamefont
  {Melchiorri}}]{Debernardis2006}%
  \BibitemOpen
  \bibfield  {author} {\bibinfo {author} {\bibfnamefont {F.}~\bibnamefont {{De
  Bernardis}}}, \bibinfo {author} {\bibfnamefont {E.}~\bibnamefont
  {Giusarma}},\ and\ \bibinfo {author} {\bibfnamefont {A.}~\bibnamefont
  {Melchiorri}},\ }\href {https://doi.org/10.1142/S0218271806008486} {\bibfield
   {journal} {\bibinfo  {journal} {Int. J. Mod. Phys. D}\ }\textbf {\bibinfo
  {volume} {15}},\ \bibinfo {pages} {759} (\bibinfo {year} {2006})}\BibitemShut
  {NoStop}%
\bibitem [{\citenamefont {Holanda}\ \emph {et~al.}()\citenamefont {Holanda},
  \citenamefont {Busti},\ and\ \citenamefont {Alcaniz}}]{Holanda2016}%
  \BibitemOpen
  \bibfield  {author} {\bibinfo {author} {\bibfnamefont {R.}~\bibnamefont
  {Holanda}}, \bibinfo {author} {\bibfnamefont {V.}~\bibnamefont {Busti}},\
  and\ \bibinfo {author} {\bibfnamefont {J.}~\bibnamefont {Alcaniz}},\ }\href
  {https://doi.org/10.1088/1475-7516/2016/02/054} {\bibfield  {journal}
  {\bibinfo  {journal} {J. Cosmol. Astropart. Phys.}\ }\textbf {\bibinfo
  {volume} {2016}}\bibinfo  {number} { (02)},\ \bibinfo {pages}
  {054}}\BibitemShut {NoStop}%
\bibitem [{\citenamefont {Ma}\ and\ \citenamefont {Corasaniti}(2018)}]{Ma2018}%
  \BibitemOpen
\bibfield  {number} {  }\bibfield  {author} {\bibinfo {author} {\bibfnamefont
  {C.}~\bibnamefont {Ma}}\ and\ \bibinfo {author} {\bibfnamefont {P.-S.}\
  \bibnamefont {Corasaniti}},\ }\href
  {https://doi.org/10.3847/1538-4357/aac88f} {\bibfield  {journal} {\bibinfo
  {journal} {Astrophys. J.}\ }\textbf {\bibinfo {volume} {861}},\ \bibinfo
  {pages} {124} (\bibinfo {year} {2018})}\BibitemShut {NoStop}%
\bibitem [{\citenamefont {McGaugh}(2004)}]{McGaugh2004}%
  \BibitemOpen
  \bibfield  {author} {\bibinfo {author} {\bibfnamefont {S.~S.}\ \bibnamefont
  {McGaugh}},\ }\href {https://doi.org/10.1086/421895} {\bibfield  {journal}
  {\bibinfo  {journal} {Astrophys. J.}\ }\textbf {\bibinfo {volume} {611}},\
  \bibinfo {pages} {26} (\bibinfo {year} {2004})}\BibitemShut {NoStop}%
\bibitem [{\citenamefont {McGaugh}(2015)}]{McGaugh2015}%
  \BibitemOpen
  \bibfield  {author} {\bibinfo {author} {\bibfnamefont {S.~S.}\ \bibnamefont
  {McGaugh}},\ }\href {https://doi.org/10.1139/cjp-2014-0203} {\bibfield
  {journal} {\bibinfo  {journal} {Can. J. Phys.}\ }\textbf {\bibinfo {volume}
  {93}},\ \bibinfo {pages} {250} (\bibinfo {year} {2015})}\BibitemShut
  {NoStop}%
\bibitem [{\citenamefont {Douspis}\ \emph {et~al.}(2018)\citenamefont
  {Douspis}, \citenamefont {Salvati},\ and\ \citenamefont {Aghanim}}]{Douspis}%
  \BibitemOpen
  \bibfield  {author} {\bibinfo {author} {\bibfnamefont {M.}~\bibnamefont
  {Douspis}}, \bibinfo {author} {\bibfnamefont {L.}~\bibnamefont {Salvati}},\
  and\ \bibinfo {author} {\bibfnamefont {N.}~\bibnamefont {Aghanim}},\
  }\href@noop {} {\bibfield  {journal} {\bibinfo  {journal} {PoS Proc. Sci.}\
  }\textbf {\bibinfo {volume} {335}},\ \bibinfo {pages} {037} (\bibinfo {year}
  {2018})}\BibitemShut {NoStop}%
\bibitem [{\citenamefont {Akaike}(1974)}]{Akaike1974}%
  \BibitemOpen
  \bibfield  {author} {\bibinfo {author} {\bibfnamefont {H.}~\bibnamefont
  {Akaike}},\ }\href {https://doi.org/10.1109/TAC.1974.1100705} {\bibfield
  {journal} {\bibinfo  {journal} {IEEE Trans. Automat. Contr.}\ }\textbf
  {\bibinfo {volume} {19}},\ \bibinfo {pages} {716} (\bibinfo {year}
  {1974})}\BibitemShut {NoStop}%
\bibitem [{\citenamefont {Schwarz}(1978)}]{Schwarz1978}%
  \BibitemOpen
  \bibfield  {author} {\bibinfo {author} {\bibfnamefont {G.}~\bibnamefont
  {Schwarz}},\ }\href {https://doi.org/10.1214/aos/1176344136} {\bibfield
  {journal} {\bibinfo  {journal} {Ann. Stat.}\ }\textbf {\bibinfo {volume}
  {6}},\ \bibinfo {pages} {461} (\bibinfo {year} {1978})}\BibitemShut {NoStop}%
\bibitem [{\citenamefont {Schwarz}(2011)}]{Schwarz2011}%
  \BibitemOpen
  \bibfield  {author} {\bibinfo {author} {\bibfnamefont {C.~J.}\ \bibnamefont
  {Schwarz}},\ }\href@noop {} {\bibinfo {title} {{Sampling, Regression,
  Experimental Design and Analysis for Environmental Scientists, Biologists,
  and Resource Managers}}},\ \bibinfo {howpublished}
  {\url{http://people.stat.sfu.ca/\%7Ecschwarz/Stat-650/Notes/PDF/MLE-AIC.pdf}}
  (\bibinfo {year} {2011}),\ \bibinfo {note} {{date accessed: 23 Nov.
  2018}}\BibitemShut {NoStop}%
\bibitem [{\citenamefont {Link}\ and\ \citenamefont {Barker}(2010)}]{Link2010}%
  \BibitemOpen
  \bibfield  {author} {\bibinfo {author} {\bibfnamefont {W.~A.}\ \bibnamefont
  {Link}}\ and\ \bibinfo {author} {\bibfnamefont {R.~J.}\ \bibnamefont
  {Barker}},\ }\href
  {https://books.google.com.mt/books?redir_esc=y&id=hecon2l2QPcC&q=BIC#v=snippet&q=BIC&f=false}
  {\emph {\bibinfo {title} {{Bayesian Inference: With Ecological
  Applications}}}}\ (\bibinfo  {publisher} {Academic Press},\ \bibinfo
  {address} {London},\ \bibinfo {year} {2010})\BibitemShut {NoStop}%
\bibitem [{\citenamefont {Kuha}(2004)}]{Kuha2004}%
  \BibitemOpen
  \bibfield  {author} {\bibinfo {author} {\bibfnamefont {J.}~\bibnamefont
  {Kuha}},\ }\href {https://doi.org/10.1177/0049124103262065} {\bibfield
  {journal} {\bibinfo  {journal} {Sociol. Methods \& Res.}\ }\textbf {\bibinfo
  {volume} {33}},\ \bibinfo {pages} {188} (\bibinfo {year} {2004})}\BibitemShut
  {NoStop}%
\bibitem [{\citenamefont {Liddle}(2007)}]{Liddle2007}%
  \BibitemOpen
  \bibfield  {author} {\bibinfo {author} {\bibfnamefont {A.~R.}\ \bibnamefont
  {Liddle}},\ }\href {https://doi.org/10.1111/j.1745-3933.2007.00306.x}
  {\bibfield  {journal} {\bibinfo  {journal} {Mon. Not. R. Astron. Soc.:
  Lett.}\ }\textbf {\bibinfo {volume} {377}},\ \bibinfo {pages} {L74} (\bibinfo
  {year} {2007})}\BibitemShut {NoStop}%
\bibitem [{\citenamefont {Jordanger}\ and\ \citenamefont
  {Tj{\o}stheim}(2014)}]{Jordanger2014}%
  \BibitemOpen
  \bibfield  {author} {\bibinfo {author} {\bibfnamefont {L.~A.}\ \bibnamefont
  {Jordanger}}\ and\ \bibinfo {author} {\bibfnamefont {D.}~\bibnamefont
  {Tj{\o}stheim}},\ }\href {https://doi.org/10.1016/j.spl.2014.06.006}
  {\bibfield  {journal} {\bibinfo  {journal} {Stat. \& Probab. Lett.}\ }\textbf
  {\bibinfo {volume} {92}},\ \bibinfo {pages} {249} (\bibinfo {year}
  {2014})}\BibitemShut {NoStop}%
\bibitem [{\citenamefont {Nesseris}\ \emph {et~al.}(2010)\citenamefont
  {Nesseris}, \citenamefont {{De Felice}},\ and\ \citenamefont
  {Tsujikawa}}]{Nesseris2010}%
  \BibitemOpen
  \bibfield  {author} {\bibinfo {author} {\bibfnamefont {S.}~\bibnamefont
  {Nesseris}}, \bibinfo {author} {\bibfnamefont {A.}~\bibnamefont {{De
  Felice}}},\ and\ \bibinfo {author} {\bibfnamefont {S.}~\bibnamefont
  {Tsujikawa}},\ }\href {https://doi.org/10.1103/PhysRevD.82.124054} {\bibfield
   {journal} {\bibinfo  {journal} {Phys. Rev. D}\ }\textbf {\bibinfo {volume}
  {82}},\ \bibinfo {pages} {124054} (\bibinfo {year} {2010})}\BibitemShut
  {NoStop}%
\bibitem [{\citenamefont {Liddle}(2004)}]{Liddle2004}%
  \BibitemOpen
  \bibfield  {author} {\bibinfo {author} {\bibfnamefont {A.~R.}\ \bibnamefont
  {Liddle}},\ }\href {https://doi.org/10.1111/j.1365-2966.2004.08033.x}
  {\bibfield  {journal} {\bibinfo  {journal} {Mon. Not. R. Astron. Soc.}\
  }\textbf {\bibinfo {volume} {351}},\ \bibinfo {pages} {L49} (\bibinfo {year}
  {2004})}\BibitemShut {NoStop}%
\bibitem [{\citenamefont {Clarkson}\ \emph {et~al.}()\citenamefont {Clarkson},
  \citenamefont {Cort\^{e}s},\ and\ \citenamefont {Bassett}}]{Clarkson2007}%
  \BibitemOpen
  \bibfield  {author} {\bibinfo {author} {\bibfnamefont {C.}~\bibnamefont
  {Clarkson}}, \bibinfo {author} {\bibfnamefont {M.}~\bibnamefont
  {Cort\^{e}s}},\ and\ \bibinfo {author} {\bibfnamefont {B.}~\bibnamefont
  {Bassett}},\ }\href {https://doi.org/10.1088/1475-7516/2007/08/011}
  {\bibfield  {journal} {\bibinfo  {journal} {J. Cosmol. Astropart. Phys.}\
  }\textbf {\bibinfo {volume} {2007}}\bibinfo  {number} { (08)},\ \bibinfo
  {pages} {011}}\BibitemShut {NoStop}%
\bibitem [{\citenamefont {Virey}\ \emph {et~al.}()\citenamefont {Virey},
  \citenamefont {Talon-Esmieu}, \citenamefont {Ealet}, \citenamefont {Taxil},\
  and\ \citenamefont {Tilquin}}]{Virey2008}%
  \BibitemOpen
\bibfield  {number} {  }\bibfield  {author} {\bibinfo {author} {\bibfnamefont
  {J.-M.}\ \bibnamefont {Virey}}, \bibinfo {author} {\bibfnamefont
  {D.}~\bibnamefont {Talon-Esmieu}}, \bibinfo {author} {\bibfnamefont
  {A.}~\bibnamefont {Ealet}}, \bibinfo {author} {\bibfnamefont
  {P.}~\bibnamefont {Taxil}},\ and\ \bibinfo {author} {\bibfnamefont
  {A.}~\bibnamefont {Tilquin}},\ }\href
  {https://doi.org/10.1088/1475-7516/2008/12/008} {\bibfield  {journal}
  {\bibinfo  {journal} {J. Cosmol. Astropart. Phys.}\ }\textbf {\bibinfo
  {volume} {2008}}\bibinfo  {number} { (12)},\ \bibinfo {pages}
  {008}}\BibitemShut {NoStop}%
\end{thebibliography}%

\end{document}